\documentclass[12pt]{JHEP3}

\usepackage{amsmath,epsfig}
\usepackage{amssymb,amsfonts}
\usepackage{latexsym}
\usepackage[latin1]{inputenc}
\usepackage{slashed}
\usepackage{empheq}
\numberwithin{equation}{section}
\usepackage{dsfont}
\usepackage{cancel}
\usepackage{overpic}
\usepackage{subcaption}
\usepackage{subeqnarray}
\usepackage{xcolor}
\let\normalcolor\relax
\usepackage{graphicx}
\usepackage{longtable}
\usepackage{color}
\usepackage{bbm}

\newcommand{\exclude}[1]{}
\def\L{\mathcal{L}}

\def\<{\langle}
\def\>{\rangle}

\def\a#1{\alpha_{#1}}

\def\beq{\begin{equation}}
\def\eeq{\end{equation}}
\def\be{\begin{equation}}
\def\ee{\end{equation}}
\def\bea{\begin{eqnarray}}
\def\eea{\end{eqnarray}}
\def\bal{\begin{align}}
\def\eal{\end{align}}

\def\2b2[#1,#2][#3,#4]{\left( \begin{array}{cc} #1 & #2 \\ #3 & #4 \end{array}
\right)}
\def\3b3[#1,#2,#3][#4,#5,#6][#7,#8,#9]{\left( \begin{array}{ccc} #1 & #2 #3 \\
#4 & #5 & #6\\#7&#8&#9\end{array} \right)}

\newcommand\fverb{\setbox\pippobox=\hbox\bgroup\verb}
\newcommand\fverbdo{\egroup\medskip\noindent%
                        \fbox{\unhbox\pippobox}\ }
\newcommand\fverbit{\egroup\item[\fbox{\unhbox\pippobox}]}

\newcommand{\bear}{\begin{eqnarray}}

\newcommand{\eear}{\end{eqnarray}}

\newcommand{\bsea}{\begin{subeqnarray}}
\newcommand{\esea}{\end{subeqnarray}}
\newbox\pippobox

\def\f{\varphi}

\def\6{\partial}
\def\a{\alpha}

\def\m{\mu}
\def\n{\nu}

\def\sq
\def\a{\alpha}

\def\hri#1#2{\href{http://arxiv.org/abs/#1}{[ArXiv:#1]#2}}
\def\hre#1#2{\href{http://arxiv.org/abs/#1/#2}{[ArXiv:#1/#2]}}

\def\hree#1#2{\href{https://doi.org/#1}{#2}}

\def\L{\Lambda}

\title{Back-reaction in massless de Sitter QFTs: holography, gravitational DBI action and f(R) gravity}

\author{Jewel K.~Ghosh$^\star$, Elias Kiritsis$^\natural$$^\flat$, Francesco Nitti$^\natural$, Lukas T.~Witkowski$^\dagger$
~\\
$^\star$ \href{https://www.icts.res.in/}{International Centre for Theoretical Sciences}, Tata Institute of Fundamental Research, Shivakote, Bengaluru 560089, India\\
~\\
$^\natural$ \href{http://www.apc.univ-paris7.fr}{APC, AstroParticule et Cosmologie}, Universit\'e de Paris, CNRS/IN2P3, CEA/IRFU,
Observatoire de Paris,\\
 10, rue Alice Domon et L\'eonie Duquet, 75205 Paris
Cedex 13, France\\
~\\
$^\flat$ \href{http://hep.physics.uoc.gr}{Crete Center for Theoretical Physics}, Institute for Theoretical and Computational Physics,
Department of Physics, Voutes University Campus,\\
GR-70013, Vasilika Vouton, Heraklion, GREECE\\
~\\
$^\dagger$ \href{http://www.iap.fr/}{Institut d'Astrophysique de Paris}, GReCO, UMR 7095 du CNRS et de Sorbonne Universit\'e, 98bis boulevard Arago, 75014 Paris, France
}

\preprint{CCTP-2020-3\\
ITCP-IPP 2020/3}

\abstract{We employ gauge-gravity duality to study the backreaction
  effect of 4-dimensional large-$N$ quantum field theories on
  constant-curvature backgrounds, and in particular de Sitter space-time. The field theories considered are holographic QFTs, dual to RG flows between UV and IR CFTs. We compute the holographic QFT contribution to the gravitational effective action for 4d Einstein manifold backgrounds. We find that for a given value of the cosmological constant $\lambda$, there generically exist two backreacted constant-curvature solutions, as long as $\lambda < \lambda_{\textrm{max}} \sim M_p^2 / N^2$, otherwise no such solutions exist.
Moreover, the backreaction effect  interpolates between that of the UV and IR CFTs.
We also find that, at finite cutoff, a holographic theory always reduces the bare cosmological constant, and this is the consequence of thermodynamic properties of the partition function of holographic QFTs on de Sitter.
}

\begin{document}
\maketitle

\section{Introduction and summary}

Defining  a quantum field theory (QFT) on de Sitter space, and in particular
answering questions about QFT backreaction on the geometry, is a notoriously
subtle issue. Although perturbative field quantization (and
renormalization) on fixed classical curved space-times is text-book material,
answering concrete questions about observable effects of quantum
fields  backreaction on the classical geometry is not
straightforward.  This concerns particularly theories which are gapless
in the infrared, due to the presence of infra-red divergences.

There are many different issues with QFTs on de Sitter or approximately de Sitter backgrounds relevant for inflation.
In  \cite{AIT,AM,Sasa,TW2} a divergence of scalar correlators was found at large times.
This was addressed in \cite{Staro,STW,GS} using a stochastic approach. The case of interacting massive scalars has been treated thoroughly more recently in \cite{Marolf,Hollands2}.  A systematic  approach  to compute corrections in the massless case is lacking and  the problem remains still open.

The accumulation of long wavelength fluctuations in an expanding universe
is another issue that  is being studied,  starting with \cite{Mukha,AWoo}. This analysis was extended further in \cite{unruh}.
The expectation that QFTs in de Sitter render the
manifold unstable has been entertained since a long time, \cite{Mottola,TW,Polyakov}. In particular, destabilizing effects were most important from massless particles, and a gravity two-loop computation in \cite{TW} suggested such an instability. Similar calculations with massless scalars suggested similar effects, \cite{Mukha,AWoo}.
Another issue concerned the fact that the two-point function of a scalar in de Sitter had to break de Sitter symmetry, due to the presence of a zero mode, \cite{MM,BAllen}. This issue, however, is of a different nature and is more similar to the fact that in two dimensions a massless scalar is IR singular.
The resolution of this issue may be therefore similar: massless scalars are not good acceptable fields on de Sitter space\footnote{We will later see in this paper that in all the theories we examine and which are all gapless, there is no breaking of de Sitter invariance, and the de Sitter invariant vacuum is chosen.} as argued in \cite{Hollands1}.

The topic of quantum effects was revived after cosmological (CMB) data became precise enough, and in \cite{Weinberg,Cha,Sloth} it was argued that large time-dependent logs from quantum effects of spectator fields may give large corrections to inflationary observables. A different approach in \cite{Sena} had on the other hand provided different results.
Therefore, a still controversial question concerns the consequences  of the
secular terms (growing with time) which arise in perturbation theory
of a massless scalar field in the cosmological patch of de Sitter: do
these contributions  indicate an instability of de
Sitter space against quantum perturbations? Or is this conclusion an artifact of
 finite orders in perturbation theory, which is expected to disappear  once an
 appropriate resummation is performed (as it is the case for infra-red
 effects in thermal perturbation theory)?
A review of these developments and additional references can be found in \cite{Seery}.

The question of QFT backreaction on de Sitter space can be cast in the
language of an effective action for gravity. Suppose we couple a QFT
to a classical background metric.  Generically, integrating out the QFT will generate new terms in the effective action for
the (classical) metric. This will generically change the ``bare''
gravity   theory to an effective theory,
\be \label{intro0}
S
[g] = S_0[g] \, + \, \log \, Z_{\textrm{QFT}}[g] \, ,
\ee
where $g$ denotes the metric,  $S_0[g]$ is the action for gravity in
the absence of the QFT, and  $Z_{\textrm{QFT}}[g]$ is the quantum partition
function of the  field theory coupled to the metric $g$. The
effective action (\ref{intro0}) will generically be a  complicated (and non-polynomial) function of curvature
invariants $R$, $R_{\mu\nu}R^{\mu\nu}$, etc.

A significant simplification  arises if we are interested only in {\em covariantly constant
curvature} backgrounds, as we will be in this work.
Such backgrounds have a covariantly constant Ricci tensor
\be
\nabla_{\rho}R_{\m\n}=0 \, ,
\ee
and the Bianchi identity implies that the scalar curvature is constant.
In that case, the metric is Einstein
\be
R_{\m\n}=\kappa ~g_{\m\n} \, ,
\label{einst} \ee
with $\kappa$ constant.
This class of metrics contains also the maximally symmetric ones, namely the Minkowski, de Sitter (dS) and Anti-de-Sitter (AdS) metrics.
For such metrics it can be shown, \cite{R,F}, that the
(quantum) contribution from the QFT action, $Z_{\textrm{QFT}}[g]$ in the infinite  coupling and large-$N$ limit, will be only  a function of the Ricci
scalar.
In a sense, the constant scalar curvature metrics are the analogue of constant gauge field strengths in the case of abelian DBI actions in string theory, \cite{book}. In that case the DBI action is considered to be the result of integrating out open strings in constant background field strengths, as well as other bulk fields, like the metric etc.
Similarly, here  $Z_{\textrm{QFT}}[g]$ is the result of integrating out the QFT defined on background metrics that are Einstein as in  (\ref{einst}).

Therefore, the full action will take the form of a so-called {\em
  $f(R)$-theory of gravity}:
\be \label{intro1}
S
[g] = \int d^d x \sqrt{-g} \, f(R) \, ,
\ee
where $f$ is a function (to be determined) containing both the
``bare'' gravitational Lagrangian and the QFT contribution. Theories of $f(R)$ gravity have been widely
considered in the modified gravity literature, \cite{fR}. They propagate two tensor degrees of
freedom plus an additional scalar mode.\footnote{With a change of
variables, the action (\ref{intro1})  can be rewritten as Einstein
gravity coupled to a scalar field with a specific potential whose form
depends on the function $f$. See appendix \ref{app:constRfromFofR} for details.}
In much of the literature, the form of the function $f(R)$ has
been chosen rather arbitrarily, and without any  compelling guiding
principle. Notable exceptions are the original Starobinsky model\footnote{Note that this model is different from what is today called Starobinsky inflation, driven by $R^2$ terms.}
\cite{St}, where the form of $f(R)$ is determined by the Weyl anomaly
of the QFT, at least if the QFT is a conformal theory, and its holographic generalization, \cite{BB}.

Unlike $f(R)$ gravity theories described in the literature, \cite{fR}, here the function $f(R)$ is not arbitrary, but uniquely determined\footnote{Up to the usual scheme dependence originating from renormalization.}  by the renormalized parameters of the gravity action $S_0$ and the QFT: The $f(R)$-modification will come from the quantum effects of a non-gravitational QFT that is coupled to the dynamical metric. Thus, if one can integrate out the
QFT exactly, this will determine the form
of $Z_{\textrm{QFT}}[g]$ and therefore $f(R)$. One will then be able to determine, for
example, the fate of the de  Sitter solutions of the ``bare''  theory:
do they still exist after QFT backreaction is included ? If yes, how is the de
Sitter Hubble parameter modified due to the presence of the QFT?

Note that for a general QFT, $Z_{\textrm{QFT}}[g]$ will be a non-local functional of the (Riemann) curvature tensor and its covariant derivatives. If the QFT is gapped, then for length scales longer that the inverse gap, $Z_{\textrm{QFT}}[g]$ has a well-defined derivative expansion in the Riemann tensor and its covariant derivatives. For gapless theories such an expansion does not exist.
Despite this, evaluating $Z_{\textrm{QFT}}[g]$ in constant-curvature (Einstein) metrics
is enough if we are after a constant-curvature solution to the quantum corrected equations of motion, like that of dS space. 
In such a case $Z_{\textrm{QFT}}[g] \to f(R)$ and the strategy then is to compute $f(R)$ by integrating out the quantum fields keeping the metric fixed but arbitrary, and then extremizing the effective action to find the solution for the  metric.

This procedure has been carried out in the past in special cases. The first is the calculation of $Z_{\textrm{QFT}}[g]$ when the QFT is that of a free massive scalar field quadratically coupled to the background curvature scalar, \cite{MM}.
In such a case, the calculation is equivalent to computing the determinant
of the massive scalar Laplacian in de Sitter space. In \cite{MM} the
dS extremum was found to be unstable\footnote{This calculation is not conclusive as the constant-curvature effective action cannot reliably capture the physics of fluctuations generically.} and a breaking of the dS symmetry
was advocated. More recently, a similar calculation was carried out for the case of $\phi^4$-theory minimally coupled to Einstein gravity with a positive cosmological constant $\lambda$, \cite{julien}. There, the tool used to integrate out the QFT is the truncated exact Renormalization Group (RG), with a scalar mass term as RG scale and infrared cutoff.  Starting with the ``quantum-corrected classical'' de Sitter solution with $R \approx 4 \lambda$ at a given RG scale, it was studied how this solution is affected once super-horizon modes are backreacted on the geometry as the infrared cutoff is lowered. The observation in \cite{julien} is that while the backreaction of super-horizon modes acts to lower $R$, the de Sitter solution persists even as the infrared cutoff is taken all the way to zero.

In this paper, we  consider induced $f(R)$ theories that arise from integrating out a holographic QFT, using the techniques of gauge/gravity duality.
Such theories are large-$N$, infinitely  strongly-coupled theories. 
As such they have a gravitational dual description in terms of a gravity theory in $d+1$ space-time dimensions (that we shall henceforth call the bulk).
 As we shall see, this is a setup where the function
$f(R)$ can be fully  determined via a semiclassical (holographic) calculation of
the QFT partition function once the two-derivative bulk gravitational action is known.

The holographic gauge/gravity duality relates the $d$-dimensional QFT to a
higher-dimensional gravitational theory. This theory in the (holographic) limit
of large $N$, and strong coupling, can
be treated semiclassically.\footnote{We will work in the holographic limit throughout this work.} The QFT can be coupled to an arbitrary
background metric by imposing suitable boundary conditions on the
metric of the higher dimensional dual space-time. In particular, we
can take the field theory to  live on a constant-curvature space-time.

In  the special case of a constant-curvature metric, which is the case
of interest in this work, the holographic calculation of the QFT
partition function can be performed using recent results about curved space
RG flows \cite{R,F}. Concretely,  the QFT partition function is
computed by the gravity-dual on-shell action for an asymptotically
AdS$_{d+1}$ solution whose radial slices have a de Sitter
geometry. 
We will focus in
particular on two types of holographic theories:
\begin{enumerate}
\item Theories at a conformal fixed point, i.e.~Conformal Field Theories (CFTs);
\item Renormalization group flows driven by a relevant operator,  from a UV to an IR fixed point. Such QFTs are characterized by a single mass scale given by the relevant coupling.
\end{enumerate}
In the former case,  the bulk metric, dual to the ground state of the CFT, is simply AdS$_{d+1}$ written in a
de Sitter radial slicing. The effective action for $f(R)$ gravity can
be calculated exactly and it can be understood in terms of
the CFT Weyl anomaly. This is to be expected, as  the boundary theory  is a
CFT on de Sitter space, which is conformally flat. Hence, any
curvature dependence of the quantum partition function must be given
by the conformal anomaly.

The second case, that of a QFT, is richer.
The QFT should be considered as a RG flow between a nontrivial UV CFT and another non-trivial IR CFT. As such, it is a gapless theory where the gapless
degrees of freedom are those of the IR (holographic) CFT. We will
consider the simplest case of a theory with a single relevant
coupling, corresponding to a mass scale. Large-$N$ YM theory is an
example of such a theory with $\Lambda_{\textrm{YM}}$ as its only
scale (but has a mass gap in the IR). A gapless theory of this kind is $\mathcal{N}=4$ sYM at infinite $N$ and t'Hooft coupling,  with a mass for a hypermultiplet.  The relevant coupling is associated to  a scalar operator in the QFT. In the bulk theory, this scalar operator is dual to a scalar field.
The stress tensor of the QFT is dual to the graviton in the bulk theory. To describe therefore the ground state of the holographic QFT, it is enough to keep the graviton and a scalar in the bulk theory.
In the strong coupling and large-$N$ limit, the bulk theory is
described by a  two-derivative action. A non-trivial potential for the
dilaton, with different extrema, allows for the existence of holographic RG flow
solutions which in the field theory language connect UV and IR conformal fixed points.

We stress that, in both the conformal and the non-conformal case, the
QFTs we consider are gapless in the IR. This is
the  feature which  gives rise to IR
divergences in perturbation theory in de Sitter space-time. As we
will summarize below, the holographic calculation allows in both cases
to explore the space of constant-curvature solutions. While in the
first case the curvature of the metric is the only dimensionful
parameter, in the second
case there is an additional mass scale $m$ which controls  the
deformation  of the  CFT away  from the UV fixed point.

An advantage of the holographic approach is that full de Sitter
invariance  is manifest at all steps of the calculation. Therefore,
the resulting backreaction effect cannot be ascribed to an artifact of
the breaking of de Sitter invariance.
This is to be contrasted with the secular  time-dependent terms which arise in
perturbation theory around the cosmological de Sitter patch, where
only a part of de Sitter invariance is manifestly preserved.

In the remainder  of this introduction we briefly summarize our results.

\subsection{Setup and summary of results}

The  effective $f(R)$ theory resulting from QFT backreaction  depends
on the way one treats the UV regime and the associated UV divergences. More specifically, the  QFT
itself may be treated as a Wilsonian effective
field theory, or as an UV complete theory. This distinction does not introduce
particular difficulties, as handling UV divergences is well-understood
in holography. It may however lead to different possible scenarios for
the resulting $f(R)$ theory. In
this paper we distinguish two different  approaches to treating the
$d$-dimensional  gravity+QFT system (which we consider both for a CFT
and an RG flow QFT) :

\begin{enumerate}
\item The QFT is UV complete, and the ultimate cut-off of the theory
  is the quantum gravity scale $M_p$ (the effective Planck scale).
\item The QFT is treated as an effective theory with cut-off $\Lambda
  \ll M_p$. 
\end{enumerate}
In the first case (treated in section \ref{sec:ResultsQFTren}), one can effectively send the QFT cut-off to
infinity, and use the {\em renormalized} generating functional as the
QFT contribution to the gravitational effective action. This will in
particular renormalize the bare cosmological constant and Planck
scale as well as introduce
new  curvature-dependent terms.

In the second case (treated in section \ref{sec:ResultsQFTcutoff}), the theory is
considered at a finite cut-off, and there is no need to
renormalize. The theory depends explicitly on the cut-off scale
$\Lambda$, which appears as an additional parameter, but no scheme-dependent counterterms appear.

For both the CFT and the RG flow theory, and in both scenarios
described above, we first calculate $f(R)$ and then 
determine the constant-curvature solutions of the resulting $f(R)$
theories. We eventually  compare with the solutions of the ``bare'' gravity
theory. For reference, we take the latter to be defined
by an action admitting constant-curvature solutions, with the Ricci scalar given
(in $d=4$) by
\be \label{intro2}
R = 4\lambda \, ,
\ee
where $\lambda$ is a parameter which represents the cosmological
constant of the gravity theory in the absence of the QFT. More
specifically, in case 1 (UV complete QFT), $\lambda$ is the
renormalized cosmological term of the pure gravity theory; in case 2
(cut-off QFT) it is the bare cosmological term.

In both cases, we can parametrize the ``size'' of the effect of the
QFT by a parameter,  denoted by $\tilde{a}$, which measures the number of
QFT degrees of the freedom, and it is essentially the conformal
anomaly coefficient. In large-$N$ holographic QFTs in four dimensions,
we have $\tilde{a}  \propto N^2$ up to a numerical $\mathcal{O}(1)$ coefficient.

\begin{enumerate}
\item {\bf UV complete QFT}\\

In this case we absorb  the UV divergences in a redefinition of the Planck
scale and cosmological constant, and define the theory in terms of
physical renormalized parameters. Then,  all other finite backreaction
effects  can be understood in terms of the conformal
anomalies of the QFT. The finite  contribution to  $f(R)$ from
  the QFT backreaction takes a simple analytic form, schematically
  given by
\be \label{intro3}
f_{\textsc{qft}}(R) \sim  \tilde{a}  R^2 \log R  +   \beta (j) C(R).
\ee
The first term, which is the only one present if the theory is a CFT, comes
from the Weyl anomaly, \cite{duff}, as  it is well known since the the work of
Starobinsky \cite{St}. In a large-$N$ QFT, $\tilde{a} \sim N^2$.
 The second  term appears in  a
  non-conformal QFT,  due to the extra breaking of
  conformal invariance by the source $j$ of a relevant operator. In
  equation (\ref{intro3}), $\beta(j)$ is the beta-function of a
  relevant operator deforming the UV CFT, and  $C(R)$  is the curvature-dependent vev of
  such an operator. While equation (\ref{intro3}) is expected to hold
  in a generic field theory, the function $C(R)$ is generically not
  known.  Holography allows a non-perturbative calculation of  this
  quantity, thereby giving access to the full $f(R)$ function.
The  results about the backreacted solutions  are  summarized below
  , and we refer the reader to section \ref{sec:ResultsQFTren}  for details.

\begin{enumerate}
\item In the backreacted $f(R)$ theory, there is an upper bound
  $\lambda_{\textrm{max}}$ to the positive
  values of the (renormalized) cosmological constant $\lambda$ for
  which de Sitter solutions exist. In other words, backreaction
  obstructs the existence of de Sitter solutions unless
\be \label{intro3b}
-\infty < \lambda < \lambda_{\textrm{max}} \approx {M_p^2 \over \tilde{a}}
\approx   {M_p^2 \over N^2}.
\ee
This is to
  be contrasted with the bare gravity theory, in which constant-curvature solutions exist for all $\lambda$, see equation
  (\ref{intro2}). 
\item Due to the nonlinearities in the $f(R)$ theory, for  $\lambda <
  \lambda_{\textrm{max}}$ there are always two branches of  constant-curvature
  solutions.  On the ``regular'' branch, the curvature has the
  same sign as the cosmological constant. On the ``exotic''  branch,
  which arises purely due to backreaction,  the space-time is always de
  Sitter regardless of the sign of the cosmological constant. In
  particular, there is a de Sitter solution also for $\lambda = 0$.

\item  On both the standard and the exotic branch, regardless of the
  value of $\lambda$, the de Sitter curvature is parametrically
  smaller (in the large-$N$ limit) than both the cosmological constant and the Planck
  scale. More explicitly, we  show that the curvature is at most
\be \label{intro3c}
R_{\textrm{max}} \sim \left\{\begin{array}{ll} \displaystyle{{M_p^2\over \tilde{a}} \approx   {M_p^2 \over
      N^2} } \, , &  \displaystyle{\quad |\lambda| \lesssim {M^2_p \over
      \tilde{a}}} \\ & \\
   \displaystyle{  { M_p \sqrt{\lambda} \over \sqrt{\tilde{a}}} \approx   {M_p
       \sqrt{\lambda} \over N}} \, , & \displaystyle{\quad |\lambda| \gtrsim  {M^2_p \over
       \tilde{a}} }\end{array} \right.
      \ee

 \item  These results  are already  known in the conformal case,
   since they can be obtained from the universal Weyl anomaly term in
   equation (\ref{intro3}).  Here,
  we show that they extend to generic holographic non-conformal QFTs,
  subject to a non-trivial RG flow from a UV to an IR fixed point. From numerical  examples, we observe that
  the bound $\lambda_{\textrm{max}}$ as  well as the value of the curvature of the backreacted solution
  depend on  the  value of the deformation parameter  $j$.\footnote{More precisely, as $j$ is dimensionful, they depend on its dimensionless ratio with other scales of the problem like the curvature scale.} The bound
  can get relaxed or tightened depending on the dimension of the
  relevant operator.  In  fact,  our numerical examples clearly show
  that the
  solutions for an RG flow QFT interpolate continuously between those of the  the UV CFT ($j=0$) and those of
  the IR CFT ($j\to \infty$).

\end{enumerate}

The most significant lesson from these results is the fact that
backreaction of a large $N$ QFT prevents the existence of a de Sitter
solution with an arbitrarily large curvature:  in the bare theory, $R$
can be as high as the Planck scale (if $\lambda \sim M_p$) whereas
here it can never exceed one of the bounds (\ref{intro3c}), which for
large $N$ are {\em hierarchically smaller} than all other scales
of the  gravity theory.\footnote{Although the results for the backreaction of a
CFT are known (the expression for the backreacted curvature goes
back to the work of Starobinsky \cite{St}), their large-$N$ scaling
is  perhaps not always appreciated,  and it can be important for phenomenology.}

In the terms of our original question about robustness of de Sitter
space against QFT backreaction, these results indicate that a large-$N$ QFT with a high UV cut-off coupled to dynamical gravity cannot be consistently defined on a large-curvature de Sitter space.
Rather, any backreacted
de Sitter solution  has a parametrically smaller curvature than  its
pure gravity  counterpart, except when the latter is already close to
$\lambda =0$ (more precisely, if  $\lambda \ll M_p^2/N^2$). This means
that backreaction does not destroy {\em any} de Sitter background,
however it may drastically lower its curvature. As we discuss below,
the same trend exists  (with some important differences) when
considering the backreaction of a Wilsonian QFT with an explicit
cut-off $\Lambda \ll M_p$.

It is interesting to speculate what happens when the starting cosmological constant is large enough so that no solution exists. 
It is plausible that effects as the amplification of vacuum fluctuations may drive the space-time geometry away from de Sitter. We cannot make any statement in that direction based on our results.
It is also true that calculations in the literature that claim departures from de Sitter are perturbative, whereas our bound cannot be perturbative. It remains however a very interesting question to investigate this issue further.

\item {\bf Cut-off QFT}
\vskip 0.3cm

In this case, the QFT is treated as an effective field theory and
integrated out from a UV cut-off $\Lambda$, which we assume to be  smaller than the
physical Planck scale. The resulting $f(R)$ theory will depend
explicitly on $\Lambda$. This makes the expressions somewhat more complicated
than in the first case. We find the following universal results:
\begin{enumerate}
\item Integrating out a holographic QFT always
  results in {\em decreasing} the space-time curvature with respect to
  the ``bare'' value  in equation (\ref{intro2}). The effect is
  the more pronounced, the higher the QFT cut-off.\footnote{Because at high cutoff, the leading contributions is proportional to the fourth power of the cutoff.} This effect can be
  traced back to the fact that, in holography, the QFT contribution to
  the vacuum energy at zero or finite four-dimensional curvature is always negative (when the space-time curvature is non-negative), \cite{R,F},
  and therefore {\em always} reduces the
  effective cosmological constant.

  This can also be given an entropic interpretation. We consider gravity and now add the effects of a QFT. The QFT adds entropy to the system increasing the overall entropy.
In de Sitter, this increase in entropy amounts to a reduction of the curvature via the standard Bekenstein-Hawking argument for the static de Sitter patch.
This is closely related to the observations in \cite{F} that the holographic partition function of a QFT on de Sitter\footnote{This remains true if we analytically continue de Sitter to a four-sphere.} has a thermal structure. The log of the partition function can be written in the form $E-TS$, \cite{F}, and as shown in section \ref{sec:ResultsQFTcutoff}, only the energy part $E$ of the partition function backreacts on the cosmological constant.

\item By a careful choice of the cut-off $\Lambda$ as a function of
  the bare cosmological constant,  we can isolate the
  effect of integrating out only the {\em super-horizon} modes, on which much
  of the literature about de Sitter backreaction has focused. Again, we find
  that the ``naive'' de Sitter curvature scale is reduced in the
  backreacted theory. However in this case the backreacted solution is always a
  space-time with positive curvature.
\item In the case of a CFT, we find an analytic expression for the
  full $f(R)$ theory. For a QFT with a non-trivial deformation
  parameter $j\neq 0$, no analytic expression is available but  the results
  are qualitatively similar to those of the conformal case. More
  precisely, when the QFT is an RG flow interpolating between an UV
  and an IR CFT, the QFT $f(R)$ function also interpolates between those
  of the UV CFT ($j\to 0$) and  the IR CFT ($j\to \infty$) when $j$ is varied.

\end{enumerate}

\end{enumerate}

As in case 1 of a UV-complete QFT, perhaps the most important point is
that backreaction always acts by reducing the de Sitter curvature
compared to the un-backreacted  gravity theory. If backreaction is
strong enough this may result in the non-existence of de Sitter
solutions. If we consider the  backreaction of  super-horizon
modes only, their effect  is also to decrease the curvature, but this
never results in the disappearance of the de Sitter solution.

This  is an intriguing property that we found in this work, and it requires some further understanding. At weak coupling, in a QFT with a finite cutoff, the contribution to the cosmological constant can be either positive or negative. At zero coupling,  a boson contributes negatively while a fermion positively to the vacuum energy.
Therefore the sign of the backreaction effect of the QFT to the vacuum energy is given by the sign of $(n_F-n_B)$, the number of fermionic minus the number of bosonic degrees of freedom. At weak coupling, the sign remains the same by continuity, but in general it can change at strong coupling as interactions contribute to the vacuum energy.
Intuition tells us that attractive interactions will tend to produce extra negative contributions (the same intuition as the Casimir effect).
It is not unreasonable to think that as the couplings are driven to infinity, eventually all corrections to the vacuum energy will turn negative.
This is precisely what we find for all holographic theories, on flat space or on spaces of positive curvature like de Sitter  or the sphere.

This may be correlated to the fact that minus the logarithm of the partition function\footnote{Defined as the renormalized, or cutoff on-shell gravitational action.} of a holographic theory on a space of (non-negative) constant curvature, has a thermodynamic structure, with
\be
\log Z_{\textrm{QFT}} \sim E - TS \, ,
\ee
where $E$ is the thermodynamic energy, $T$ is the ``de Sitter temperature'', related to the curvature and $S$ is the de Sitter entropy. Moreover various thermodynamic relations hold, which will play an important role in the analysis and which is discussed in appendix \ref{app:thermo}. What we found in section \ref{sec:ResultsQFTcutoff} is the backreaction to the cosmological constant is minus $E$ which,  as stated, always reduces the bare cosmological constant. It is possible that this result has a thermodynamic explanation and if this is the case, it would be interesting to uncover it.

The paper is organized as follows. In section \ref{sec:strategy} we present our setup,
show how to  recast the backreaction problem into an $f(R)$ theory  and discuss
 the general properties of the resulting effective theory.
In section \ref{sec:holo} we  present  the calculation of the effective
$f(R)$ theory in holographic QFTs.
In sections \ref{sec:ResultsQFTren} and \ref{sec:ResultsQFTcutoff} we apply the general framework to  simple
examples of  holographic theories whose gravity side consists of
Einstein-dilaton theory with quartic potential, and we compute
numerically (and when possible, analytically) the de Sitter
backreaction. In particular, section \ref{sec:ResultsQFTren} presents the results for a
UV-complete QFT, whereas section \ref{sec:ResultsQFTcutoff} deals with a cut-off QFT.
Several technical results, including a summary of the holographic
dictionary, are presented in the Appendix.
\\

{\it {\bf Note:} Should a reader not be interested in the technical details
of the holographic calculation, he or she may skip section 3. The
paper can still be read self-consistently provided such reader accepts
for granted some of the expressions in sections  4 and 5.}

\section{Strategy and setup}
\label{sec:strategy}

In this section we lay out in detail the setup we use to couple QFTs
to gravity and obtaining $f(R)$ theories from performing the quantum
path integral, including the issue of  regularization/renormalization. The discussion is general, and only refers very schematically
to holography in subsection \ref{sec:intoutholo}. In particular, the general form of
the effective action and the resulting $f(R)$ theory we discuss in
sections \ref{sec:UVdiv} and \ref{sec:fofR} are independent of holography, although we will
use holographic methods to compute them in section \ref{sec:holo}.

\subsection{Back-reaction of a QFT in a maximally symmetric geometry}
Consider a gravitational theory in $d$ dimensions described by the action
\begin{align}
\label{eq:S0def} S_0[g] &= \frac{M_{0}^{d-2}}{2}  \int d^d x \sqrt{|g|} \left(R - 2 \lambda_0  + M_0^2 \sum_{n=2}^{\lfloor \frac{d}{2} \rfloor} a_{n} \, \big( M_0^{-2} R \big)^n \right)  \, ,
\end{align}
where $g_{\mu \nu}$ is the metric and $R$ the corresponding curvature
scalar. Here $M_0$ is the gravitational coupling constant, $\lambda_0$
is the cosmological constant (with dimensions of energy-squared) and
$a_n$ are dimensionless couplings.\footnote{Here we use lower case
  symbol $\lambda$ to denote the cosmological constant instead of the
  more conventional upper case symbol $\Lambda$, as the latter will
  later be employed to describe an energy cutoff.} Note that in
\eqref{eq:S0def} we included terms with more than two derivatives of
the metric, as long as they correspond to relevant operators in $d$
dimensions. Further terms are suppressed by $M_0$, which we regard to
be of the same order of the ultimate UV cut-off of the theory, where
stringy or quantum gravity effects become important.


As we shall be exclusively interested in constant-curvature solutions, we also restrict attention to theories where the metric-dependence of the Lagrangian density can be written exclusively in terms of $R$.\footnote{In appendix \ref{app:higherderivative}, we show explicitly for the 4-dimensional case that, as long as we are exclusively interested in constant-curvature solutions, we do not miss solutions by already restricting the 4-derivative terms in the action to be just given by $\sim R^2$. The argument is given for backgrounds without boundary.} More precisely, we restrict attention to Einstein backgrounds with
\begin{align}
\label{eq:Rmunuansatz} R_{\mu \nu} = \kappa \, g_{\mu \nu} \, , \quad \kappa = \textrm{const.} \, ,
\end{align}
already at the level of the action. While this will simplify the
analysis for the existence of backreacted constant-curvature
solutions, the modified action will not exhibit the same dynamics away
from constant-curvature solutions as the full theory. This will
unfortunately preclude an analysis of stability of the
constant-curvature solutions, which will be left for a future
work\footnote{This question has been studied recently in holographic
  CFTs in \cite{2003.05501}}.

In this work we shall be mainly interested in theories in $d=4$, in which case \eqref{eq:S0def} becomes
\begin{align}
\label{eq:S0def4d} S_0[g] = \frac{M_{0}^{2}}{2}  \int d^4 x \sqrt{|g|} \left(R - 2 \lambda_0 + a M_0^{-2} R^2  \right)  \, ,
\end{align}
where we dropped the subscript on the numerical parameter $a$. This system then permits a constant-curvature solution with
\begin{align}
\label{eq:RdSfromS0} R = 4 \, \lambda_0 \, ,
\end{align}
for any value of $a$ in \eqref{eq:S0def4d}. This is because, in $d=4$,
the  $R^2$ term does not contribute to Einstein equations in the case
of a maximally symmetric ansatz, as we will see in more detail in
section \ref{sec:fofR}.

We now return to general $d$ and couple a QFT to the gravitational theory in \eqref{eq:S0def}. In this work, we shall distinguish the following two contexts:
\begin{enumerate}
\item \textbf{The QFT is UV complete:} The combined system of
  gravitational theory described by action $S_0$ coupled to the QFT is
  taken to be a valid description up to arbitrarily high energy
  scales. In practice, the scale where quantum gravitational effects
  become important (which we identify with the physical Planck scale), will provide an ultimate cutoff, but here we shall ensure that this scale is well above all other scales in the system, so that the UV cutoff can effectively taken to be infinite.
\item \textbf{The QFT is an effective theory valid up to a scale $\Lambda$:} The other possibility is that the QFT is only an effective theory valid up to an energy scale $\Lambda$, at which quantum-gravitational effects can still safely be ignored, but above which a different description of our system is needed.
\end{enumerate}
This distinction will become important later when we shall discuss the renormalization of UV divergences.

The $d$-dimensional QFT to be coupled to the gravitational system in \eqref{eq:S0def} is taken to have the following properties: It describes a RG flow between a UV CFT and an IR CFT. The RG flow is driven by a scalar operator $O$. The fundamental set of fields in the path-integral description of the QFT will be collectively denoted by $\Phi$. The UV values of the corresponding coupling constant (i.e.~the source of the operator $\mathcal{O}$) will be denoted by $j$. The combined action for the gravity-QFT system can then be written as
\begin{align}
\label{eq:Sfull} S[g, j, \Phi] &= S_0[g] + S_{\textrm{QFT}}[g, j, \Phi] \, .
\end{align}
The generating functional for the combined system is thus given by
\begin{align}
\nonumber Z[j] &=  \int d[g] \, d[\Phi] \, e^{i S[g, j, \Phi]} =  \int d[g] \, d[\Phi] \, e^{iS_0[g] +i S_{\textrm{QFT}}[g, j, \Phi]} \\
\label{eq:Zfull} &= \int d[g] \, e^{iS_0[g]} \left( \int d[\Phi] \, e^{i S_{\textrm{QFT}}[g, j, \Phi]} \right) \, .
\end{align}
We can integrate out the QFT by performing the path integral over $\Phi$. To this end, we introduce the effective action $S_{\textrm{QFT}}^{\textrm{eff}}$ as
\begin{align}
\label{eq:SQFTeffdef} e^{i S_{\textrm{QFT}}^{\textrm{eff}}[g,j]} = \int d[\Phi] \, e^{i S_{\textrm{QFT}}[g, j, \Phi]} \, .
\end{align}
Inserting this into the expression \eqref{eq:Zfull} for the generating functional, we obtain
\begin{align}
\label{eq:Zfull2} Z[j] &=  \int d[g] \, e^{iS_0[g] +i S_{\textrm{QFT}}^{\textrm{eff}}[g,j]} =  \int d[g] \, e^{i S_{\textrm{tot}}[g,j]} \, .
\end{align}
where we introduced
\begin{align}
\label{eq:Stot} S_{\textrm{tot}}[g,j] = S_0[g] + S_{\textrm{QFT}}^{\textrm{eff}}[g,j]
\end{align}
We are hence left with a theory of the metric $g_{\mu \nu}$ (for some choice of sources $j$). This is a theory of $g_{\mu \nu}$ with the backreaction effects due to the QFT already included. By studying its constant-curvature solutions we can hence determine how these are affected by backreacting a QFT, which is the primary objective of this work. To find these solutions we simply have to extremize the action $S_{\textrm{tot}}$ in \eqref{eq:Stot} with respect to the metric $g_{\mu \nu}$.

\subsection{Integrating out a QFT using holography}
\label{sec:intoutholo}
To proceed we need an explicit expression for
$S_{\textrm{QFT}}^{\textrm{eff}}$. This is typically the main obstacle
in this analysis, in particular if the QFT is interacting. In this
case the path integral in \eqref{eq:SQFTeffdef} cannot be evaluated in
all generality and an analytical expression for
$S_{\textrm{QFT}}^{\textrm{eff}}$ is beyond reach. To proceed one may
work in perturbation theory or turn to numerical studies on a
lattice. Here we shall follow a different direction which will allow
us to make progress. In particular, by considering QFTs which possess
weakly coupled gravity duals we shall be able to integrate out the QFT
using the dual description. In this and the next subsection  we
explain schematically the idea behind the holographic computation, and we refer the interested reader to section 3 for details.

We assume that the $d$-dimensional QFT admits a dual description in
terms of a gravitational theory. This gravity dual is defined on a
$(d+1)$-dimensional space-time $\mathcal{M}$ with a dynamical metric
$G_{ab}$. In addition, for every scalar operator $\mathcal{O}$ on the
QFT side, there will be a scalar field $\f$ on the gravity side.  The
manifold $\mathcal{M}$ possesses an (AdS$_{d+1}$) boundary, which
under the duality
corresponds the fact that the QFT reaches a conformal fixed point in
the UV.  The holographic dictionary then corresponds to an identification between the generating functional $Z_{\textrm{QFT}}$ on the field theory and the corresponding expression $Z_{\textrm{grav},(d+1)}$ on the dual gravity side, i.e.
\begin{align}
\label{eq:duality} Z_{\textrm{QFT}}[g,j] = Z_{\textrm{grav},(d+1)}[g,j] \, ,
\end{align}
with
\begin{align}
\label{eq:ZQFTdef} Z_{\textrm{QFT}}[g,j] &= \int d[\Phi] \, e^{i S_{\textrm{QFT}}[g, j, \Phi]} = e^{i S_{\textrm{QFT}}^{\textrm{eff}}[g,j]} \, , \\
Z_{\textrm{grav},(d+1)}[g,j] &= \int_{\substack{G|_{\partial \mathcal{M}} = g \\ \f|_{\partial \mathcal{M}} = j}} d[G] \, d[\f] \, e^{i S_{\textrm{grav},(d+1)}[G, \f]} \, .
\end{align}
Roughly speaking, the holographic dictionary identifies\footnote{These identifications, which here we simplified  for
illustrational purposes, will be made precise in Section 3.} the metric $G$
restricted to the boundary $\partial \mathcal{M}$ with the metric $g$
on which the field theory is defined, and the value of the scalars
$\f$ on $\partial \mathcal{M}$ with the values of the sources
$j$.

The duality is most useful when the dual gravitational theory is dominated by Einstein gravity, which occurs when the field theory is strongly coupled, with a gauge group of large rank. In this case one can approximate
\begin{align}
\label{eq:ZgravwithSonshell} Z_{\textrm{grav},(d+1)}[g,j] = e^{iS_{\textrm{grav},(d+1)}^{\textrm{on-shell}}[G|_{\partial \mathcal{M}} = g, \f|_{\partial \mathcal{M}} = j]} \, ,
\end{align}
where $S_{\textrm{grav},(d+1)}^{\textrm{on-shell}}$ is the action evaluated on the classical solution with the appropriate boundary conditions. Using this and \eqref{eq:ZQFTdef} the duality statement \eqref{eq:duality} implies
\begin{align}
\label{eq:effduality} S_{\textrm{QFT}}^{\textrm{eff}}[g,j] = S_{\textrm{grav},(d+1)}^{\textrm{on-shell}}[G|_{\partial \mathcal{M}} = g, \f|_{\partial \mathcal{M}} = j] \, .
\end{align}
As a result, we can write the total action \eqref{eq:Stot} as
\begin{align}
\label{eq:Stot2} S_\textrm{tot}[g,j] = S_0[g] + S_{\textrm{grav},(d+1)}^{\textrm{on-shell}}[G|_{\partial \mathcal{M}} = g, \f|_{\partial \mathcal{M}} = j] \, .
\end{align}
Recall that to arrive at this, we had to assume that the QFT is a strongly-coupled large-rank (ie. holographic) gauge theory. Therefore, all results obtained from \eqref{eq:Stot2} will be valid when these assumptions hold.

\subsection{A subset of QFTs with simple gravitational duals}
\label{sec:QFTdef}
While the strategy laid out above is valid for any QFT that allows for
a gravity dual, for simplicity we shall henceforth restrict our
attention to a particular subset of QFTs, whose  gravity dual is  a
$d+1$  Einstein-dilaton theory with dilaton potential. All details regarding this gravitational dual will be postponed until section \ref{sec:holosetup}, while here we describe the properties of the QFTs considered in field theory language.

In particular, we shall exclusively consider QFTs that can be defined in terms of a UV CFT with $N_{\textsc{uv}}$ degrees of freedom, perturbed by a single scalar operator $\mathcal{O}$ of dimension $\Delta$ with $0 < \Delta < d$. The corresponding coupling constant (the `source') has dimension $d-\Delta$ and we shall denote its UV value by $j$. We can hence associate a mass scale $m$ with this theory, which we define as
\begin{align}
\label{eq:mdef} m \equiv |j|^{1/(d-\Delta)} \, .
\end{align}
We further restrict to QFTs where the renormalization group flow induced by the operator $\mathcal{O}$ terminates at an IR fixed point associated with a corresponding IR CFT with $N_{\textsc{ir}}$ degrees of freedom. The arrival at this IR fixed point can be understood as a perturbation of the IR CFT by an irrelevant scalar operator of dimension $\Delta^{\textsc{ir}} > d$.

Thus, in the following, whenever we refer to the backreacting QFT we refer to a theory with the properties laid out above. A particular case that we shall study is given by the above theory with $j=0$ (and hence $m=0$), in which case the QFT reduces to the (UV) CFT.

\subsection{UV divergences and renormalized parameters}
\label{sec:UVdiv}

The calculation of the effective action (or equivalently the on-shell
action) is sensitive to ultraviolet physics and as a result the
quantity $S_{\textrm{QFT}}^{\textrm{eff}}$ is  typically UV divergent. As a first step, we regulate UV divergences by the introduction of an energy cutoff $\Lambda$. The UV cutoff used here will be consistent with the (maximal) symmetry of the space-time described by $g_{\mu \nu}$ (in contrast to a simple momentum cutoff). This can be easily achieved in the holographic formulation, by appropriately cutting off the bulk geometry before the boundary is reached. This will be described in detail in section \ref{sec:onshell}.

The effective action $S_{\textrm{QFT}}^{\textrm{eff}}$ (or equivalently the on-shell action $S_{\textrm{grav},(d+1)}^{\textrm{on-shell}}$) will hence depend on the three dimensionful parameters $\Lambda, R, m$. Then, without loss of generality we can write
\begin{align}
\label{eq:Seffcutoffreg}  S_{\textrm{QFT}}^{\textrm{eff}} = \tilde{a}_{\textsc{uv}} \int d^d x \sqrt{|g|} \, \Lambda^d \, \mathcal{F} \Big( \frac{m}{\Lambda}, \frac{R}{\Lambda^2} \Big) \, ,
\end{align}
with $\mathcal{F}$ a function of the dimensionless ratios
$\tfrac{m}{\Lambda}$ and $\tfrac{R}{\Lambda^2}$. The entire dependence
of $S_{\textrm{QFT}}^{\textrm{eff}}$ on the number of UV degrees of
freedom can be written as an overall multiplicative factor, which we
label by $\tilde{a}_{\textsc{uv}}$. For a large-$N$ gauge theory
possessing a gravitational dual described by classical gravity,  one has
\begin{align}
\tilde{a}_{\textsc{uv}} \sim N_{\textsc{uv}}^2 \, \qquad \textrm{with} \qquad N_{\textsc{uv}} \rightarrow \infty \, ,
\end{align}
where $N_{\textsc{uv}}$ has the interpretation as the rank of the UV
gauge theory. The precise proportionality constant is determined by
the precise string compactification providing the gravity dual.  In
practice, to arrive at numerical results we shall set
$N_{\textsc{uv}}$ to be finite but large. We stress that any result
thus obtained will miss contributions suppressed by negative powers of
$N_{\textsc{uv}}$, which corresponds to quantum (bulk) gravity effects.

Isolating in the expression  (\ref{eq:Seffcutoffreg})  terms which are UV-divergent in the limit $\Lambda \to \infty$ we expect  on general
grounds that,  in a local QFT,  these come with integer powers of $R$:
\begin{align}
\label{eq:Seffdivergences} S_{\textrm{QFT}}^{\textrm{eff}} =  \tilde{a}_{\textsc{uv}} \int d^4 x \sqrt{|g|} \, \bigg[ & \ \hphantom{+} \Lambda^d \mathcal{F}_0\big( \tfrac{m}{\Lambda} \big) + R \Lambda^{d-2} \mathcal{F}_1 \big( \tfrac{m}{\Lambda} \big) + R^2 \Lambda^{d-4} \mathcal{F}_2 \big( \tfrac{m}{\Lambda} \big) \\
\nonumber &+ \ldots + \delta_{d/2 \, \textrm{mod} \, 2} \, R^{\tfrac{d}{2}} \log \big( \tfrac{m^2}{\Lambda^2} \big) \mathcal{F}_{d/2} \big( \tfrac{m}{\Lambda} \big) + \textrm{finite for } \Lambda \rightarrow \infty \, ,
\end{align}
where all the functions $\mathcal{F}$'s are finite in the limit. As we
will see in section 3, the holographic calculation does reproduce this structure.

Note that the term $\sim R^{d/2}$ is only present for even $d$. There typically is further non-trivial (and non-analytic) dependence on $R$ in the finite terms which we currently suppress in this discussion of divergent pieces. Adding $S_{\textrm{QFT}}^{\textrm{eff}}$ to the action $S_0$ one then can make the following observations:
\begin{itemize}
\item The term $\sim \Lambda^d \mathcal{F}_0 (\tfrac{m}{\Lambda})$ acts as a contribution to the overall cosmological constant.
\item The term $\sim R \Lambda^{d-2} \mathcal{F}_1 (\tfrac{m}{\Lambda})$ modifies the coefficient of $R$ in the action and hence contributes to the Planck scale.
\item The remaining divergent terms modify the higher derivative terms $\sim a_n R^{n}$ in \eqref{eq:S0def}.
\end{itemize}
The physical relevance of these observations now crucially depends on
whether we assume the QFT to be UV complete,  or not. We discuss these
two possibilities separately.

\vspace{0.3cm}

\noindent \textbf{The QFT is UV complete.} In this case the description of our physical system is valid to arbitrarily high energies and we should hence take the cutoff to infinity, i.e.~$\Lambda \rightarrow \infty$.\footnote{Again, note that the quantum gravity scale, expected to be of the order of the physical Planck scale $M_p$, will provide an effective cutoff. However, we shall be mainly interested in situations where all dimensionful parameters are small compared to this scale, i.e.~$m/M_p \ll 1$ and $R/ M_p^2 \ll 1$. In this case we expect that we can safely take the cutoff to infinity without missing large effects.} In this limit the contribution from the QFT to the combined action diverges, but this is to be treated using standard renormalization.

Consider for example the overall cosmological constant of the combined
system. This consists of the bare cosmological constant $\lambda_0$
introduced in \eqref{eq:S0def} and a UV-divergent contribution from
the QFT which we  denote by $\Delta \lambda$. The important point is
that for an observer in the IR only the combined cosmological constant
$\lambda_{\textrm{ren}} \equiv \lambda_0 + \Delta \lambda$ is
observable and hence physical, but not the individual contributions
$\lambda_0$ and $\Delta \lambda$. Thus, from an IR point of view it is sensible to replace $\lambda_0 + \Delta \lambda$ by $\lambda_{\textrm{ren}}$ and consider $\lambda_{\textrm{ren}}$ as a
(finite) parameter to be eventually determined by experiment. The
UV-divergent contribution has been subsumed into a renormalized
parameter, at the expense of the calculability of this parameter
within the theory. Then we can set $\Lambda \rightarrow \infty$
without harm. This is just the usual procedure of renormalization of a
relevant coupling (c.f.~the renormalization of the Higgs mass), and
can be made precise by introducing counterterms as we discuss below.

In this work we are only interested in the IR point of view and hence divergent contributions will be subsumed into appropriately defined renormalized quantities. In particular, all divergences can be absorbed by defining the quantities:
\begin{align}
M_{\textrm{ren}} \, , \qquad \lambda_{\textrm{ren}} \, , \qquad \textrm{and} \qquad a_n^{\textrm{ren}}
\end{align}
which can be defined by combining the bare parameters $M_0$, $\lambda_0$ and $a_n$ in \eqref{eq:S0def} with the corresponding UV-divergent expressions from \eqref{eq:Seffdivergences}. While this can be performed explicitly using the cutoff-regulated effective action \eqref{eq:Seffcutoffreg}, it will be technically cleaner to introduce one further intermediate step before defining $M_{\textrm{ren}}$, $\lambda_{\textrm{ren}}$ and $a_n^{\textrm{ren}}$.

In particular, this step consists of removing the UV divergences in $S_{\textrm{QFT}}^{\textrm{eff}}$ explicitly by adding suitable counterterms. That is, we define a renormalized effective action
\begin{align}
\label{eq:Seffrendef} S_{\textrm{QFT}}^{\textrm{eff}, \, \textrm{ren}} \equiv \lim_{\Lambda \rightarrow \infty} \bigg( S_{\textrm{QFT}}^{\textrm{eff}} + S_{ct} \bigg) \, ,
\end{align}
where $S_{ct}$ denotes the necessary counterterms to ensure that
$S_{\textrm{QFT}}^{\textrm{eff}, \, \textrm{ren}}$ is UV finite. For
systems with a gravity dual in terms of an Einstein-dilaton theory, the precise form of the counterterm action is known \cite{1106.4826} and hence this procedure can be performed explicitly. Details can be found in section \ref{sec:onshell} and here we summarise the most important points.
\begin{itemize}
\item The counterterm action can be written as a series in integer powers of $R$, just like the UV divergent terms in \eqref{eq:Seffdivergences} come as a series in powers of $R$:
\begin{align}
\label{eq:Sctexpansion} S_{ct} 
=
\tilde{a}_{\textsc{uv}}  \int d^d x\, \sqrt{|g|} \left[
\sum_{0\leq n < [d/2]}  s_{ct,n}\big( \tfrac{m}{\Lambda} \big) \,
\Lambda^{d-2n} R^{2n} + s_{ct,d/2}\big( \tfrac{m}{\Lambda} \big) \,
\log {\Lambda \over m}  R^{[d/2]} \right]\, .
\end{align}
where the last term is only present in even dimensions.  The functions
$s_{ct,n}$ are finite in the limit $\Lambda \to \infty$.  The counterterm at order $R$ then removes the UV divergent piece at that order in $R$. Thus, to remove all divergences we need to include counterterms up to and including $\mathcal{O}(R^{\lfloor d/2 \rfloor})$.
\item Further, each counterterm $S_{ct,n}$ at a given order in $R$ is only defined up to one arbitrary numerical parameter, which we  denote by $c_{ct,n}$. These parameters $c_{ct,n}$ can be chosen freely, with a particular choice corresponding to a particular renormalization scheme.
\end{itemize}
One can then expect (and the holographic calculation will confirm this explicitly) that the renormalized effective action \eqref{eq:Seffrendef} takes the form:
\begin{align}
\nonumber S_{\textrm{QFT}}^{\textrm{eff}, \, \textrm{ren}} =  \int d^4 x \sqrt{|g|} \, \bigg[ &  \tilde{a}_{\textsc{uv}} c_{ct,0} \, m^d +  \tilde{a}_{\textsc{uv}} c_{ct,1} \, R m^{d-2} + \ldots +  \tilde{a}_{\textsc{uv}} c_{ct, \lfloor d/2 \rfloor} R^{\lfloor d/2 \rfloor} m^{d-2 \lfloor d/2 \rfloor} \\
\label{eq:Seffrenexpression} & + \delta_{d/2 \, \textrm{mod} \, 2} \, \tfrac{1}{96} \, \tilde{a}_{\textsc{uv}} \, R^{\tfrac{d}{2}} \Big( 1+ \log \big( \tfrac{R}{48 m^2}\big) \Big) + \tilde{a}_{\textsc{uv}} \, m^d \, \mathcal{G} \big(\tfrac{R}{m^2} \big) \bigg] \, .
\end{align}
The first line contains all scheme-dependent pieces while on the
second line we collected all scheme-independent pieces, i.e.~all
finite terms that do not come with a factor $c_{ct,n}$. The second
line contains the term $\sim R^{d/2} \log R$, which  is only present
for even $d$.  This term  was explicitly separated from the
(generically unknown but finite) function $\mathcal{G}(R/m^2)$ because
 it reproduces the Weyl anomaly in a constant-curvature background,
 which reads:
\be \label{anomaly}
\<T^{\textrm{ren}, \mu}_\mu \>  = - {a \over 48} R^2
\ee
By calculating the renormalized  stress tensor from the definition
\be
\<T^{\textrm{ren}}_{\mu\nu}\> = - {2\over \sqrt{|g|}} {\delta S_{\textrm{QFT}}^{\textrm{eff}, \textrm{ren}} \over \delta g^{\mu\nu}}
\ee
and comparing with equation (\ref{anomaly}),
we can identify the anomaly coefficient $a$ (of the UV CFT) with the
parameter $\tilde{a}_{\textsc{uv}}$. The  factors $\tfrac{1}{96}$ and
$\tfrac{1}{48}$, as well as the extra finite $R^{d/2}$ term  in the
second line  of equation (\ref{eq:Seffrenexpression}) are included for
later convenience. The remaining finite pieces are collected into the
term $\tilde{a}_{\textsc{uv}} m^d \mathcal{G} (Rm^{-2})$ with
$\mathcal{G}$ some function that will need to be determined for every
QFT separately. As will be confirmed later by the holographic calculation, the term $m^d \mathcal{G} (Rm^{-2})$ satisfies
\begin{align}
m^d \mathcal{G} (Rm^{-2}) \underset{m \rightarrow 0}{\longrightarrow} 0 \, ,
\end{align}
i.e.~this term disappears in the UV conformal limit of the QFT.

Adding $S_{\textrm{QFT}}^{\textrm{eff}, \, \textrm{ren}}$ in \eqref{eq:Seffrenexpression} to $S_0$ in \eqref{eq:S0def} the action for the combined system is then given by:
\begin{align}
\nonumber S_{\textrm{tot}} = \int d^4 x \sqrt{|g|} \, \bigg[ & \hphantom{+} \Big( - M_0^{d-2} \lambda_0 + \tilde{a}_{\textsc{uv}} \, c_{ct,0} \, m^d \Big) + R \Big( \tfrac{1}{2} \, M_0^{d-2} + \tilde{a}_{\textsc{uv}} \, c_{ct, 1}  \, m^{d-2}\Big) \\
\nonumber & + \sum_{n=2}^{\lfloor d/2 \rfloor} R^n \Big( \tfrac{1}{2} \, a_n \, M_0^{d-2n} + \tilde{a}_{\textsc{uv}} \, c_{ct,n} \, m^{d-2n} \Big) \\
\label{eq:Stotren} & + \delta_{d/2 \, \textrm{mod} \, 2} \, \tfrac{1}{96} \, \tilde{a}_{\textsc{uv}} \, R^{\tfrac{d}{2}} \Big( 1+ \log \big( \tfrac{R}{48 m^2}\big) \Big) + \tilde{a}_{\textsc{uv}} \, m^d \, \mathcal{G} \big(\tfrac{R}{m^2} \big) \bigg] \, .
\end{align}
Here it is apparent that $c_{ct,0}$ controls the renormalization of the cosmological constant, $c_{ct,1}$ is related to the renormalization of the Planck scale, with the remaining coefficients $c_{ct,n}$ associated with the renormalization of the parameters $a_n$.

We can then define renormalized parameters as follows. For one, we choose to renormalize the Planck scale and cosmological constant to reproduce the standard flat space limit of classical gravity. That is, we define the cosmological constant as the constant piece of the Lagrangian density for $R \rightarrow 0$ and the Planck scale as the coefficient of the linear piece in $R$ for $R \rightarrow 0$.\footnote{Alternatively, what is referred to as the `physical Planck scale' is sometimes also defined in terms of the gravitational coupling constant of graviton fluctuations about a given background solution. This cannot be unambiguously defined without finding a solution first.} Writing
\begin{align}
\nonumber S_{\textrm{tot}} = \int d^4 x \sqrt{|g|} f(R) \, ,
\end{align}
this definition implies
\begin{align}
\label{eq:Mrendef1} \tfrac{1}{2} M_{\textrm{ren}}^{d-2} & \equiv \left. \tfrac{\partial f(R)}{\partial R} \right|_{R=0} = \tfrac{1}{2} M_0^{d-2} + \tilde{a}_{\textsc{uv}} \, c_{ct,1} \, m^{d-2} +  \tilde{a}_{\textsc{uv}} \, m^{d-2} \, \mathcal{G}'(0) \, , \\
\label{eq:Mrendef2} M_{\textrm{ren}}^{d-2} \lambda_{\textrm{ren}} & \equiv -f(0) =M_0^{d-2} \lambda_0 - \tilde{a}_{\textsc{uv}} \, c_{ct,0} \, m^d \, - \tilde{a}_{\textsc{uv}} \, m^d \, \mathcal{G}(0) \, ,
\end{align}
where ${}'$ implies a derivative with respect to the argument of $\mathcal{G}(Rm^{-2})$. Regarding the parameters $a_n^{\textrm{ren}}$ which will multiply higher powers of $R$, there is no reason to define them at any particular value of $R$. Hence there we define the renormalized parameters simply as the combination of the bare parameters $a_n$ and the corresponding arbitrary constants $c_{ct,n}$:
\begin{align}
\label{eq:Mrendef3} \tfrac{1}{2} a_n^{\textrm{ren}} M_{\textrm{ren}}^{d-2n}& \equiv \tfrac{1}{2} a_n M_0^{d-2n} + \tilde{a}_{\textsc{uv}} \, c_{ct,n} \, m^{d-2n}  \, , \qquad 2 \leq n \leq \lfloor d/2 \rfloor \, .
\end{align}
Using these renormalized parameters the total action \eqref{eq:Stotren} can be written as
\begin{align}
\nonumber S_{\textrm{tot}} = \int d^d x \sqrt{|g|} \, \bigg[ & \hphantom{+} \frac{M_{\textrm{ren}}^{d-2}}{2}  \Big( R - 2 \lambda_{\textrm{ren}}  + M_{\textrm{ren}}^2 \sum_{n=2}^{\lfloor \frac{d}{2} \rfloor} a_{n}^{\textrm{ren}} \, \big( M_{\textrm{ren}}^{-2} R \big)^n \Big) \\
\nonumber & + \delta_{d/2 \, \textrm{mod} \, 2} \, \tfrac{1}{96} \, \tilde{a}_{\textsc{uv}} \, R^{\tfrac{d}{2}} \Big( 1+ \log \big( \tfrac{R}{48 m^2}\big) \Big) \\
\label{eq:Stotrenfinal} & + \tilde{a}_{\textsc{uv}} \, m^d \, \Big(  \mathcal{G} \big(\tfrac{R}{m^2} \big) - \mathcal{G}(0) - \tfrac{R}{m^2} \mathcal{G}'(0) \Big) \bigg] \, ,
\end{align}
Note that the first line of \eqref{eq:Stotrenfinal} takes the same
form as the bare action $S_0$ in \eqref{eq:S0def}, but with the bare
parameters $M_0$, $\lambda_0$, $a_n$ replaced by the renormalized
quantities $M_{\textrm{ren}}$, $\lambda_{\textrm{ren}}$ and
$a_n^{\textrm{ren}}$. The second line contains a contribution from the
conformal anomaly of the form $\sim R^{d/2} \log R$, which only exists
in even dimensions and is not renormalized. On the third line we have a contribution from integrating out the QFT which exists in any dimension and is present as long as $m \neq 0$. Finally, by definition, for $R \rightarrow 0$ we observe that
\begin{align}
S_{\textrm{tot}} \mathrel{\mathop{=}^{}_{R \rightarrow 0}} \int d^dx \sqrt{|g|} \, \bigg[ - M_{\textrm{ren}}^{d-2} \lambda_{\textrm{ren}} + \frac{M_{\textrm{ren}}^{d-2}}{2} \, R + \mathcal{O}(R^2) \bigg] \, ,
\end{align}
i.e.~the total action reduces to that of standard Einstein theory with Planck scale $M_{\textrm{ren}}$ and cosmological constant $\lambda_{\textrm{ren}}$.

\vspace{0.3cm}

\noindent \textbf{The QFT is an effective theory valid up to a scale $\Lambda$.} In this case the bare action $S_0$ in \eqref{eq:S0def} and the cutoff-regulated action $S_{\textrm{QFT}}^{\textrm{eff}}$ in \eqref{eq:Seffcutoffreg} have the following interpretation. Here the action $S_0$ is the action for a gravitational system, with all effects above the energy scale $\Lambda$ already included.
At this scale $\Lambda$ we then couple the QFT and study the ensuing effects on the geometry. Here $\Lambda$ is a finite physical scale and hence $S_{\textrm{QFT}}^{\textrm{eff}}$ is finite. We hence simply add it to $S_0$ and study the resulting combined system. The total action in this case is given by
\begin{align}
\label{eq:Stotcutoff} S_{\textrm{tot}} = \int d^d x \sqrt{|g|} \, \bigg[ \frac{M_{0}^{d-2}}{2} \Big(R - 2 \lambda_0  + M_0^2 \sum_{n=2}^{\lfloor \frac{d}{2} \rfloor} a_{n} \, \big( M_0^{-2} R \big)^n \Big) +  \tilde{a}_{\textsc{uv}} \Lambda^d \, \mathcal{F} \Big( \frac{m}{\Lambda}, \frac{R}{\Lambda^2} \Big) \bigg] \, .
\end{align}

One special case within this class of models will receive our particular attention. Consider a maximally symmetric space-time with some value $R$ for the curvature scalar. One question then is how this particular geometry would be affected if a QFT is coupled to this curved background. One way of understanding the backreaction due to the presence of the QFT is in terms of particle creation in a curved background. However, from all degrees of freedom described by the QFT only modes with momenta $p$ with $p^2 \lesssim R$ are sensitive to the curvature of the background, while modes with $p^2 \gtrsim R$ perceive the background to be effectively flat. Hence, if backreaction is mainly due to particle creation in a curved background, only the modes with $p^2 \lesssim R$ should be important for evaluating backreaction on the geometry.

In our setting we can capture this situation by choosing $\Lambda$
self-consistently, to the value $\Lambda_*$ which satisfies the equation
\be
\Lambda_*^2 = R(\lambda_0, M_0, \Lambda_*)
\ee
where the right hand side is the {\em solution} of the $f(R)$ gravity
equation of motion with parameters $\lambda_0, M_0$ and
$\Lambda_*$.

\vspace{0.3cm}

\subsection{The combined system as a $f(R)$-theory of gravity}
\label{sec:fofR}

The systems of study in this work fall into the category of higher-derivative theories of gravity, as the action contains terms proportional to $R^n$ with $n \geq 1$, which lead to terms with more than two derivatives in the Lagrangian density. Arbitrary higher-derivative theories commonly suffer from ghosts, i.e.~degrees of freedom with the wrong sign of kinetic term.

In our case the effective action will depend on the curvature via various curvature invariants involving the scalar curvature, the Ricci tensor of the Riemann tensor. However, we will be working with metrics that are Einstein,
and such metrics have constant scalar curvature.
Moreover, as we have shown in \cite{R,F}, the effective action after integrating out a holographic QFT is a functional {\em only} of the scalar curvature. One of the reasons is that for Einstein metrics Ricci invariants are related to scalar curvature invariants. However, Kretschmann-like invariants can be different, but they do not appear to leading order in $1/N_c$ and in the strong coupling expansion.\footnote{References \cite{R}  and \cite{F} focused on the Einstein manifold being a space of maximal symmetry. However all results are valid for any Einstein manifold.}

In the following we shall argue that in the systems of interest in this work the appearance of ghosts in the gravitational action can easily be avoided.
 The central observation, made also above, is that in the systems described by the action \eqref{eq:Stotrenfinal} and \eqref{eq:Stotcutoff} all higher-derivative terms can be written in terms of the scalar curvature $R$ alone, without requiring terms coming from contractions of Ricci and Riemann tensors like $R_{\mu \nu} R^{\mu \nu}$ or $R_{\mu \nu \rho \sigma}R^{\mu \nu \rho \sigma}$. Higher-derivative theories of this kind are also referred to as $f(R)$ theories of gravity, indicating that the Lagrangian density can be entirely written as a function of the scalar curvature $R$, i.e.
\begin{align}
\label{eq:fofR1} S_{\textrm{tot}}= \int d^d x \sqrt{|g|} \, f(R) \, .
\end{align}
Theories of $f(R)$ gravity have been studied extensively in the literature, e.g.~as possible alternatives to $\Lambda$CDM cosmology (see e.g.~the reviews \cite{fR}).

The stability of $f(R)$ theories has also been addressed in the past. In particular, for the graviton not to be a ghost one requires:
\begin{align}
\label{eq:fofR2} f_R \equiv \frac{\partial f}{\partial R} > 0 \, .
\end{align}
If this is satisfied $f(R)$ theories are ghost-free (see e.g.~the
subsection on ghosts of $f(R)$ theories in the review by Sotiriou and
Faraoni \cite{fR} and references therein).\footnote{As described in
  the review by De Felice and Tsujikawa \cite{fR}, if $f(R)$ theories
  are employed as models for the observed cosmological evolution the
  condition \eqref{eq:fofR2} only has to be satisfied for $R > R_0$
  with $R_0$ the Ricci scalar today. As we do not apply our results to
  cosmology in this work,  we shall require the stricter condition of
  \eqref{eq:fofR2} holding for all values of $R$.} As we argue below,
for all cases of interest we shall always be able to satisfy the
stability condition \eqref{eq:fofR2}. 

Treating our systems of interest in \eqref{eq:Stotrenfinal} and \eqref{eq:Stotcutoff} as a $f(R)$ theory of gravity also has the following advantage: The equations of motion for the metric can be written very compactly with the help of the function $f(R)$. In particular, for a theory given by \eqref{eq:fofR1} the equations of motion are
\begin{align}
\label{eq:fofR3} f_R(R)R_{\m\n}-\nabla _{\m}\nabla_{\n} f_{R}+g_{\m\n}\square f_R(R)-{1\over 2}f(R)g_{\m\n}=0 \, , \quad \textrm{with} \quad f_R\equiv \frac{\partial f}{\partial R} \, .
\end{align}
For constant curvature solutions the 2nd and 3rd term on the LHS of \eqref{eq:fofR3} vanish identically. Contracting with $g^{\mu \nu}$ the above can then be simplified to
\begin{align}
\label{eq:fofR4} d f(R) -2 R f_R(R) = 0 \, .
\end{align}
This is the equation that we shall solve in the following for obtaining constant-curvature solutions. The relevant expression for $f(R)$ can simply be read off from \eqref{eq:Stotrenfinal} and \eqref{eq:Stotcutoff}.

Given a $f(R)$ theory of gravity with action \eqref{eq:fofR1}, an equivalent formulation exists in terms of a scalar-tensor theory of a scalar $\phi$ with a non-trivial potential coupled to Einstein gravity, i.e.\footnote{We use the `mostly plus' signature for the metric.}
\begin{align}
\label{eq:scalartensor1} S_{\textrm{tot}}= \int d^d x \sqrt{|\tilde{g}|} \, \bigg[ \frac{M_p^{d-2}}{2} \tilde{R} - \frac{1}{2} \tilde{g}^{\mu \nu} \partial_{\mu} \phi \partial_{\nu} \phi - V(\phi) \bigg] \, .
\end{align}
The metric $\tilde{g}_{\mu \nu}$ is related to $g_{\mu \nu}$ by a conformal rescaling and $M_p$ and $V(\phi)$ can be obtained from $f(R)$ and its derivatives. Details for this reformulation can be found in appendix \ref{app:constRfromFofR}.  A condition for this map to exist is that $f_R$ is invertible. This is always the case when \eqref{eq:fofR2} is satisfied.
The physical system described by the action \eqref{eq:fofR1} is classically equivalent to the system described by \eqref{eq:scalartensor1}. As the latter is just the theory of a scalar with a potential coupled to Einstein gravity it is manifestly ghost-free. In terms of the tensor-scalar formulation of the theory constant-curvature solutions have a very intuitive counterpart. As can be shown (see appendix \ref{app:constRfromFofR}), solutions to equation \eqref{eq:fofR4} correspond to extremal points of the potential $V(\phi)$:
\begin{align}
\label{eq:fofR5} d f(R) -2 R f_R(R) = 0 \, , \quad \Leftrightarrow \quad \frac{\partial V(\phi)}{\partial \phi} = 0 \, .
\end{align}

To close this section, we now argue that for all relevant cases the no-ghost condition \eqref{eq:fofR2} can always be satisfied. In particular, we shall be interested in solutions to \eqref{eq:fofR4} for $d=4$, but the following argument can always be made for any even value for $d$. The observation is that for $d$ even the function $f(R)$ considered here generically contains a term $a_{d/2} R^{d/2} \subset f(R)$. This term does not contribute to equation \eqref{eq:fofR4} and hence the value of $a_{d/2}$ does not affect constant-curvature solutions. However, this term gives a contribution $\tfrac{d}{2} a_{d/2} R^{d/2-1}$ to $f_R$. Thus, as along as we consider a constant-curvature solution with $R \neq 0$, we can always ensure that  $f_R > 0$ by choosing $a_{d/2}$ appropriately. Hence, for all cases of interest graviton ghosts can be avoided.

The quantum effective action for the curvature that we calculate holographically should be thought of as a kind of DBI action for gravity, in that it is applicable in cases where the scalar curvature is slowly varying.
This action was calculated from first principles in holographic
theories for the first time  in  \cite{R} and its strong and weak curvature asymptotics were analysed.

\section{Holographic dual}
\label{sec:holo}

In this section we set up the holographic dual to the $d$-dimensional
QFT to be integrated out, and calculate the corresponding contribution
to the gravitational effective action. The latter is computed in the
gravity dual by the classical {\em on-shell action}, i.e. the action
governing the $(d+1)$-dimensional gravitational theory evaluated on a solution which
is dual to a field theory RG flow on a constant-curvature   $d$-dimensional space-time.

First, in subsections 3.1 and 3.2  we briefly review the features of
the holographic geometries dual to RG flows in curved space-times.
We refer the reader to  \cite{R,F} for details.  Based
on these results the calculation of the effective action is
presented in some detail in sections 3.4 and 3.5, in which we recast the result of
\cite{F} in a form which is useful for the problem at hand.  A
summary of the holographic dictionary can be found in Appendix
\ref{app:dictionary}.\\

{\em Note: The reader who wishes to skip the details of the
  holographic calculation can skip this section and go directly to
  Section 4.} \\

\subsection{Holographic setup}
\label{sec:holosetup}

As stated before, we assume the QFT to be a
strongly coupled large-$N$ gauge theory, such that it corresponds to a
classical gravitational theory in $(d+1)$ dimensions. A general QFT
can be described by a CFT perturbed in the UV by a set of
operators. For simplicity, here we assume that there is just one
relevant scalar operator $\mathcal{O}$ present in the UV, which in the
dual theory requires the existence of exactly one scalar field $\f$
with a potential $V(\f)$. The scalar field is interpreted as the
running coupling corresponding to the operator  $\mathcal{O}$. The fact that the operator is relevant then
implies that the UV of the QFT is identified with a maximum of that
potential at $\f=\f_{UV}$. We  also assume that the QFT (when defined on flat
Minkowski space-time)  flows to a conformal fixed
point in the IR, which in the dual geometry is identified with a
minimum of $V(\f)$, at  $\f=\f_{IR}$

We then consider an Einstein-dilaton theory with action given by
\begin{align}
\label{eq:Sgrav} S_{\textrm{grav},(d+1)} = M^{d-1} \int d^{d+1} x \sqrt{|G|} \left( R^{(G)} -\frac{1}{2} \partial_a \f \partial^a \f - V(\f) \right) + S_{\textrm{GHY}} \, ,
\end{align}
where $G_{ab}$ is the metric of the $(d+1)$-dimensional bulk
space-time, $R^{(G)}$ the corresponding scalar curvature and $M$ is
the $(d+1)$-dimensional gravitational coupling constant, and
$S_{\textrm{GHY}}$  is the Gibbons-Hawking-York boundary term.

We shall be interested in solutions of the form:
\begin{align}
\label{eq:ansatz} x^a=(u, x^{\mu}) \, , \quad ds^2 = G_{ab} \, dx^a dx^b = du^2 + e^{2 A(u)} g_{\mu \nu} dx^{\mu} dx^{\nu} \, , \quad \f = \f(u) \, ,
\end{align}
which is the ansatz corresponding to holographic RG flows.
Here we
introduced a scale factor $A(u)$ and a $d$-dimensional metric $g_{\mu
  \nu}$ with scalar curvature $R$.  By shifting $A(u)$ by a constant
we can always ensure that $g_{\mu \nu}$ is identified with the metric
on which the dual field theory is defined. As we are exclusively
interested in QFTs on a  maximally symmetric space-time, we take
$g_{\mu \nu}$ to have  constant positive, negative or vanishing
curvature $R$. For non-zero $R$, the corresponding (A)dS$_d$ radius
$\alpha$ is defined by $R = \pm \tfrac{d(d-1)}{\alpha^2}$.

For {\em
  regular} geometries, the scalar
field $\f$ interpolates between the UV value, corresponding to the
maximum of $V(\f)$ (which we set to $\f_\textsc{UV}=0$ without loss of
generality), and an IR value $\f_0$.

Independently of the  value of $R$, as $\f \to 0$, the
bulk geometry approaches the boundary region of an asymptotically
AdS$_{d+1}$ space-time, with AdS radius given by the relation
\be  \label{elluv}
V(0) = - {d(d-1) \over \ell^2_\textsc{uv}}.
\ee
We  choose the boundary to be reached for $u\rightarrow - \infty$.

While the near-UV boundary region  of the geometry is universal, the
deep interior (corresponding to the IR) crucially
depends on the value of $R$. Regularity in the interior requires the
following features:
\begin{itemize}
\item For $R=0$ the IR endpoint of the flow is the value $\f_\textrm{IR}$
  corresponding to the minimum
  of the potential , and the interior asymptotes an interior AdS region
  as $u \to +\infty$, where the  scale factor $e^A \to 0$,  with a
  different radius $\ell_\textrm{IR}$ given by
\be \label{eq:LIRdef}
V(\f_\textrm{IR}) =- {d(d-1) \over \ell^2_\textsc{ir}}.
\ee
\item For $R>0$, the flows stops before reaching the IR fixed point,
  at a value $\f_0 < \f_{\textrm{IR}}$ (which depends on $R$ and on the
  relevant deformation parameter). This value is reached at a finite value $u_0$ of the
  holographic  coordinate. At $u=u_0$ the scale factor $e^A = 0$, the Euclidean geometry smoothly
  caps off, while the Lorentzian space-time exhibits a coordinate  horizon.
\item For $R<0$ the geometry has a turning point at $\f_0$, where
  $e^A$ has a minimum. The geometry can be continued for $\f> \f_0$,
  where the flows continues towards another UV region\footnote{At this stage we note that for field theories on AdS$_d$ the ansatz given in \eqref{eq:ansatz} is only consistent if a defect is also introduced along a portion of the bulk space-time boundary.
The precise nature of this defect is a matter of speculation and it
will not be investigated further here. For details on the precise
near-boundary geometry in the case of both dS$_d$ and AdS$_d$ slices
and conditions for the defect see \cite{R}.}
\end{itemize}

It will be useful to define a new set of functions $W(\f)$, $S(\f)$ and $T(\f)$ as follows:
\begin{align}
\label{eq:defWc} W(\f) & \equiv -2 (d-1) \dot{A} \, , \\
\label{eq:defSc} S(\f) & \equiv \dot{\f} \, , \\
\label{eq:defTc}  T(\f) & \equiv R \, e^{-2A} 
 \, .
\end{align}
Written in terms of these functions the equations of motion 
are
\begin{align}
\label{eq:EOM4} S^2 - SW' + \frac{2}{d} T &=0 \, , \\
\label{eq:EOM5} \frac{d}{2(d-1)} W^2 -S^2 -2 T +2V &=0 \, , \\
\label{eq:EOM6} SS' - \frac{d}{2(d-1)} SW - V' &= 0 \, .
\end{align}
where a prime denotes a derivative with respect to $\f$.

These functions have to be thought as functions of $\f$ after
inverting the relation between $\f$ and $u$ and substituting in  the
right hand sides.

In the flat case $R=0$, the equations of motion imply  $T=0$ and $S= dW/d\f$, and  $W(\f)$
reduces to the  superpotential of the solution \cite{SkenderisTownsend1, SkenderisTownsend2}.

\vspace{0.3cm}

\noindent \textbf{A first look at the on-shell action.} \newline
The most  important quantity which will calculate from holography is
the {\em on-shell action} $S_{on-shell}$, i.e. the gravitational
action \eqref{eq:Sgrav}
 evaluated on a solution corresponding to a
holographic RG flow, i.e.~a solution satisfying the ansatz
\eqref{eq:ansatz}. According to the holographic dictionary (see
Appendix \ref{app:dictionary}),  this gives directly the quantum
contribution to the effective action  from integrating out the dual
QFT\footnote{More precisely, this statement is true when the deforming
operator has dimension $\Delta > d/2$. Otherwise, one must take a
Legendre transform of the on-shell action with respect to the
deformation parameter.}

As will be sketched at the beginning of section \ref{sec:onshell}, this can be brought into the form
\begin{align}
\label{eq:Son1st} S_{\textrm{grav},(d+1)}^{\textrm{on-shell}} = M^{d-1} \int d^dx \sqrt{|g|}  \, \bigg[- 2(d-1) {\big[e^{dA} \dot{A} \big]}_{\textrm{UV}} + \frac{2}{d} R \int_{\textrm{UV}}^{\textrm{IR}} du \, e^{(d-2)A} \bigg] \, .
\end{align}
Here, UV and IR refer to the UV fixed point and IR end point of a holographic RG flow solution, which will be made precise later.
To remove clutter, we shall drop most of the sub- and superscripts on $S_{\textrm{grav},(d+1)}^{\textrm{on-shell}}$ and refer to the on-shell action just as $S_{\textrm{on-shell}}$ in the following.

It will be also convenient to rewrite \eqref{eq:Son1st} entirely in terms of quantities evaluated in the UV. This can be done by introducing a function $U$ as follows:
\begin{align}
\label{eq:UfromA}
U (u) = - \frac{2}{d} \, e^{-(d-2) A(u)} \int_{\textrm{IR}}^u \, d \tilde{u} \, e^{(d-2) A(\tilde{u})} \, ,
\end{align}
where the lower limit of the integration will be the value of $u$ coinciding with the IR end point. With the help of this the on-shell action can be written as
\begin{align}
\label{eq:Son2nd} S_{\textrm{on-shell}} = M^{d-1} \int d^dx \sqrt{|g|}  \, {\Big[ e^{dA} {\big( W + T U \big)} \Big]}_{\textrm{UV}} \, ,
\end{align}
where in addition to \eqref{eq:UfromA} we also used the definitions \eqref{eq:defWc} and \eqref{eq:defTc}.

Last, note that we can calculate the function $U$ also directly from the functions $W$ and $S$. As one can confirm explicitly, $U$ as defined in \eqref{eq:UfromA} is equivalent to the solution to the  following differential equation:
\begin{align}
\label{eq:Uequation} S U' - \frac{d-2}{2(d-1)} WU = - \frac{2}{d} \, ,
\end{align}
The integration constant in this equation is fixed by  the boundary
condition that $U$ vanishes at the IR end point of a flow. With this
choice, the solution reduces to equation (\ref{eq:UfromA}).

\subsection{Holographic RG flows for QFTs on Einstein metrics}
\label{sec:holoRGreview}
Given a solution of the form (\ref{eq:ansatz}), according to the holographic
dictionary the dual field theory data can
be read-off from the expansion of the solution near the AdS
boundary. This takes the  form

\begin{align}
\label{eq:Anearboundary} A(u) &\underset{u \rightarrow -\infty}{=} - \frac{u}{\ell_{\textsc{uv}}} - \frac{\ell_{\textsc{uv}}^2 |R|}{4d(d-1)} \, e^{2u / \ell_{\textsc{uv}}} + \ldots \, , \\
\label{eq:phinearboundary} \f(u) &\underset{u \rightarrow -\infty}{=}  \f_- \ell_{\textsc{uv}}^{\Delta_-} e^{\Delta_- u / \ell_{\textsc{uv}}} + \f_+ \ell_{\textsc{uv}}^{\Delta_+} e^{\Delta_+ u / \ell_{\textsc{uv}}} + \ldots \, ,
\end{align}
where $\f_{\pm}$ are integration constants,  and
\begin{align}
\label{eq:DeltapmUVdef} \Delta_\pm \equiv \frac{1}{2} \Big( d \pm \sqrt{d^2 + 4 m^2 \ell_{\textsc{uv}}^2} \Big) \, .
\end{align}
We fixed a further additive  integration constant in $A(u)$ to vanish,
so that the fixed metric $g_{\mu\nu}$ on each radial slice can be
identified with the QFT metric \cite{R}.
Note that we are near  a maximum of the potential, therefore $0 < \Delta_- \leq d/2$ and $\Delta_+ \geq d/2$.

In this work we shall employ the standard holographic dictionary (also referred to as `standard quantisation'). According to this, the value $\Delta_+$ is identified with the dimension of the scalar operator $\mathcal{O}$ dual to the bulk field $\f$. Furthermore, the quantity $\f_-$ corresponds to the source $j$ of $\mathcal{O}$ and has dimension $\Delta_-$. The parameter $\f_+$ is related to the vev $\langle \mathcal{O} \rangle$, i.e.
\be
\label{eq:jandvevdef} j= \f_- \, , \qquad \langle \mathcal{O} \rangle = (M \ell_{\textsc{uv}})^{d-1} \, (2\Delta_+ -d) \, \f_+ \, .
\ee
As $e^A \rightarrow +\infty$  at the AdS boundary, it follows
from \eqref{eq:defTc} that $T \rightarrow 0$ when approaching a UV
fixed point. Hence, in the UV we can also use $T$ write  $W$, $S$ and
$U$ as a perturbative expansion in $T$,
\begin{align}
\label{eq:WnearUVphiT} W(\f,T) &= \sum_{n=0}^{\infty} (\ell_{\textsc{uv}}^2 T)^n W_n(\f) = W_0(\f) + \ell_{\textsc{uv}}^2 T W_1 (\f) + \ell_{\textsc{uv}}^4 T^2 W_2(\f) + \mathcal{O}(\ell_{\textsc{uv}}^6 T^3) \, , \\
\label{eq:UnearUVphiT} U(\f,T) &= \sum_{n=0}^{\infty} (\ell_{\textsc{uv}}^2 T)^n U_n(\f) = U_0(\f) + \ell_{\textsc{uv}}^2 T U_1 (\f) + \mathcal{O}(\ell_{\textsc{uv}}^4 T^2) \, .
\end{align}
This form will be particularly advantageous later for the
regularisation and renormalization of the on-shell
action.\footnote{Ultimately, we can also derive a near-UV expansion
  for $T$ in (non-analytic) powers of $\f$, which in turn can be used
  to write the near-UV expressions of $W$ and $S$ as expansions in
  powers of $\f$ only. This has been done in e.g.~\cite{R,F}.} Inserting this ansatz into
the equations of motion one can then
solve for the functions $W_n$, $S_n$ and $U_n$. The analysis is shown
in appendix \ref{app:nearboundary} and here we only collect the
results\footnote{There is a second branch of solution which corresponds to
  flows driven by a vev, but with the source $\f_-$ set to zero  \cite{R}. To
  obtain the corresponding expressions, replace
  $\Delta_- \rightarrow \Delta_+$ in the above and also set $C=0$. We
  will not discuss this branch in the present work. }:
\begin{align}
W_0(\f) &= \frac{1}{\ell_{\textsc{uv}}} \bigg[2(d-1) + \frac{\Delta_{-}}{2} \f^2 + C \, |\f|^{\frac{d}{\Delta_-}} +  \mathcal{O}(\f^3) \bigg] \, , \\
W_1(\f) &= \frac{1}{\ell_{\textsc{uv}}} \bigg[\frac{1}{d} -\frac{(d-2) \Delta_-}{4d(d-1)(d-2-2\Delta_-)} \f^2 + \mathcal{O}(\f^3) + \mathcal{O}\big(C^2 \, |\f|^{\frac{2d}{\Delta_-}-2}  \big) \bigg] \, , \\
W_2(\f) &\underset{d=4}{=} - \frac{1}{192 \, \ell_{\textsc{uv}}} \, , \\
\nonumber \\
\nonumber U_0(\f) &= \ell_{\textsc{uv}} \bigg[\frac{2}{d(d-2)} + \frac{\Delta_-}{2d(d-1)(2 \Delta_- +2 -d)} \, \f^2 + \mathcal{O}\big(C^2 |\f|^{\frac{2d}{\Delta_-}-2} \big) \\
\label{eq:U0sol} & \hphantom{AAA} + B \, |\f|^{\frac{d-2}{\Delta_-}} \Big(1 + \mathcal{O}(\f) + \mathcal{O}\big(C|\f|^{\frac{d}{\Delta_-}-2}\big) + \mathcal{O}\big(B |\f|^{\frac{d-2}{\Delta_-}}\big) \Big) \bigg] \, , \\
\label{eq:U1sol} U_1(\f) &\underset{d=4}{=} \ell_{\textsc{uv}} \bigg[ \frac{1}{48 \Delta_-} \, \ln \f + \mathcal{O}(\f) + \mathcal{O}\big(C |\f|^{\frac{4}{\Delta_-}-2} \big)  + \mathcal{O}\big(B |\f|^{\frac{2}{\Delta_-}}\big) \bigg] \, ,
\end{align}
with $\Delta_-$  defined in \eqref{eq:DeltapmUVdef}.

 A few  observations are in order:
\begin{itemize}
\item Here we collected all the results which will be used later in
  $d=4$, as this will be the most relevant case for applications to
  our  universe. For this,  we shall need terms up to order
  $\mathcal{O}(T^2)$ in the near-boundary expansion of $W$, but for $U$ terms up to order $\mathcal{O}(T)$ will be sufficient.
\item The functions $W_2(\f)$ and $U_1(\f)$ have a different form for $d=4$ vs.~$d \neq 4$. For example, the appearance of the term $\log \f$ in the expression \eqref{eq:U1sol} for $U_1$ is specific to $d=4$. Here, whenever there is an ambiguity, we only display the result for $d=4$.
\item The parameter $C$ appearing in the  expansions above is one of
  the two integration constants of the  system of equations \eqref{eq:EOM4}--\eqref{eq:EOM6}.  It corresponds to the vev of $\mathcal{O}$ in units of the source $\f_-$,
\begin{align}
\label{eq:vevdef} C = (M \ell_{\textsc{uv}})^{-(d-1)} \frac{\Delta_-}{d} \, \langle \mathcal{O} \rangle \, |\f_-|^{- \frac{\Delta_+}{\Delta_-}} \, .
\end{align}
While $C$ appears  as a free parameter in the UV expansions, in a
complete solution it is
fixed once the IR end point $\f_0$ is specified. 
\item A  second integration constant is contained in $T$. To make this
  explicit one must write $T$ in terms of its near-boundary expansion
  in powers of $\f$. This can be done using the definition
  (\ref{eq:defTc}) and the boundary expansion \eqref{eq:Anearboundary}
  and \eqref{eq:phinearboundary},  which in turn is obtained from $W$ and $S$ by
  integrating  equations  \eqref{eq:defWc}
  and \eqref{eq:defSc}. This leads to: 
\begin{align}
T(\f) = \ell_{\textsc{uv}}^{-2} \, \mathcal{R} \, {|\f|}^{\frac{2}{\Delta_-}} \bigg(1 + \mathcal{O}(\f) + \mathcal{O} \big( C \, |\f|^{\frac{d}{\Delta_-}-2} \big) \bigg) \, ,
\end{align}
where
\begin{align}
\label{eq:calRdef} \mathcal{R} \equiv R \, |\f_-|^{-2 / \Delta_-} \, ,
\end{align}
The parameter $\mathcal{R}$  expresses the  the UV boundary curvature in units of the source.
It is  the second integration constant besides $C$ characterising the
solutions to equations \eqref{eq:EOM4}--\eqref{eq:EOM6}. Just like
$C$, while $\mathcal{R}$ appears as a free parameter in the above UV
expansions, it is  fixed by regularity of the complete solution by a
choice of the IR endpoint $\f_0$.
\item As $U(T, \f)$ is obtained by solving the ordinary differential equation \eqref{eq:Uequation}, this leads to the appearance of a further integration constant, which we denoted $B$ in \eqref{eq:U0sol}. Again, while $B$ enters as a free parameter in the UV, it will be determined once we consider a complete RG flow. In particular, it will be fixed by the condition that $U$ vanishes at the IR end point of a flow.\footnote{The parameter $B$ can be related to a holographic entanglement entropy across a suitably chosen entangling surface. For more details see e.g.~\cite{R,F}.}
\end{itemize}

\vspace{0.1cm}

To summarize,  on a complete flow the parameters $B$, $C$ and
$\mathcal{R}$ are fixed by regularity plus a choice of the endpoint
$\f_0$, i.e.  {\em regular} solutions of the system
\eqref{eq:EOM4}--\eqref{eq:EOM6} form a one-parameter family
parametrized by $\f_0$. Notice that this characterization is
coordinate-invariant, since $\f$ is a scalar and $W,S,T$ and $U$ are
scalar functions.  Each such solution of the system
\eqref{eq:EOM4}--\eqref{eq:EOM6} in turn corresponds to a
one-parameter family of solutions of the form (\ref{eq:ansatz}),
parametrized by a choice of the source $\f_-$: indeed, notice that the
latter does not appear in $W,S$ and $T$, but it appears in the
metric-dilaton solution which requires further integration.

\begin{align}
\textrm{Choice of } \f_0 + \textrm{regularity} 
\underset{\textrm{fixes}}{\longrightarrow} \mathcal{R} (\f_0), \, C(\f_0), \, B(\f_0) \, .
\end{align}
Hence, by scanning over all possible end points $\f_0$ that
 can be reached by a flow from a particular UV fixed point, one can
 build up the complete space of solutions, with parameters $\mathcal{R}
 (\f_0), \, C(\f_0), \, B(\f_0)$, attainable from that UV fixed
 point.
 
Next, for every interval over which $\mathcal{R}(\f_0)$ is monotonic, we can invert this to obtain $\f_0(\mathcal{R})$. Inserting this into $C(\f_0), \, B(\f_0)$, these can be interpreted as functions of $\mathcal{R}$, i.e.
\begin{align}
\mathcal{R} (\f_0), \, C(\f_0), \, B(\f_0) \longrightarrow \mathcal{R}, \, C(\mathcal{R}), \, B(\mathcal{R}) \, .
\end{align}
This way of presenting the results is more intuitive from the dual
field theory perspective, where one specifies  the UV value of the
boundary curvature $R$ and  of the operator source $\f_-$.

The parameters $ C(\mathcal{R})$ and  $B(\mathcal{R})$  
will appear explicitly in the on-shell action. 

\subsection{Solutions for $W$ and $U$: conformal case} \label{sec:WU-CFT}

For the theory at the fixed point theory,  which corresponds to  $\f_-=0$ and constant dilaton,
analytic expressions can be obtained for $W$
and $U$ defined in \eqref{eq:defWc} and \eqref{eq:UfromA}. In this case
the space-time is a foliation of AdS$_{d+1}$ by (A)dS$_{d}$ slices,
with scale factor given by (see e.g.~\cite{KarchRandall}):
\begin{align}
\label{eq:AdSScaleFactor}
e^{A(u)} = \left\{
  \begin{array}{l l l}
   \displaystyle{\frac{\ell}{\alpha} \sinh \left(-\frac{u-u_0}{\ell}\right)}, & \qquad
   -\infty < u \leq u_0,  &
   \qquad \textrm{dS}_d \, , \\
& \\
   \displaystyle{\exp\left(- \frac{u-u_0}{\ell}\right)}, &  \qquad -\infty < u <
   +\infty, &  \qquad \textrm{Minkowski}_d \, , \\
& \\
 \displaystyle{
   \frac{\ell}{\alpha} \cosh
   \left(\frac{u-u_0}{\ell}\right)}, & \qquad -\infty < u <
   +\infty, &  \qquad \textrm{AdS}_d \, . \\
  \end{array} \right.
\end{align}
Here we introduced the AdS${}_{d+1}$ length $\ell$ which is defined via
$\ell^2=-\frac{d(d-1)}{V_{\textrm{CFT}}}$ with $V_{\textrm{CFT}}$ the value of the potential for $V(\f)= \textrm{const}.$ or $V_{\textrm{CFT}} = V(\f_{\textrm{ext}})$ for a CFT associated with an extremum of the potential. The length scale
$\alpha$ was introduced in section \ref{sec:holosetup} and is related to the curvature $R$ of the (A)dS$_d$ slices as $R = \pm \tfrac{d(d-1)}{\alpha^2}$.

The quantity $u_0$ is an integration constant, which will be fixed by implementing the boundary condition $A(u \rightarrow - \infty) = - u / \ell$. This implies
\begin{align}
\label{eq:u0cases}
e^{2 u_0 / \ell} = \left\{
  \begin{array}{l l l}
   \displaystyle{\frac{4 \alpha^2}{\ell^2} = \hphantom{-} \frac{4 d (d-1)}{\ell^2 R}} \, , &  \qquad \textrm{dS}_d \, , \\
& \\
   \displaystyle{ \ \, 0} \, , &  \qquad \textrm{Minkowski}_d \, , \\
& \\
 \displaystyle{\frac{4 \alpha^2}{\ell^2} = - \frac{4 d (d-1)}{\ell^2 R}} \, , & \qquad \textrm{AdS}_d \, . \\
  \end{array} \right.
\end{align}
Hence choosing $u_0$ is equivalent to choosing the curvature $R$ of the (A)dS$_d$ slices.

Since we have $\f=\textrm{const},$   the inversion $u(\f)$ is not
possible. However $W$ and $U$ can still be defined by (\ref{eq:defWc},\ref{eq:UfromA}) but
now  treated as be functions of $T$ after inverting $u=u(T)$.

To calculate $W$ we start with its definition \eqref{eq:defWc} and insert the corresponding expressions for $A(u)$ from \eqref{eq:AdSScaleFactor}, which gives us $W$ as a function of $u$. Similarly, from \eqref{eq:AdSScaleFactor} and \eqref{eq:defTc} we obtain $T$ as a function of $u$. This can be inverted to give $u$ as a function of $T$. For dS$_d$ slicings $e^A$ and hence $T$ is monotonic in $u$ and thus the inversion can always be performed unambiguously. Inserting into $W$ we arrive at the desired expression:
\begin{align}
\label{eq:WCFT}
W(T) = \frac{2(d-1)}{\ell} \sqrt{1+ \frac{\ell^2 T}{d(d-1)}} \, ,
\end{align}
with $0 \leq T < \infty$. For AdS$_d$ slicings the geometry consists of the two halves $u \leq u_0$ and $u \geq u_0$ glued at $u=u_0$. On every half $e^A$ and hence $T$ is monotonic in $u$ and can be inverted, allowing us to drive an expression for $W(T)$ defined on either half of the geometry. Inserting the relevant results one again finds \eqref{eq:WCFT}, but now $T_{\textrm{min}} < T \leq 0$ with $T_{\textrm{min}} \equiv T(u_0) = -\tfrac{d(d-1)}{\ell^2}$.

To calculate $U(T)$ we use \eqref{eq:UfromA} and
\eqref{eq:AdSScaleFactor} to get an expression for $U$ as a function
of $u$, which we in turn eliminate in favour of $T$. The expression
\eqref{eq:UfromA} only determines $U$ up to a constant, which is fixed
by imposing  $U(u_0)=0$.
After some algebra,
for $d=4$ one obtains:
\begin{align}
\label{eq:UCFTdS} \textrm{dS}_4\textrm{:} \quad &U(T) \underset{d=4}{=} \frac{\ell}{48} \Bigg[ \ell^2 T \, \log \bigg( \sqrt{1 +  \frac{12}{\ell^2 T} } - \sqrt{ \frac{12}{\ell^2 T} } \bigg) + 12 \sqrt{1 + \frac{\ell^2 T}{12}} \, \Bigg] \, , \\
\label{eq:UCFTAdS} \textrm{AdS}_4\textrm{:} \quad &U(T) \underset{d=4}{=} \frac{\ell}{48} \Bigg[- \ell^2 T \, \log \bigg(\sqrt{-\frac{12}{\ell^2 T}} + \sqrt{-\frac{12}{\ell^2 T}-1}  \bigg) + 12 \sqrt{1 + \frac{\ell^2 T}{12}} \, \Bigg] \, .
\end{align}

As one can check explicitly, the expressions for $W(T)$ and $U(T)$ in \eqref{eq:WCFT} and (\ref{eq:UCFTdS}, \ref{eq:UCFTAdS}) satisfy the relation
\begin{align}
\label{eq:thermoidCFT} \frac{\partial W}{\partial T} = \frac{d-2}{2} U - T \frac{\partial U}{\partial T} \, ,
\end{align}
which will be helpful later when seeking for constant curvature
solutions. For dS$_d$ slicings, the identity \eqref{eq:thermoidCFT}
can be rephrased as the first law of thermodynamics by describing
dS$_d$ as a thermal system \cite{F}. Interestingly, we find that \eqref{eq:thermoidCFT} also holds in the case of AdS$_d$ slices where a thermodynamic interpretation is not readily available.

\subsection{The cutoff-regulated on-shell action}
\label{sec:onshell}

We are now ready to evaluate the on-shell action, starting from
equation (\ref{eq:Son2nd}),
\begin{align}
\label{eq:SonWTU}  S_{\textrm{on-shell}} &= M^{d-1} \int d^dx \sqrt{|g|}  \, {\Big[e^{dA} \Big( W + T U \big)  \Big]}_{\textrm{UV}} \, ,
\end{align}
and using the UV expansions we found for $W$ and $U$. As
the expression above is divergent in the UV, we will have to
regularize the expressions with a UV cut-off, and eventually defined a
renormalized on-shell action. A general result in holography is that these divergences are in one-to-one
correspondence with the UV divergences (and related counterterms) on the
field theory side.

\vspace{0.3cm}

\noindent \textbf{On-shell action for a cutoff CFT} \newline
We start with the simplest case of a  CFT, with the functions $W$ and
$U$  written  in terms  of $T$, as described in subsection
\ref{sec:WU-CFT}.

As discussed above, we have to introduce a UV cut-off to evaluate
equation (\ref{eq:SonWTU}). We hence introduce a cut-off coordinate
$u_\Lambda$ and  an associated   quantity $\Lambda$ with units
of energy,
\begin{align}
\label{eq:LambdaUVdef} \left. \Lambda \right|_{\textrm{UV CFT}} \equiv \frac{e^{A(u_{\Lambda})}}{\ell_\textsc{uv}} \, .
\end{align}

Then, from the definitions of $T$ and $\Lambda$ in \eqref{eq:defTc} and \eqref{eq:LambdaUVdef} it follows that $T$ evaluated at the UV cutoff surface is just the boundary curvature $R$ in units of the cutoff, i.e.
\begin{align}
\label{eq:TasRoverCutoffUVCFT} \ell_{\textsc{uv}}^2 T(u_\L) = \ell_{\textsc{uv}}^2 R \, e^{-2 A(u_\L)} = \frac{R}{\Lambda^2} \, .
\end{align}
Thus, $\ell_{\textsc{uv}}^2 T(u_\L)$ can be replaced everywhere by $R/\Lambda^2$. As the full expression will not be very useful in further chapters we hence refrain from giving it here. What is important to note that the on-shell action takes the form
\begin{align}
\label{eq:SCFTschematic} S_{\textrm{on-shell}}^{\textrm{CFT}} = \tilde{a}_{\textsc{uv}} \int d^dx \sqrt{|g|}  \, \Lambda^d \mathcal{F} \Big( \frac{R}{\Lambda^2} \Big) \, ,
\end{align}
where the function $\mathcal{F}$ can be reconstructed from $W$ and $T$.

In equation \eqref{eq:SCFTschematic} we extracted  the numerical factor $\tilde{a}_{\textsc{uv}}$, which we defined as the anomaly coefficient, i.e.~the coefficient of the term $\sim R^{d/2} \log R$ in the Lagrangian density, c.f.~\eqref{eq:Stotrenfinal}. Given our explicit expression for the on-shell action for the UV CFT, we are now in a position to determine $\tilde{a}_{\textsc{uv}}$, relating it to the parameters $M$ and $\ell_{\textsc{uv}}$ characterising the gravity dual. To this end we expand the on-shell action for the UV CFT in powers of $R / \Lambda^2$. To be specific, consider a field theory on $dS_4$. Then, from \eqref{eq:SonWTU}, \eqref{eq:WCFT}, \eqref{eq:UCFTdS} on can show that that
\begin{align}
\label{eq:SCFTexpanded} S_{\textrm{on-shell}}^{\textrm{CFT}} \underset{d=4}{=} (M \ell_{\textsc{uv}})^3 \int d^dx \sqrt{|g|} \, \left[ 6 \Lambda^4 + \frac{\Lambda^2 R}{2} + \frac{R^2}{192} + \frac{R^2}{96} \log \Big( \frac{R}{48 \Lambda^2} \Big) + \mathcal{O} \Big( \frac{R^3}{\Lambda^2} \Big) \right] .
\end{align}
Then, comparing e.g.~with \eqref{eq:Stotrenfinal} we can read off that
\begin{align}
\label{eq:anomalyUVdef} \tilde{a}_{\textsc{uv}} \underset{d=4}{=} (M \ell_{\textsc{uv}})^3 \, .
\end{align}

So far we have exclusively discussed the UV CFT associated with the UV fixed point of a holographic RG flow. We can equally consider the IR CFT associated with the corresponding IR fixed point at a minimum of the bulk potential. All expressions derived for the UV CFT also hold for the IR CFT as long as one replaces everywhere
\begin{align}
\ell_{\textsc{uv}} \rightarrow \ell_{\textsc{ir}} \, ,
\end{align}
where $\ell_{\textsc{ir}}$ was given in \eqref{eq:LIRdef} and defines the energy cutoff as
\begin{align}
\label{eq:LambdaIRdef} \left. \Lambda \right|_{\textrm{IR CFT}} \equiv \frac{e^{A(u_{\Lambda})}}{\ell_\textsc{ir}} \, .
\end{align}
We label the corresponding anomaly coefficient by $\tilde{a}_{\textsc{ir}}$. For $d=4$ it is given by
\begin{align}
\tilde{a}_{\textsc{ir}} \underset{d=4}{=} (M \ell_{\textsc{ir}})^3 \, .
\end{align}

\vspace{0.3cm}

\noindent \textbf{On-shell action for a cutoff QFT with RG flow} \newline
Next, we consider the more general case considered here which is a QFT obtained by adding a relevant deformation to a UV CFT. The on-shell action can still be written in the form \eqref{eq:SonWTU}, but now the functions $W$ and $U$ depend on $\f$, which exhibits a non-trivial flow. Evaluating $W$ and $U$ at the UV cutoff is then equivalent to evaluating them at
\begin{align}
\f_\L \equiv \f(u_\L) \, .
\end{align}
We begin by giving a physical interpretation to $\f_\L$.

Initially, consider defining the UV cutoff $\Lambda$ as in the case of the UV CFT \eqref{eq:LambdaUVdef}. Now choose $u_\L \rightarrow - \infty$ approaching the UV boundary such that $\f_\L \rightarrow 0$ approaches the UV fixed point. Then, from the UV expansions in \eqref{eq:Anearboundary} and \eqref{eq:phinearboundary} it follows that
\begin{align}
\label{eq:phiUVmotivation} \f_\L \underset{u_\L \rightarrow - \infty}{=} \f_- \ell_{\textsc{uv}}^{\Delta_-} e^{\Delta_- u_\L / \ell_{\textsc{uv}}} \underset{u_\L \rightarrow - \infty}{=} \f_- \ell_{\textsc{uv}}^{\Delta_-} e^{- \Delta_- A(u_\L)} = \frac{\f_-}{\Lambda^{\Delta_-}} \, ,
\end{align}
i.e.~for $u_\L \rightarrow - \infty$ the quantity $\f_\L$ is just the UV source $\f_-$ in units of the UV cutoff. Once we move the UV cutoff away from the boundary, this is no longer true. However, for any value of the UV cutoff we can always interpret $\f_\L$ as the source of the cut-off theory in units of the UV cutoff. Thus we define
\begin{align}
\label{eq:mfromphiUV} m \equiv |\f_\L|^{1/ \Delta_-} \Lambda \, ,
\end{align}
with $m$ having the interpretation as the mass scale associated with the source of the cut-off theory. We take this as the holographic equivalent of the mass scale introduced in \eqref{eq:mdef}. From \eqref{eq:phiUVmotivation} it follows that
\begin{align}
\label{eq:mUVlimit} m \underset{u_\L \rightarrow - \infty}{\longrightarrow} |\f_-|^{1/ \Delta_-} \, ,
\end{align}
i.e.~$m$ becomes the mass scale associated with the UV source.

In addition to the source $m$, the functions $W$ and $U$ also depend on the metric source $R$. Recall that in section \ref{sec:holoRGreview} we developed solutions for $W$ and $U$ in a double expansion in $\f$ and $T$. When evaluated at the UV cutoff, the expansion in $\f_\L$ will become an expansion in powers of $m / \Lambda$, while the expansion in $T(\f_\L)$ will later be related to an expansion in $R \Lambda^{-2}$.

Now consider a holographic RG flow solution for $W(\f)$, $U(\f)$ and $T(\f)$ with a fixed value of IR end/ turning point $\f_0$. Picking a value $\f_\L$ is equivalent to choosing a value $m / \Lambda$. By letting our choice of $\f_\L$ vary over the allowed range ($\f_{\textsc{uv}} < \f_\L < \f_0$) we can then dial $m/ \Lambda$ to any value allowed on this flow.\footnote{Here we implicitly assume that an RG flow away from the UV is always in the positive $\f$-direction.} However, choosing a value for $\f_\L$ also fixes the corresponding value $T(\f_\L)$ and hence $R/ \Lambda^2$. Thus, when restricting to a single flow (i.e.~fixed $\f_0$) we cannot choose the sources $m$ and $R$ independently. To do so, we need to consider the whole family of RG flows associated with the UV fixed point at $\f_{\textsc{uv}}$, i.e.~all flows with any value of $\f_0$ that can be reached from $\f_{\textsc{uv}}$.

Over this family of flows it is then convenient to consider $W$ and $U$, when evaluated at the cutoff, as functions of both $\f_\L$ and $T(\f_\L)$. This can be seen as follows. First, we introduce the following shorthand notation for quantities evaluated at the UV cutoff:
\begin{align}
\label{eq:TWULdef} T_\L \equiv T(\f_\L) \, , \qquad W_\L \equiv W(\f_\L) \, , \qquad U_\L \equiv U(\f_\L) \, .
\end{align}
Next, choose a value for $\f_\L$ with $\f_\L \in [\f_{\textsc{uv}}, \f_{\textsc{ir}}]$.\footnote{To be specific, here we assume that $\f_{\textsc{ir}} >  \f_{\textsc{uv}}$.} For this $\f_\L$ we then consider all RG flow solutions with IR end point/ turning point $\f_0 \in [\f_\L, \f_{\textsc{ir}}]$, recording the corresponding values $T_\L$, $W_\L$ and $U_\L$. This is repeated for all values $\f_\L \in [\f_{\textsc{uv}}, \f_{\textsc{ir}}]$. The results can be collected in two tables of triplets of numbers
\begin{align}
\big(\f_\L \, , \ T_\L \, , \ W_\L \big) \, , \qquad \big(\f_\L \, , \ T_\L \, , \ U_\L \big) \, ,
\end{align}
From these tables we can then write $W_\L$ and $U_\L$, when considered over the whole space of holographic RG flow solutions, as functions of $\f_\L$ and $T_\L$:
\begin{align}
\label{eq:WLULasfuncofTL} W_\L= W_\L\big( \f_\L, T_\L \big) \, , \qquad U_\L= U_\L\big( \f_\L, T_\L \big) \, .
\end{align}

It is these expressions $W_\L\big( \f_\L, T_\L \big)$ and $U_\L\big( \f_\L, T_\L \big)$ which are inserted into the expression for the on-shell action \eqref{eq:SonWTU}. Using \eqref{eq:mfromphiUV} we can trade $\f_\L$ for $m/ \Lambda$ and, as we shall argue shortly, the value of $T_\L$ determines $R / \Lambda^2$. Thus for the QFTs considered here the on-shell action given in \eqref{eq:SonWTU} will take the following schematic form
\begin{align}
\label{eq:SQFTschematic} S_{\textrm{on-shell}}= \tilde{a}_{\textsc{uv}} \int d^dx \sqrt{|g|}  \, \Lambda^d \mathcal{F} \Big( \frac{m}{\Lambda} , \frac{R}{\Lambda^2} \Big) \, .
\end{align}
For convenience, we have factored out $\tilde{a}_{\textsc{uv}}$, i.e.~the anomaly coefficient of the UV CFT which is given by \eqref{eq:anomalyUVdef}.

So far we defined the UV cutoff $\Lambda$ as in \eqref{eq:LambdaUVdef}, but this definition will turn out not to be very useful or even problematic for the QFTs considered here. Recall that the QFTs studied here are defined in terms of a UV CFT perturbed by a relevant deformation, leading to a non-trivial RG flow. In flat space this RG flow then ends at an IR fixed point associated with an IR CFT. In holography this choice of QFT corresponds to a bulk potential $V(\f)$ with a maximum which we always pick to be at $\f=\f_{\textsc{uv}}=0$, and a minimum at some $\f=\f_{\textsc{ir}}$. For certain values of the parameter $\f_\L$ we expect the QFT to be mainly dominated by its UV ($\f_\L \rightarrow 0$) or IR ($\f_\L \rightarrow \f_{\textsc{ir}}$) fixed point, i.e.~by either the UV and IR CFT. As argued before, we have to define the UV cutoff differently for the UV and IR CFTs, \eqref{eq:LambdaUVdef} vs.~\eqref{eq:LambdaIRdef} to get universal expressions for CFTs. As the QFTs considered here interpolate between a UV and a IR CFT a good definition of UV cutoff should also interpolate between the definitions \eqref{eq:LambdaUVdef} and \eqref{eq:LambdaIRdef} when going from $\f_\L =0$ to $\f_\L= \f_{\textsc{ir}}$. One way of realising this is to define the UV energy cutoff $\Lambda$ as
\begin{align}
\label{eq:LambdaQFTdef} \Lambda \equiv \frac{e^{A(u_\L)}}{\tilde{\ell}(\f_\L)} \, ,
\end{align}
with
\begin{align}
\label{eq:lambdatildedef} \tilde{\ell}^{-2} (\f_\L) \equiv - \frac{V(\f_\L)}{d(d-1)} \, .
\end{align}
This is what we shall use in the following. Furthermore, we keep the
definition of $m$ given in \eqref{eq:mfromphiUV}, but now take
$\Lambda$ to be defined as in \eqref{eq:LambdaQFTdef}. For holographic
RG flow solutions with UV fixed point at a maximum $\f_{\textsc{uv}}$
of the potential $V$ and IR end/ turning point $\f_0$ in the interval
between that maximum and a neighbouring minimum at $\f_{\textsc{ir}}$,
the cutoff $\Lambda$ as defined in \eqref{eq:LambdaQFTdef} decreases
monotonically as $\f_\L$ is varied from $\f_{\textsc{uv}}$ to $\f_0$
for fixed $u_\L$. This would not be the case for a holographic RG flow
with UV fixed point at a minimum of $V$ or for RG flows that skip
extrema along the flow \cite{exotic}. In these situations multiple values of $\f_\L$ would give rise to the same value of $\Lambda$ and \eqref{eq:LambdaQFTdef} cannot be regarded as a suitable UV cutoff for the corresponding field theory.\footnote{In these cases one can instead define $$\Lambda =\frac{e^{A(u_\L)}}{\hat{\ell}(\f_\L, T_\L)} \, , \quad \textrm{with} \quad \hat{\ell}^{-2} (\f_\L, T_\L) \equiv \frac{W_\L(\f_\L, T_\L)}{4 (d-1)^2} - \frac{T_\L}{d(d-1)} \, ,$$ which does not suffer from the non-monotonicity issues of \eqref{eq:LambdaQFTdef}.} However, in this work we shall exclusively consider flows with UV fixed point at a maximum that do not skip other extrema, so that these problems do not occur.

Next, we examine how $T_\L$ is related to $R \Lambda^{-2}$. From the definitions \eqref{eq:lambdatildedef} and \eqref{eq:LambdaQFTdef} it follows that
\begin{align}
\label{eq:relateRoverLambdatoT} R \Lambda^{-2} = \tilde{\ell}^2 \, R \, e^{-2A(u_\L)} = \tilde{\ell}^2 \, T_\L = - \frac{d(d-1) \, T_\L}{V(\f_\L)} \, .
\end{align}
Thus a variation in $T_\L$ indeed induces a change in $R \Lambda^{-2}$. However, unlike in the case of a CFT where $R \Lambda^{-2}$ was determined by $T_\L$ alone, c.f.~\eqref{eq:TasRoverCutoffUVCFT}, here $R \Lambda^{-2}$ is also affected by a change of $\f_\L$ due to the  $\f_\L$-dependence in $V(\f_\L)$.

Using \eqref{eq:LambdaQFTdef} together with \eqref{eq:lambdatildedef} in the expression for the on-shell action \eqref{eq:SonWTU} this can be written as
\begin{align}
\label{eq:SonWTUnewLambda}  S_{\textrm{on-shell}} &= (M \tilde{\ell})^{d-1} \int d^dx \sqrt{|g|}  \, \Lambda^d \, \tilde{\ell} \, \big(W_\L + T_\L U_\L \big) \, .
\end{align}
Comparing with \eqref{eq:SQFTschematic} we can read off an explicit expression for $\mathcal{F}$ from \eqref{eq:SonWTUnewLambda}, finding:
\begin{align}
\mathcal{F} = {\left( \tfrac{\tilde{\ell}}{\ell_{\textsc{uv}}} \right)}^{d-1} \, \tilde{\ell} \, \big(W_\L + T_\L U_\L \big) \, ,
\end{align}
where we also used \eqref{eq:anomalyUVdef}.

We close this subsection with an explicit expression for the cutoff-regulated on-shell action for $\Lambda \rightarrow \infty$, $m \Lambda^{-1} \rightarrow 0$, $R \Lambda^{-2} \rightarrow 0$. In this regime, from \eqref{eq:relateRoverLambdatoT} one finds:
\begin{align}
R \Lambda^{-2} = \ell_{\textsc{uv}}^2 T_\L \Big( 1 + \mathcal{O} (\f_\L^2) \Big) \, ,
\end{align}
confirming that an expansion in small $m \Lambda^{-1}$, $R \Lambda^{-2}$ is consistent with an expansion in small $\f_\L$, $\ell^2_{\textsc{uv}} T_\L$. For concreteness, we then restrict to $d=4$ and insert our near-boundary expansions for $W_\L(\f_\L, T_\L)$ and $U(\f_\L, T_\L)$ in \eqref{eq:WnearUVphiT} and \eqref{eq:UnearUVphiT} into the expression for the on-shell action \eqref{eq:SonWTUnewLambda}. After some algebra one obtains:
\begin{align}
\nonumber S_{\textrm{on-shell}} \underset{d=4}{=} (M \ell_{\textsc{uv}})^3 \int d^4x \sqrt{|g|} \, \bigg[ & 6 \Lambda^4 \Big( 1 + \mathcal{O} \big( (\tfrac{m}{\Lambda})^{2 \Delta_-} \big) \Big) + \frac{1}{2} R \Lambda^2 \Big( 1 + \mathcal{O} \big( (\tfrac{m}{\Lambda})^{2 \Delta_-} \big) \Big) \\
\nonumber &+ \frac{R^2}{96} \log \Big( \frac{m^2}{\Lambda^2} \Big) - \frac{R^2}{192} + m^4 C \big(\tfrac{R}{m^2} \big) + R m^2 B \big( \tfrac{R}{m^2} \big) \\
\nonumber & + \textrm{ vanishing for } \Lambda \rightarrow \infty \, .
\end{align}
This indeed takes the schematic form of the cutoff-regulated action as asserted in \eqref{eq:Seffdivergences}, i.e.~all UV-divergent terms come in an expansion in integer powers of $R$. One can also confirm that this reduces to the on-shell action for the UV CFT in \eqref{eq:SCFTexpanded} in the limit $m \rightarrow 0$. To do so we use the analytic expressions for $C$ and $B$ for large $Rm^{-2}$ given in appendix \ref{app:largesmallR}. Here the relevant results are
\begin{align}
m^4 C \big(\tfrac{R}{m^2} \big) &\underset{m \rightarrow 0}{\longrightarrow} 0 \, , \\
R m^2 B \big(\tfrac{R}{m^2} \big) &\underset{m \rightarrow 0}{\longrightarrow} \frac{R^2}{96} \Big( 1+ \log \tfrac{R}{48 m^2} \Big) \, .
\end{align}

\subsection{The renormalized on-shell action}
\label{sec:Sonren}
If we want to send the cut-off $\Lambda$ to infinity, we have to
define a renormalized on-shell action.
This can be obtained from the cutoff-regulated on-shell action by the addition of counterterms, with the subsequent application of the limit $\Lambda \rightarrow \infty$. For this procedure the subtleties with the definition of the UV cutoff $\Lambda$ discussed in the previous section will not be important. Thus, here $\Lambda$ will be defined as for the UV CFT, i.e.~it is given by \eqref{eq:LambdaUVdef}. With this definition the regulated on-shell action in \eqref{eq:SonWTU} can just be written as
\begin{align}
\label{eq:Son2}  S_{\textrm{on-shell}} &= M^{d-1} \ell_{\textsc{uv}}^d \int d^dx \sqrt{|g|}  \, \Lambda^d \Big( W + T U \big)_{\textrm{UV}} \, .
\end{align}
The renormalized on-shell action is thus given by
\begin{align}
\label{eq:Sonrendef} S_{\textrm{on-shell}}^{\textrm{ren}} = \lim_{\Lambda \rightarrow \infty} \Big( S_{\textrm{on-shell}} + \sum_{n=0}^{\lfloor d/2 \rfloor} S_{ct}^{(n)} \Big) \, .
\end{align}
The counterterms can be written in covariant form as
\begin{align}
\label{eq:Sct0def} S_{ct}^{(0)} & = - M^{d-1} \int d^dx \sqrt{|\gamma (\f) |} \, W_{ct} (\f) \Big|_{\f_\L} \, , \\
\label{eq:Sct1def} S_{ct}^{(1)} & = - M^{d-1} \int d^dx \sqrt{|\gamma (\f) |} \, R^{(\gamma)} (\f) \, U_{ct} (\f) \Big|_{\f_\L} \, , \\
\label{eq:Sct2def} S_{ct}^{(2)} & = - M^{d-1} \int d^dx \sqrt{|\gamma (\f) |} \, \big(R^{(\gamma)}(\f) \big)^2 \, Y_{ct} (\f) \Big|_{\f_\L} \, , \\
\nonumber & \vdots
\end{align}
where $\gamma_{\mu\nu}$ is the induced metric on radial slices,
\be\label{eq:gammadef}
\gamma_{\mu\nu} = e^{2A(u)} g_{\mu\nu}
\ee
 and
the functions $W_{ct} (\f)$, $U_{ct} (\f)$ and $Y_{ct} (\f)$ are solutions to the following equations:
\begin{align}
\label{eq:Wcteq} W_{ct}' W_{ct}' - \frac{d}{2(d-1)} W_{ct} W_{ct} &= 2V \, , \\
\label{eq:Ucteq} W_{ct}' U_{ct}' - \, \frac{(d-2)}{2(d-1)} W_{ct} U_{ct} &= - 1 \, , \\
\label{eq:Ycteq} W_{ct}' Y_{ct}' - \, \frac{(d-4)}{2(d-1)} \, W_{ct} Y_{ct} &= \frac{(d-2)^2}{d^2(d-1)} (U_{ct})^2 - \frac{2}{d}(U_{ct}')^2 \, .
\end{align}
We shall be mainly interested in the case $d=4$ in which case the three counterterms \eqref{eq:Sct0def}--\eqref{eq:Sct2def} are all we need.

In the vicinity of a maximum of $V$ at $\f=0$ one can solve for $W_{ct}$ and $U_{ct}$ as an expansion in powers of $\f$ as follows:
\begin{align}
\label{eq:Wctnearboundary} W_{ct} &= \frac{1}{\ell_\textsc{uv}} \, \Big[ \, 2(d-1) + \frac{\Delta_-}{2} \f^2 + C_{ct} \, |\f|^{\frac{d}{\Delta_-}} + \ldots \Big] \, , \\
\label{eq:Uctnearboundary}U_{ct}  &= \, \ell_\textsc{uv} \ \bigg[ \, \frac{1}{d-2} + B_{ct} \, |\f|^{\frac{d-2}{\Delta_-}} + \ldots \bigg] \, ,
\end{align}
where $C_{ct}$ and $B_{ct}$ are integration constants. Similarly, given the solutions for $W_{ct}$ and $U_{ct}$ one can solve \eqref{eq:Ycteq} for $Y_{ct}$ in the vicinity of $\f=0$. However, one has to distinguish the cases of $d \neq 4$ and $d = 4$ in this analysis: In the former case the expression for  $Y_{ct}$ is an expansion in powers of $\f$ while in the latter case a term $\sim \ln \f$ appears. Here we only record the result for $d=4$,in which case we find:
\begin{align}
\label{eq:Yctnearboundary4d} Y_{ct}  &\overset{d=4}{=} \ell_{\textsc{uv}}^3 \, \bigg[ \, \frac{1}{48 \Delta_-} \, \ln(\f) + \mathcal{A}_{ct} + \ldots \bigg] \, ,
\end{align}
with $\mathcal{A}_{ct}$ an integration constant.

In the end we want to write the renormalized on-shell action in terms of an integral with volume form $d^4x \, \sqrt{|g|}$ and UV curvature $R$. To this end note that, given the definition of the induced metric in \eqref{eq:gammadef} and that of the UV cutoff $\Lambda$ in \eqref{eq:LambdaUVdef}, we can write
\begin{align}
\int d^d x \sqrt{|\gamma (\f_\L) |} &= \ell_{\textsc{uv}}^d \quad \, \int d^d x \sqrt{| g |} \, \Lambda^d \, , \\
\int d^d x \sqrt{|\gamma (\f_\L) |} \, R^{(\gamma)}(\f_\L) &= \ell_{\textsc{uv}}^{d-2} \, \int d^d x \sqrt{| g |} \, R \, \Lambda^{d-2} \, , \\
\int d^dx \sqrt{|\gamma (\f_\L) |} \, \big(R^{(\gamma)}(\f_\L) \big)^2 &= \ell_{\textsc{uv}}^{d-4} \int d^d x \sqrt{| g |} \, R^2 \, \Lambda^{d-4} \, .
\end{align}

We now have all the ingredients to derive an expression for the renormalized on-shell action in $d=4$. Starting with \eqref{eq:Sonrendef} and substituting for the counterterms with the results above as well as using \eqref{eq:Son2} for $S_{\textrm{on-shell}}$, we find
\begin{align}
\nonumber S_{\textrm{on-shell}}^{\textrm{ren}, d=4} &= \lim_{\Lambda \rightarrow \infty} \bigg( S_{\textrm{on-shell}}^{d=4} + S_{ct}^{(0)} + S_{ct}^{(1)} + S_{ct}^{(2)} \bigg) \\
\nonumber &= \lim_{\Lambda \rightarrow \infty} \bigg( M^3 \int d^4 x \sqrt{| g |} \, \Big[ \ell^4 \Lambda^4 \big(W-W_{ct} \big) + \ell^2 R \Lambda^2 \big(U -U_{ct} \big) - R^2 \, Y_{ct} \Big]_{\f_\L} \bigg) \, .
\end{align}
Inserting the near-boundary expansions for $W$ and $U$ from appendix \ref{app:nearboundary} and for $W_{ct}$, $U_{ct}$, $Y_{ct}$ from \eqref{eq:Wctnearboundary}, \eqref{eq:Uctnearboundary}, \eqref{eq:Yctnearboundary4d} we arrive at
\begin{align}
\nonumber S_{\textrm{on-shell}}^{\textrm{ren}, d=4} &= (M \ell)^3 \, |\f_-|^{4 / \Delta_-} \int d^4 x \sqrt{| g |} \, \Big[ \big(C(\mathcal{R})-C_{ct} \big) + \mathcal{R} \big(B(\mathcal{R})-B_{ct} \big)  \\
& \hphantom{AAAAAAAAAAAAAAAAAAAAAAAAlllll} + \mathcal{R}^2 \big(-\tfrac{1}{192} - \mathcal{A}_{ct} \big) \Big] \, .
\end{align}
As discussed in section \ref{sec:strategy}, the terms involving the arbitrary constants $C_{ct}$, $B_{ct}$ and $\mathcal{A}_{ct}$ capture the effect of renormalisation of the cosmological constant, the Einstein-Hilbert term and the $R^2$ term. We can group the term $\tfrac{1}{192}$ multiplying $\mathcal{R}^2$ with $\mathcal{A}_{ct}$, as this also just shifts the coefficient of the $R^2$-term. New effects due to the backreaction of the QFT are contained in the terms containing $C(\mathcal{R})$ and $B(\mathcal{R})$, i.e.
\begin{align}
\label{eq:Srenuniversal} S_{\textrm{on-shell}}^{\textrm{univ}, d=4} &= (M \ell_{\textsc{uv}})^3 \, |\f_-|^{4 / \Delta_-} \int d^4 x \sqrt{| g |} \, \Big[ C(\mathcal{R}) + \mathcal{R} B(\mathcal{R}) \Big] \, .
\end{align}
We refer to this as the `universal' contribution to the on-shell action, as this is the part that is manifestly independent of the arbitrary constants $C_{ct}$, $B_{ct}$ and $\mathcal{A}_{ct}$.

We can now make contact with the notation introduced in section \ref{sec:strategy}. There the `universal' contribution due to the renormalized QFT was written as (c.f.~\eqref{eq:Stotrenfinal}):
\begin{align}
\label{eq:Srenuniversal2} S_{\textrm{on-shell}}^{\textrm{univ}, d=4} &= \tilde{a}_{\textsc{uv}} \int d^4 x \sqrt{| g |} \, \bigg[ \tfrac{1}{96} R^2 \Big( 1+ \log \big( \tfrac{R}{48 m^2}\big) \Big) + m^4 \, \mathcal{G} \Big( \tfrac{R}{m^2}\Big) \bigg] \, ,
\end{align}
where we specialised to $d=4$. As $m = |\f_-|^{1 / \Delta_-}$ we identify $\mathcal{R} = R m^{-2}$. Then, comparing \eqref{eq:Srenuniversal} and \eqref{eq:Srenuniversal2} we find
\begin{align}
\label{eq:calGdef} \mathcal{G}(R m^{-2}) \overset{d=4}{=} C(\mathcal{R}) + \mathcal{R} B(\mathcal{R}) - \tfrac{1}{96} \mathcal{R}^2 \Big( 1+ \log \big( \tfrac{\mathcal{R}}{48}\big) \Big) \, .
\end{align}
where we also used \eqref{eq:anomalyUVdef}. As described in section \ref{sec:holoRGreview}, we typically have to resort to numerical methods to determine $C(\mathcal{R})$ and $B(\mathcal{R})$. However, for $\mathcal{R} \rightarrow 0$ and $\mathcal{R} \rightarrow \infty$ one can obtain analytic expressions. The calculation and results are collected in appendix \ref{app:largesmallR}. Using these results, one can show that $\mathcal{G}(R m^{-2})$ as given by \eqref{eq:calGdef} satisfies
\begin{align}
m^4 \mathcal{G}(R m^{-2}) \underset{m \rightarrow 0}{\longrightarrow} 0 \, ,
\end{align}
as stated in section \ref{sec:strategy}.

\subsection{Backreaction on constant-curvature solutions}
\label{sec:QFTcontrib}
In this section we analyse how backreaction due to the QFT affects the equation for constant-curvature solutions.

As noted in section \ref{sec:fofR}, the system of study in this work takes the form of a $f(R)$ theory of gravity. In the following, we shall focus on the contribution to this function due to the QFT, which we  denote by $f_{\textrm{QFT}}(R)$. This is related to the on-shell action as
\begin{align}
S_{\textrm{on-shell}} = \int d^d x \, \sqrt{|g|} \, f_{\textrm{QFT}}(R) \, .
\end{align}
Here $S_{\textrm{on-shell}}$ denotes collectively the cutoff-regulated on-shell action \eqref{eq:SonWTUnewLambda}, or the universal part of the renormalized on-shell action \eqref{eq:Srenuniversal}, depending on which system we are interested in.

Constant-curvature solutions are solutions to the equation \eqref{eq:fofR4}, which can also be written as
\begin{align}
\mathcal{D}_R f(R) =0 \, , \qquad \textrm{with} \qquad \mathcal{D}_R \equiv \bigg(d - 2 R \frac{\partial}{\partial R}  \bigg) \, .
\end{align}
The contribution to the LHS of this equation due to the presence of the QFT is
\begin{align}
\mathcal{D}_R f_{\textrm{QFT}} (R) \, .
\end{align}
In the following we derive explicit expressions for this contribution.

\vspace{0.3cm}

\noindent \textbf{CFT with a cutoff}: To be specific, here we consider the UV CFT. In this case $f_{\textrm{QFT}}$ is given by:
\begin{align}
\label{eq:fQFTcutoff} f_{\textrm{QFT}} =  (M \ell_{\textsc{uv}})^{d-1} \, \Lambda^d \, \ell_{\textsc{uv}}  \,\Big[ W_{\Lambda} (T_\L) + T_\L U_{\L}(T_\L) \Big] \, .
\end{align}
This can be read off from the on-shell action in \eqref{eq:SonWTU}. The UV energy cutoff $\Lambda$ is defined as in \eqref{eq:LambdaUVdef}. For the UV CFT $W_{\Lambda} (T_\L)$ and $U_{\Lambda} (T_\L)$ are given by \eqref{eq:WCFT} and (\ref{eq:UCFTdS}, \ref{eq:UCFTAdS}) with the substitution $\ell \rightarrow \ell_{\textsc{uv}}$.

We now act on $f_{\textrm{QFT}}$ with the differential operator $\mathcal{D}_R$. Recall that for a CFT $\ell_{\textsc{uv}}^2T_\L = R \Lambda^{-2}$. Thus we can trade the $R$-derivative at constant $\Lambda$ in $\mathcal{D}_R$ for a derivative with respect to $T_\L$:
\begin{align}
\mathcal{D}_R = \bigg(d - 2 R \frac{\partial}{\partial R}  \bigg) = \bigg(d - 2 T_\L \frac{\partial}{\partial T_\L}  \bigg) \, .
\end{align}
Applying this to $f_{\textrm{QFT}}$ one finds
\begin{align}
\label{eq:DRfQFT1} \mathcal{D}_R f_{\textrm{QFT}} = d (M \ell_{\textsc{uv}})^{d-1} \, \Lambda^d \, \ell_{\textsc{uv}}W_\L - 2 (M \ell_{\textsc{uv}})^{d-1} \, \Lambda^d \, \ell_{\textsc{uv}}T_\L \bigg( \frac{\partial W_\L}{\partial T_\L} - \frac{d-2}{2} U_\L + T_\L \frac{\partial U_\L}{\partial T_\L} \bigg) \, .
\end{align}
As we now show, this can be simplified drastically. Recall that $W_\L$ and $U_\L$ satisfy the differential identity \eqref{eq:thermoidCFT}. Using this in the above, the second term in \eqref{eq:DRfQFT1} vanishes and we are left with
\begin{align}
\label{eq:DRfQFT2} \mathcal{D}_R f_{\textrm{QFT}} = d (M \ell_{\textsc{uv}})^{d-1} \, \Lambda^d \, \ell_{\textsc{uv}} W_\L \, .
\end{align}
For a CFT it is hence the function $W_\L$ alone that enters the equation for constant curvature solutions. We can repeat the analysis for the IR CFT. The result is that given in \eqref{eq:DRfQFT2} with all instances of $\ell_{\textsc{uv}}$ replaced by $\ell_{\textsc{ir}}$.

We now introduce notation that will be helpful in section \ref{sec:ResultsQFTcutoff}, where we study constant-curvature solutions. In particular, we define a function $\mathcal{W}(R \Lambda^{-2})$ as
\begin{align}
\label{eq:DRfQFT2b} \mathcal{W}(R \Lambda^{-2}) \equiv 2(d-1) \, \sqrt{1 + \frac{R \Lambda^{-2}}{d(d-1)}} \, .
\end{align}
The observation is that for a CFT we can always write $W_\L$ in terms of $\mathcal{W}$. For example, for the UV and IR CFTs we find
\begin{align}
\ell_{\textsc{uv}} \left. W_\L(T_\L) \right|_{\textrm{UV-CFT}} &= \mathcal{W}(R \Lambda^{-2}) \, , \qquad \textrm{with} \qquad R \Lambda^{-2} = \ell_{\textsc{uv}}^2 T_\L \, , \\
\ell_{\textsc{ir}} \left. W_\L(T_\L) \right|_{\textrm{IR-CFT}} &= \mathcal{W}(R \Lambda^{-2}) \, , \qquad \textrm{with} \qquad R \Lambda^{-2} = \ell_{\textsc{ir}}^2 T_\L \, ,
\end{align}
for the expression $W_\L(T_\L)|_{\textrm{CFT}}$ given by \eqref{eq:WCFT}. Thus, the contribution to the condition for constant-curvature solutions due to the UV CFT \eqref{eq:DRfQFT2} can be written as
\begin{align}
\label{eq:DRfQFT2c} \mathcal{D}_R f_{\textrm{QFT}} = d (M \ell_{\textsc{uv}})^{d-1} \, \Lambda^d \, \mathcal{W}(R \Lambda^{-2}) \underset{d=4}{=} 4 \tilde{a}_{\textsc{uv}} \, \Lambda^4 \, \mathcal{W}(R \Lambda^{-2}) \, ,
\end{align}
where in the last step \eqref{eq:anomalyUVdef} has been used. The expression for the IR CFT is obtained upon replacing $\ell_{\textsc{uv}} \rightarrow \ell_{\textsc{ir}}$ and $\tilde{a}_{\textsc{uv}} \rightarrow\tilde{a}_{\textsc{ir}}$.

\vspace{0.3cm}

\noindent \textbf{QFT with a cutoff}: For the theory perturbed by a relevant operator we can read off $f_{\textrm{QFT}}(R)$ from the expression for the on-shell action \eqref{eq:SonWTUnewLambda}, which gives
\begin{align}
\label{eq:DRfQFT3} f_{\textrm{QFT}}(R) &= M^{d-1} \, \Lambda^d \, \tilde{\ell}^d \big(W_\L + T_\L U_\L \big) \, ,
\end{align}
where $\tilde{\ell}$ was defined in \eqref{eq:lambdatildedef}.

The $R$-derivative in $\mathcal{D}_R$ is to be performed at constant $m$ and $\Lambda$, or equivalently, at constant $\f_\L$ and $\Lambda$. Note that for fixed $\f_\L$ and $\Lambda$ the quantity $\tilde{\ell}$ as defined in \eqref{eq:lambdatildedef} is constant. Hence it is unaffected by a derivative with respect to $R$ at fixed $m$ and $\Lambda$. Thus, at constant $m$ and $\Lambda$, we can once more trade an $R$-derivative for a derivative with respect to $T_\L$:\begin{align}
\label{eq:DRfQFT4} R \left. \frac{\partial}{\partial R} \right|_{\f_\L, \Lambda} &= R \left. \frac{\partial T_\L}{\partial R} \right|_{\f_\L, \Lambda} \frac{\partial}{\partial T_\L} = R \Lambda^{-2} \tilde{\ell}^{-2} \frac{\partial}{\partial T_\L} = T_\L \frac{\partial}{\partial T_\L} \, .
\end{align}
where we used \eqref{eq:relateRoverLambdatoT} and \eqref{eq:LambdaQFTdef}. Using this, one can show that the contribution to the equation for constant curvature solutions due to \eqref{eq:DRfQFT3} is
\begin{align}
\label{eq:DRfQFT5} \mathcal{D}_R f_{\textrm{QFT}} = d (M \tilde{\ell})^{d-1} \, \Lambda^d \, \tilde{\ell} W_\L - 2 (M \tilde{\ell})^{d-1} \, \Lambda^d \, \tilde{\ell} T_\L \bigg( \frac{\partial W_\L}{\partial T_\L} - \frac{d-2}{2} U_\L + T_\L \frac{\partial U_\L}{\partial T_\L} \bigg) \, ,
\end{align}
i.e.~this takes the same form as the expression \eqref{eq:DRfQFT1} for the UV CFT, but with $\ell_{\textsc{uv}}$ replaced by $\tilde{\ell}$.

At this stage we cannot make much further progress using analytical methods. However, we  find that in all numerical examples considered the second term in \eqref{eq:DRfQFT5} vanishes in virtue of the expression in brackets vanishing, i.e.
\begin{align}
\label{eq:DRfQFT6} \frac{\partial W_\L}{\partial T_\L} - \frac{d-2}{2} U_\L + T_\L \frac{\partial U_\L}{\partial T_\L} =0 \, .
\end{align}
This is just like in the case of the UV CFT discussed above. 
 \eqref{eq:DRfQFT6} can be proven even if  the theory is not a CFT,
 and coincides  with  a thermodynamic relation in de
Sitter space, that was proven in \cite{F} and is further discussed appendix \ref{app:thermo}.
Overall, we are left with
\begin{align}
\label{eq:DRfQFT7} \mathcal{D}_R f_{\textrm{QFT}} = d (M \tilde{\ell})^{d-1} \Lambda^d \, \tilde{\ell} W_\L \, .
\end{align}

To make contact with the discussion in section \ref{sec:ResultsQFTcutoff}, we rewrite this as follows. First, we introduce a (dimensionless) function $\mathcal{W}(m \Lambda^{-1}, R\Lambda^{-2})$, which is defined as
\begin{align}
\label{eq:DRfQFT8} \mathcal{W} \big( \tfrac{m}{\L}, \tfrac{R}{\L^2} \big) \equiv \tilde{\ell}(\f_\L) \, W_\L(\f_\L, T_\L) \, ,
\end{align}
together with the identifications \eqref{eq:mfromphiUV} and \eqref{eq:relateRoverLambdatoT}. We now specialise to $d=4$. We then make the observation that the quantity $(M \tilde{\ell})^3$ interpolates between the anomaly coefficients $\tilde{a}_{\textsc{uv}}$ and $\tilde{a}_{\textsc{ir}}$ of the UV and IR CFTs as $\f_\L$ is varied from $\f_\L \rightarrow 0$ to $\f_\L \rightarrow \f_{\textsc{ir}}$, i.e.
\begin{align}
(M \tilde{\ell})^3 & \underset{\f_\L \rightarrow 0}{\longrightarrow} (M \ell_\textsc{uv})^3 = \tilde{a}_{\textsc{uv}} \, , \\
(M \tilde{\ell})^3 & \underset{\f_\L \rightarrow \f_{\textsc{ir}}}{\longrightarrow} (M \ell_\textsc{ir})^3 = \tilde{a}_{\textsc{ir}} \, .
\end{align}
Hence we find it convenient to introduce a running anomaly coefficient, denoted by $\tilde{a}(m \Lambda^{-1})$, and defined as as
\begin{align}
\label{eq:DRfQFT9} \tilde{a}\big( \tfrac{m}{\L} \big) \underset{d=4}{\equiv} \big(M \tilde{\ell}(\f_\L) \big)^3 \, ,
\end{align}
together with the identification \eqref{eq:mfromphiUV}. Putting everything together, the expression in \eqref{eq:DRfQFT7} can be written as
\begin{align}
\label{eq:DRfQFT10} \mathcal{D}_R f_{\textrm{QFT}} \underset{d=4}{=} 4 \tilde{a} \big( \tfrac{m}{\Lambda} \big) \,  \Lambda^4 \, \mathcal{W} \big(\tfrac{m}{\Lambda} , \tfrac{R}{\Lambda^2} \big) \, .
\end{align}

\vspace{0.3cm}

\noindent \textbf{UV complete 4d QFT}: In the renormalized case we only derived an expression for the on-shell action for $d=4$ and hence we restrict to the choice in the following analysis. We write the contribution to $f(R)$ from the QFT as $f_{\textrm{QFT}}^{\textrm{ren}}$. This can be read off from \eqref{eq:Stotrenfinal}:
\begin{align}
\nonumber f_{\textrm{QFT}}^{\textrm{ren}} &= \tilde{a}_{\textsc{uv}} \, \Big[ \tfrac{1}{96} R^2 \Big(1 + \log \big( \tfrac{1}{48} R m^{-2} \big)  \Big) + m^4 \mathcal{G}(Rm^{-2}) - m^4 \mathcal{G}(0) - Rm^{2} \mathcal{G}'(0) \Big] \\
\label{eq:fQFTuniv} &= (M \ell_{\textsc{uv}})^3 \, |\f_-|^{4/\Delta_-} \, \Big[ C(\mathcal{R}) + \mathcal{R} B(\mathcal{R}) -C(0) - \mathcal{R} B(0) - \mathcal{R} C'(0) \Big] \, ,
\end{align}
and in the second line we used \eqref{eq:calGdef}, \eqref{eq:anomalyUVdef}, $m = |\f_-|^{1 / \Delta_-}$ and $\mathcal{R}= Rm^{-2}$.

The contribution to the equation for constant-curvature solutions is then
\begin{align}
\label{eq:DRfQFT12} \mathcal{D}_R f_{\textrm{QFT}} = \, & \, 4 (M \ell_{\textsc{uv}})^3 \, |\f_-|^{4/\Delta_-} \, \bigg( C - C(0) - \mathcal{R} \frac{B(0) + C'(0)}{2} \bigg) \\
\nonumber & - 2 (M \ell_{\textsc{uv}})^3 \, |\f_-|^{4/\Delta_-} \, \mathcal{R} \bigg(C'(\mathcal{R}) - B + \mathcal{R} B'(\mathcal{R}) \bigg) \, ,
\end{align}
where here ${}^\prime = \partial / \partial \mathcal{R}$.
Again, generally we do not have explicit expressions for $C(\mathcal{R})$ and $B(\mathcal{R})$, which we extract using numerical methods as expressed in section \ref{sec:holoRGreview}. However, as in the case with a cutoff, de Sitter thermodynamics implies a relation between $C(\mathcal{R})$ and $B(\mathcal{R})$, see appendix \ref{app:thermo}. For $d=4$ this relation can be written as
\begin{align}
\label{eq:thermoidmaintext} C'(\mathcal{R}) - B + \mathcal{R} B'(\mathcal{R}) = \frac{\mathcal{R}}{96} \, .
\end{align}
While we cannot prove this relation, we confirm numerically that it is satisfied by all examples considered here. As long as $B'(\mathcal{R})$ is regular for $\mathcal{R} \rightarrow 0$, which we can again confirm numerically, the above identity implies $B(0) = C'(0)$. Using this and the identity \eqref{eq:thermoidmaintext}, the expression in \eqref{eq:DRfQFT12} becomes
\begin{align}
\nonumber \mathcal{D}_R f_{\textrm{QFT}} &= 4 (M \ell_{\textsc{uv}})^3 \, |\f_-|^{4/\Delta_-} \, \Big( C - C(0) - \mathcal{R} C'(0) \Big) - \tfrac{1}{48} (M \ell_{\textsc{uv}})^3 R^2 \\
\label{eq:DRfQFT13} &= 4 \tilde{a}_{\textsc{uv}} \, m^4 \, \Big( C(R m^{-2}) - C(0) - Rm^{-2} C'(0) \Big) - \tfrac{1}{48} \tilde{a}_{\textsc{uv}} R^2 \, ,
\end{align}
where we also employed \eqref{eq:anomalyUVdef}. For a CFT ($m=0$) the first term vanishes and only the second term contributes. As a final observation recall that the parameter $C$ is related to the vev of the operator $\mathcal{O}$ as, c.f.~\eqref{eq:vevdef}:
\begin{align}
C(Rm^{-2}) \underset{d=4}{=} (M \ell_{\textsc{uv}})^{-3} \, \frac{\Delta_-}{4} \, \langle \mathcal{O} \rangle \, |\f_-|^{- \frac{\Delta_+}{\Delta_-}} = \tilde{a}_{\textsc{uv}}^{-1} \, \frac{4 - \Delta}{4} \, \langle \mathcal{O} \rangle \, m^{-\Delta}  \, .
\end{align}
Hence a UV complete QFT gives two contributions to the equation for constant curvature solutions. Firstly, there is a term $\sim R^2$. This is always present regardless whether the QFT is a CFT or not. Secondly, if the QFT is a CFT perturbed in the UV by an operator $\mathcal{O}$, there is a contribution that involves the vev $\langle \mathcal{O} \rangle$.

\section{Constant-curvature solutions: Backreacted UV-complete 4d QFT}
\label{sec:ResultsQFTren}
In this section we study constant-curvature solutions for the combined system of the gravitational theory described by $S_0$ in \eqref{eq:S0def} coupled to a UV-complete QFT as described in section \ref{sec:strategy}. Here we exclusively consider the case $d=4$, as this will be most relevant for potential applications to our universe. Most importantly, the discussion contained in this section does not rely on the concepts and notation introduced in section \ref{sec:holo} discussing the holographic setup. This section can thus be read without having previously consulted section \ref{sec:holo} at the expense that some expressions have to be taken granted. Appendix \ref{app:dictionary} contains a `dictionary' to relate expressions used here and in section \ref{sec:strategy} to the corresponding expressions in section  \ref{sec:holo}.

We begin by collecting the most relevant results from section \ref{sec:strategy}, specialised to the case $d=4$. The action of the total system can be found in \eqref{eq:Stotrenfinal}, which for $d=4$ can be written as
\begin{align}
\label{SolRen1} S_{\textrm{tot}} = \int d^4 x \sqrt{|g|} \, f(R) \, , \qquad \textrm{with} \qquad f(R) = f_0^{\textrm{ren}}(R) + f_{\textsc{qft}}^{\textrm{ren}} (R) \, ,
\end{align}
where\footnote{In \eqref{eq:Stotrenfinal}, in addition to the term $\tfrac{1}{2} a_2^{\textrm{ren}} R^2$, one also finds at the same order in $R$ the terms $\tfrac{1}{96} \tilde{a}_{\textsc{uv}} R^2$ and $ - \tfrac{\log 48}{96} \tilde{a}_{\textsc{uv}} R^2$. Here these terms have all been subsumed into the term $\tfrac{1}{2} a^{\textrm{ren}} R^2$.}
\begin{align}
\label{SolRen1b} f_0^{\textrm{ren}}(R) &= \tfrac{1}{2} M_{\textrm{ren}}^{2} R - M_{\textrm{ren}}^{2} \lambda_{\textrm{ren}} + \tfrac{1}{2} \, a^{\textrm{ren}} R^2 \, , \\
\label{SolRen1c} f_{\textsc{qft}}^{\textrm{ren}} (R) &=   \tilde{a}_{\textsc{uv}} \, \Big( \tfrac{1}{96} \, R^{2} \log \big( \tfrac{R}{m^2} \big) + m^4 \, \mathcal{G} \big(\tfrac{R}{m^2} \big) - m^4 \, \mathcal{G} \big(0 \big) - Rm^2 \, \mathcal{G}' \big(0 \big) \Big) \, .
\end{align}
Here we have split the total action into a part containing the Einstein-Hilbert term, the cosmological constant and a $R^2$-term, and a second part containing the contributions due to the QFT which go beyond renormalizing the terms already present in the first part.

The equation for constant-curvature solutions is given in \eqref{eq:fofR4}, which for $d=4$ becomes:
\begin{align}
\label{SolRen2} 2 f(R) - R f_R(R) =0 \, .
\end{align}
We can obtain the same expression by considering the Einstein equation
\begin{align}
\label{SolRen2b} R_{\mu \nu} - \frac{1}{2} g_{\mu \nu} R + \lambda_{\textrm{ren}} \, g_{\mu \nu} = M_{\textrm{ren}}^{-2} \, \langle T_{\mu \nu}^{\textrm{ren}} \rangle \, ,
\end{align}
restricted to 4d Einstein backgrounds, i.e.~$R_{\mu \nu} = \tfrac{1}{4} R g_{\mu \nu}$, together with
\begin{align}
\label{SolRen2c} \langle T_{\mu \nu}^{\textrm{ren}} \rangle &= - \frac{2}{\sqrt{|g|}} \, \frac{\delta}{\delta g^{\mu \nu}} \, \int d^4 x \sqrt{|g|} \, f_{\textsc{qft}}^{\textrm{ren}} (R) \\
\nonumber &= \tilde{a}_{\textsc{uv}} \,  \Big( - \tfrac{1}{192} \, R^{2} + m^4 \, \mathcal{G} \big(\tfrac{R}{m^2} \big) - \tfrac{1}{2} R m^2 \, \mathcal{G}' \big(\tfrac{R}{m^2} \big) - m^4 \, \mathcal{G} \big(0 \big) - \tfrac{1}{2} R m^2 \, \mathcal{G}' \big(0 \big) \Big) g_{\mu \nu} \, .
\end{align}
Taking the trace of \eqref{SolRen2b} one finds
\begin{align}
\label{SolRen3} M_{\textrm{ren}}^{2} R - 4 M_{\textrm{ren}}^{2} \lambda_{\textrm{ren}} + \langle T_{\mu}^{\textrm{ren}, \mu} \rangle = 0 \, ,
\end{align}
with
\begin{align}
\label{SolRen3b} \langle T_{\mu}^{\textrm{ren}, \mu} \rangle = \tilde{a}_{\textsc{uv}} \,  \Big( - \tfrac{1}{48} \, R^{2} + 4 m^4 \mathcal{G} \big(\tfrac{R}{m^2} \big) - 2 R m^2 \mathcal{G}' \big(\tfrac{R}{m^2} \big) - 4 m^4 \mathcal{G} \big(0 \big) - 2 R m^2 \mathcal{G}' \big(0 \big)\Big) \, .
\end{align}
Inserting \eqref{SolRen1b}, \eqref{SolRen1c} into \eqref{SolRen2} leads to the same result.

We can then make the following observations:
\begin{itemize}
\item Constant-curvature solutions are unaffected by the term $\sim a^{\textrm{ren}} R^2$, as $a^{\textrm{ren}}$ does not appear in equation \eqref{SolRen3}. It is easy to see that any term $\propto R^2 \subset f(R)$ is annihilated by the LHS of \eqref{SolRen2} and hence does not enter the equation for constant-curvature solutions.
\item From \eqref{SolRen3} it follows that any non-trivial backreaction effects due to the QFT on constant-curvature solutions (i.e.~effects beyond renormalization of $M_{\textrm{ren}}$ and $\lambda_{\textrm{ren}}$) enter the equation via $\langle T_{\mu}^{\textrm{ren}, \mu} \rangle$, i.e.~the trace of the stress tensor associated with the matter Lagrangian \eqref{SolRen1c}. This is nothing but the conformal anomaly associated with the QFT and hence it is the conformal anomaly which completely determines non-trivial backreaction due to the QFT.
\item In this work the backreacting QFTs considered are defined holographically in terms of a 4d UV CFT deformed by a relevant scalar operator $\mathcal{O}$ of dimension $\Delta$ (and the resulting RG flow). As can be inferred using results from sections \ref{sec:holoRGreview} and \ref{sec:Sonren}, the vev $\langle \mathcal{O} \rangle$ is related to $\mathcal{G}$ as
\begin{align}
\label{SolRen3c} \langle \mathcal{O} \rangle (R,m) &= \frac{4 \, \tilde{a}_{\textsc{uv}}}{4-\Delta} m^{\Delta} \, \Big( \mathcal{G} \big(\tfrac{R}{m^2} \big) - \tfrac{1}{2} \tfrac{R}{m^2} \, \mathcal{G}' \big(\tfrac{R}{m^2} \big) \Big)  \, .
\end{align}
We can then define a subtracted vev $\overline{\langle \mathcal{O} \rangle}$ as
\begin{align}
\label{SolRen3cc} \overline{\langle \mathcal{O} \rangle} (R,m) & \equiv \langle \mathcal{O} \rangle (R,m) - \langle \mathcal{O} \rangle (0,m) - \left. \partial_R \langle \mathcal{O} \rangle \right|_{R=0,m} \\
\nonumber &= \frac{4 \, \tilde{a}_{\textsc{uv}}}{4-\Delta} m^{\Delta} \, \Big( \mathcal{G} \big(\tfrac{R}{m^2} \big) - \tfrac{1}{2} \tfrac{R}{m^2} \, \mathcal{G}' \big(\tfrac{R}{m^2} \big) - \mathcal{G} \big(0 \big) - \tfrac{1}{2} \tfrac{R}{m^2} \mathcal{G}' \big(0 \big) \Big)  \, .
\end{align}
This can be understood as the vev minus its contribution to the cosmological constant and Newton's constant for $R \rightarrow 0$. It is this subtracted vev that appears in the trace of the stress tensor \eqref{SolRen3b}, which can thus be written as
\begin{align}
\label{SolRen3d} \langle T_{\mu}^{\textrm{ren}, \mu} \rangle = - \frac{\tilde{a}_{\textsc{uv}}}{48} R^{2} - (\Delta-4) \overline{\langle \mathcal{O} \rangle} \, m^{4-\Delta}  \, .
\end{align}
This takes the expected form for the conformal anomaly for the QFT considered, see e.g.~\cite{1401.0888}. Inserting \eqref{SolRen3d} into \eqref{SolRen3} it follows that the effect of backreaction due to a QFT can be split into two contributions, one depending purely on the background curvature $R$ and one including $m$ and the (subtracted) vev. An equation with the same schematic structure has also been obtained in \cite{julien}, where backreaction on 4-dimensional de Sitter space due to a theory of scalar fields with interaction term $\phi^4$ was considered using non-perturbative renormalisation group techniques.
\end{itemize}
For the further discussion we find it convenient to define a function $C(Rm^{-2})$ as
\begin{align}
\label{SolRen4} C \big(\tfrac{R}{m^2} \big) &= \frac{4 - \Delta}{4 \, \tilde{a}_{\textsc{uv}}} \, \langle \mathcal{O} \rangle \, m^{-\Delta} = \mathcal{G} \big(\tfrac{R}{m^2} \big) - \tfrac{1}{2} \tfrac{R}{m^2} \, \mathcal{G}' \big(\tfrac{R}{m^2} \big) \, ,
\end{align}
i.e.~$C(R m^{-2})$ is proportional to the vev in units of the coupling constant $m$. The quantity $C (R m^{-2})$ is identical to the quantity $C(\mathcal{R})$ introduced in section \ref{sec:holoRGreview} with the identification $\mathcal{R} = R m^{-2}$. In that section we also explain how $C (R m^{-2})$ can be determined numerically. Then, similar to $\overline{\langle \mathcal{O} \rangle}$ given in \eqref{SolRen3cc}, we define
\begin{align}
\nonumber \overline{C} \big(\tfrac{R}{m^2} \big) &= \frac{4 - \Delta}{4 \, \tilde{a}_{\textsc{uv}}} \, \overline{\langle \mathcal{O} \rangle} \, m^{-\Delta} = \mathcal{G} \big(\tfrac{R}{m^2} \big) - \tfrac{1}{2} \tfrac{R}{m^2} \, \mathcal{G}' \big(\tfrac{R}{m^2} \big) - \mathcal{G} \big(0 \big) - \tfrac{1}{2} \tfrac{R}{m^2} \mathcal{G}' \big(0 \big) \, , \\
\label{SolRen4a} & = C \big(\tfrac{R}{m^2} \big) - C \big(0 \big) - \tfrac{R}{m^2} \, C' \big(0 \big) \, .
\end{align}
Using this, the equation for constant-curvature \eqref{SolRen3} solutions can be written as
\begin{align}
\label{SolRen4b} M_{\textrm{ren}}^{2} R - 4 M_{\textrm{ren}}^{2} \lambda_{\textrm{ren}} - \tfrac{1}{48} \, \tilde{a}_{\textsc{uv}} \, R^{2} + 4 \tilde{a}_{\textsc{uv}} \, m^4 \, \overline{C} \big(\tfrac{R}{m^2} \big) = 0 \, ,
\end{align}

In the following we will solve \eqref{SolRen4b} for $R$ for various choices of $ \lambda_{\textrm{ren}}$, $m$ and $\tilde{a}_{\textsc{uv}}$. In all these analyses we will keep $M_{\textrm{ren}}$ fixed, and hence it can be used as a reference scale for all other dimensionful quantities. It will hence be convenient to define the dimensionless parameters $\hat{R}$, $\hat{\lambda}$ and $\hat{m}$ as
\begin{align}
\label{SolRen5} \hat{R} \equiv \frac{R}{M_{\textrm{ren}}^2} \, , \qquad \hat{\lambda} \equiv \frac{\lambda_{\textrm{ren}}}{M_{\textrm{ren}}^2} \, , \qquad \hat{m} \equiv \frac{m}{M_{\textrm{ren}}} \, .
\end{align}
Using this equation \eqref{SolRen4b} becomes:
\begin{align}
\label{SolRen6} \hat{R} - 4 \hat{\lambda} - \tfrac{1}{48} \tilde{a}_{\textsc{uv}} \hat{R}^{2} + 4 \tilde{a}_{\textsc{uv}} \, \hat{m}^4 \, \overline{C} \big(\hat{R} \hat{m}^{-2} \big) = 0 \, .
\end{align}
We can also study the system at hand for arbitrary $\tilde{a}_{\textsc{uv}}$ by exploiting a scaling property of solutions to \eqref{SolRen6}. In particular, multiplying \eqref{SolRen6} by $\tilde{a}_{\textsc{uv}}$ and introducing the quantities
\begin{align}
\label{SolRen5b} \tilde{R} \equiv \tilde{a}_{\textsc{uv}} \hat{R} = \frac{\tilde{a}_{\textsc{uv}} \, R}{M_{\textrm{ren}}^2} \, , \qquad \tilde{\lambda} \equiv \tilde{a}_{\textsc{uv}} \hat{\lambda} = \frac{\tilde{a}_{\textsc{uv}} \, \lambda_{\textrm{ren}}}{M_{\textrm{ren}}^2} \, , \qquad \tilde{m}^2 \equiv \tilde{a}_{\textsc{uv}} \hat{m}^2 = \frac{\tilde{a}_{\textsc{uv}} \, m^2}{M^2_{\textrm{ren}}} \, ,
\end{align}
the equation \eqref{SolRen6} can be written as
\begin{align}
\label{SolRen6b} \tilde{R} - 4 \tilde{\lambda} - \tfrac{1}{48} \tilde{R}^{2} + 4 \tilde{m}^4 \, \overline{C} \big(\tilde{R} \tilde{m}^{-2} \big) = 0 \, ,
\end{align}
i.e.~$\tilde{a}_{\textsc{uv}}$ has disappeared from the equation.

\begin{figure}[t]
\centering
\begin{overpic}
[width=0.65\textwidth]{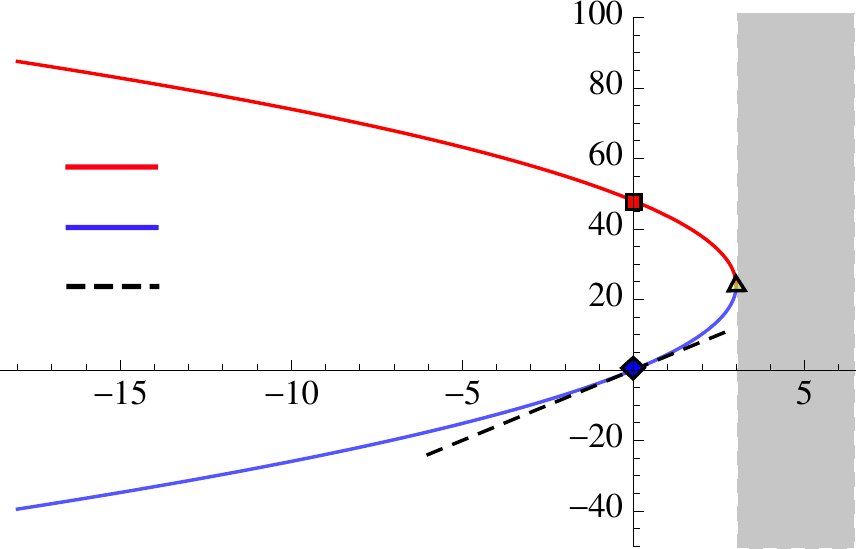}
\put(104,19){$\tilde{\lambda}$}
\put(58,61){$\tilde{R}$}
\put(20,42.5){$\tilde{R}_{+}$}
\put(20,35.5){$\tilde{R}_{-}$}
\put(20,29){$\tilde{R}=4 \tilde{\lambda}$}
\end{overpic}
\caption{Dimensionless scalar curvature $\tilde{R}$ vs.~dimensionless cosmological constant $\tilde{\lambda}$ for constant-curvature solutions including the backreaction of a UV-complete 4d CFT. There are two branches of solutions $\tilde{R}_{+}$ (shown in red) and $\tilde{R}_{-}$ (shown in blue). The solution $\tilde{R} = 4 \tilde{\lambda}$ in absence of the CFT is also shown for comparison (black, dashed). No solutions with a backreacted CFT exist for $\tilde{\lambda} >3$, which is indicated by the shaded grey region. We highlighted the point where the two branches meet as well as the two solutions for $\tilde{\lambda}=0$ by markers, to help with later comparison with the results from non-CFTs.}
\label{fig:RvsLambdaCFT}
\end{figure}
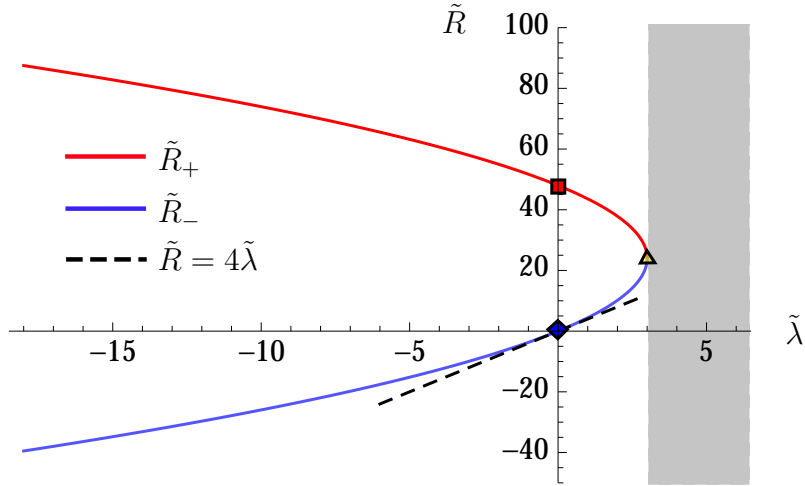

\subsection{Backreacted renormalized 4d CFT}
\label{sec:ResultsQFTrenCFT}
The case of backreaction of a CFT on constant-curvature solutions (and cosmological backgrounds) has been studied extensively in the literature (see e.g.~\cite{St, VilenkinStaro, anomalyinflation1, anomalyinflation2, anomalyinflation3, 0612068, Koksma1, anomalyinflation4}), and hence we will be brief here. The results shown in this section will be a useful benchmark against which backreaction due to non-conformal QFTs can be compared.

For a CFT on an Einstein manifold background, the conformal anomaly only consists of the curvature-term
\begin{align}
\label{SolRen8} \langle T_{\mu}^{\textsc{cft}, \textrm{ren}, \mu} \rangle = - \frac{\tilde{a}}{48} R^{2} \, .
\end{align}
Here we removed the subscript on the anomaly parameter $\tilde{a}$ as the following results will hold for any CFT and not just what we refer to as the UV CFT.
As can be seen from \eqref{eq:Mrendef1} and \eqref{eq:Mrendef2}, for the renormalization procedure employed here the Planck scale and cosmological constant are not renormalized for a CFT ($m=0$), so that we can replace ($M_{\textrm{ren}}$, $\lambda_{\textrm{ren}}$) with ($M_0$, $\lambda_0$). As a result the equation for constant-curvature solutions \eqref{SolRen3} becomes
\begin{align}
\label{SolRen9a} R - 4 \lambda_0 - \tfrac{1}{48} \tilde{a} M_0^{-2} R^2 =0 \, .
\end{align}
The fact that a conformal field theory backreacts on the background space-time via its conformal anomaly, which can be written as a local expression of curvature invariants is well-known (see e.g.~\cite{St, 0612068}), and it does not require holography to arrive at this result. In particular, for a CFT with ($N_S$, $N_F$, $N_V$) fields of spins ($0$, $1/2$, $1$) respectively, the parameter $\tilde{a}$ in \eqref{SolRen9a} will in general be a linear function of ($N_S$, $N_F$, $N_V$), see e.g.~\cite{Birrell:1982ix}. The study of the resulting effect on cosmological backgrounds is the basis for the Starobinsky model of inflation \cite{St, VilenkinStaro} and subsequent generalisations referred to as trace anomaly-induced inflation \cite{anomalyinflation1, anomalyinflation2, anomalyinflation3, anomalyinflation4}.\footnote{Also see \cite{1404.7349} for a more recent work within the framework of $f(R)$-gravity.}

Here we focus on constant-curvature solutions which correspond to solutions to equation \eqref{SolRen9a}. Using the rescaled variables defined in \eqref{SolRen5b} this becomes:\footnote{Here again we use the fact from \eqref{eq:Mrendef1}, \eqref{eq:Mrendef2} that for a CFT we have $(M_0, \lambda_0) = (M_{\textrm{ren}}, \lambda_{\textrm{ren}})$.}
\begin{align}
\label{SolRen9b} \tilde{R} - 4 \tilde{\lambda} - \tfrac{1}{48} \tilde{R}^2 =0 \, .
\end{align}
This is a quadratic equation for $\tilde{R}$ which can be solved to give
\begin{align}
\label{SolRen10} \tilde{R} &= 24 \left( 1 \pm \sqrt{1 - \frac{1}{3} \tilde{\lambda}} \right) \, .
\end{align}
In fig.~\ref{fig:RvsLambdaCFT} we plot $\tilde{R}$ vs.~$\tilde{\lambda}$ for this solution, referring to the two branches as $\tilde{R}_+$ and $\tilde{R}_-$, depending on the sign in front of the square root in \eqref{SolRen10}.\footnote{In the literature on trace anomaly-induced inflation, the $(+)$-branch is also referred to as the `quantum attractor' while the $(-)$-branch as the (quantum-corrected) `classical attractor', see e.g.~\cite{Koksma1}.} We can make the following observations:
\begin{itemize}
\item For a given CFT we thus have either two, one or zero solutions depending on the value of $\tilde{\lambda}$, i.e.
\begin{align}
\nonumber \tilde{\lambda} < 3 : \quad & \textrm{Two solutions.} \\
\nonumber \tilde{\lambda}  = 3 : \quad & \textrm{One solution with } \tilde{R} = 24 = 8 \tilde{\lambda} >0 \textrm{ (yellow marker in fig.~\ref{fig:RvsLambdaCFT})} ,  \\
 \nonumber \hat{\lambda} > 3 : \quad & \textrm{No solution.} \
\end{align}
That is, there is an upper bound $\tilde{\lambda} \leq 3$ for constant-curvature solutions to exist.
\item For a given value of $\tilde{\lambda}<3$ the corresponding solution on the $\tilde{R}_+$-branch is always a dS background (i.e.~$\tilde{R}_+ >0$) while the solution on the $\tilde{R}_-$-branch may give a dS, Minkowski or AdS background depending on the value of $\tilde{\lambda}$. The fact that backreaction of a CFT via the conformal anomaly always permits a dS solution for $\tilde{\lambda}=0$ has already been made by Starobinsky in \cite{St}.
\end{itemize}

In this work we are particularly concerned with backreaction of large-$N$ gauge theories and hence we close this section by considering the case of a CFT in the limit $N_{\textsc{uv}} \rightarrow \infty$ or, equivalently, $\tilde{a} \rightarrow \infty$. To reinstate  $\tilde{a}$ we revert to equation \eqref{SolRen9a} in terms of the un-rescaled parameters which can be solved for $R$ as
\begin{align}
\label{SolRen11} R &= 24 \, \frac{M_0^2}{\tilde{a}} \left( 1 \pm \sqrt{1 - \frac{\tilde{a}}{3} \frac{\lambda_0}{M^2_0}} \right) \, .
\end{align}
We then make the following observations for $\tilde{a} \rightarrow \infty$:
\begin{itemize}
\item For $\lambda_0 >0$ constant-curvature solutions only exist if
\begin{align}
\lambda_0 \leq \frac{3 M_0^2}{\tilde{a}} \, .
\end{align}
Thus for solutions with $\lambda_0 >0$ to exist $\lambda_0$ must be chosen small as to be parametrically suppressed as $\lambda_0 \lesssim 1 / \tilde{a}$. The corresponding values of $R$ on both the $(+)$- and $(-)$-branches are parametrically suppressed as $R_\pm \sim 1 / \tilde{a}$.
\item For $\lambda_0 < 0$ we find that
\begin{align}
R_\pm \underset{\tilde{a} \rightarrow \infty}{=} \pm 8 \sqrt{3} \, \frac{M_0 \sqrt{|\lambda_0|}}{\sqrt{\tilde{a}}} \, .
\end{align}
i.e.~the solutions on both branches are parametrically small as $|R_\pm| \sim 1 / \sqrt{\tilde{a}}$.
\end{itemize}
Thus, as long as $M_0, \lambda_0$ are independent of $\tilde{a}$ we find that constant-curvature solutions from backreacting a CFT with $\tilde{a} \rightarrow \infty$ exhibit a parametrically suppressed curvature which is at least suppressed as $|R| \lesssim 1 / \sqrt{\tilde{a}}$.

\subsection{Backreacted renormalized 4d QFT}
\label{sec:ResultsQFTrenQFT}
Once we leave the special case of conformal field theories, constant-curvature solutions are obtained by solving the full equation \eqref{SolRen4b} which receives non-trivial contributions from the dimensionless subtracted vev $\overline{C}(R m^{-2})$ defined in \eqref{SolRen4a}. For $Rm^{-2} \rightarrow 0$ and $Rm^{-2}\rightarrow \infty$ explicit expressions for $\overline{C}(R m^{-2})$ can be derived using analytical methods (see appendix \ref{app:largesmallR}). However, to derive an expression for $\overline{C}(R m^{-2})$ over its whole domain we will employ numerical methods.

As described in section \ref{sec:QFTdef}, the QFTs considered here have the following properties:
\begin{align}
\nonumber \textrm{QFT } = & \textrm { UV CFT with gauge group of rank } N_{\textsc{uv}} \, , \\
\nonumber & \textrm{ perturbed by a relevant scalar operator } \mathcal{O} \textrm{ of dimension } \Delta \textrm{ with source } j, \\
\nonumber & \textrm{ and corresponding mass scale } m \equiv |j|^{1/(d-\Delta)} \, , \\
\nonumber & \textrm{ flowing to a IR CFT with gauge group of rank } N_{\textsc{ir}} \, , \\
\nonumber & \textrm{ with the IR perturbation an irrelevant operator of dimension } \Delta^{\textsc{ir}} \, .
\end{align}
In particular, when specifying the QFT we are free to choose the parameters $\Delta$, $\Delta^{\textsc{ir}}$, $N_{\textsc{uv}}$, $N_{\textsc{ir}}$. However, rather than specifying $N_{\textsc{uv}}$, $N_{\textsc{ir}}$ we find it more convenient to choose values for the anomaly parameters $\tilde{a}_{\textsc{uv}}$, $\tilde{a}_{\textsc{ir}}$ of the corresponding UV and IR CFTs, as it is the anomaly parameters that appear in the equation for constant-curvature solutions.\footnote{Recall that for the theories considered here, i.e.~field theories with a gravity dual in terms of Einstein-dilaton gravity, the anomaly parameters depend on the numerical parameters $N_{\textsc{uv}}$ and $N_{\textsc{ir}}$ as $\tilde{a}_{\textsc{uv}/ \textsc{ir}} \sim N_{\textsc{uv}/ \textsc{ir}}^2$.} To keep our result more general, we  keep $\tilde{a}_{\textsc{uv}}$ unspecified and only fix the ratio $\tilde{a}_{\textsc{uv}} / \tilde{a}_{\textsc{ir}}$. Thus, the numerical examples presented here correspond to results obtained for different choices of the parameters
\begin{align}
\label{eq:modelparameters} \Delta \, , \quad \Delta^{\textsc{ir}} \, , \quad \tilde{a}_{\textsc{uv}} / \tilde{a}_{\textsc{ir}} \, .
\end{align}
Here we show numerical results for two example QFTs with the following choices of parameters:\footnote{In the holographic dual, both these cases correspond to a choice of bulk potential given by
$$ V(\f) = - \frac{12}{\ell_{\textsc{uv}}^2} - \frac{\Delta (4-\Delta)}{2 \ell_{\textsc{uv}}^2} \, \f^2 + \frac{\lambda}{\ell_{\textsc{uv}}} \, \f^4 \, , \quad \textrm{with} \quad \lambda = \frac{\Delta^2 (4-\Delta)^2}{192} \bigg( \Big( \frac{\tilde{a}_{\textsc{uv}}}{\tilde{a}_{\textsc{ir}}} \Big)^{2/3} -1 \bigg)^{-1} \, .
$$}
\begin{align}
\nonumber & \textrm{QFT}_1 \textrm{:} \quad \Delta=3.1 \, , \quad \Delta^{\textsc{ir}} = 4.74 \, , \quad  \tilde{a}_{\textsc{uv}} / \tilde{a}_{\textsc{ir}} = 2 \, , \\
\nonumber  & \textrm{QFT}_2 \textrm{:} \quad \Delta=2.6 \, , \quad \Delta^{\textsc{ir}} = 4.93 \, , \quad  \tilde{a}_{\textsc{uv}} / \tilde{a}_{\textsc{ir}} = 2 \, .
\end{align}
These two examples are representative for displaying the various effects of backreaction on constant-curvature solutions.

\begin{figure}[t]
\centering
\begin{subfigure}{.5\textwidth}
 \centering
   \begin{overpic}
[width=1.0\textwidth]{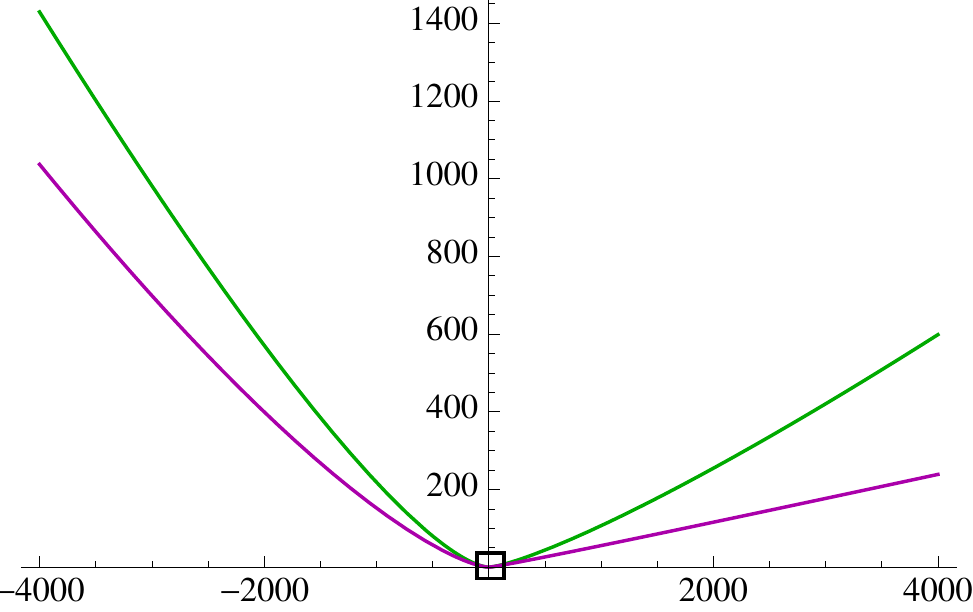}
\put(86,7){$Rm^{-2}$}
\put(53.5,58){$\overline{C}$}
\end{overpic}
\caption{\hphantom{A}}
\label{fig:CvsRplotsLarge}
\end{subfigure}%
\begin{subfigure}{.5\textwidth}
 \centering
   \begin{overpic}
[width=1.0\textwidth]{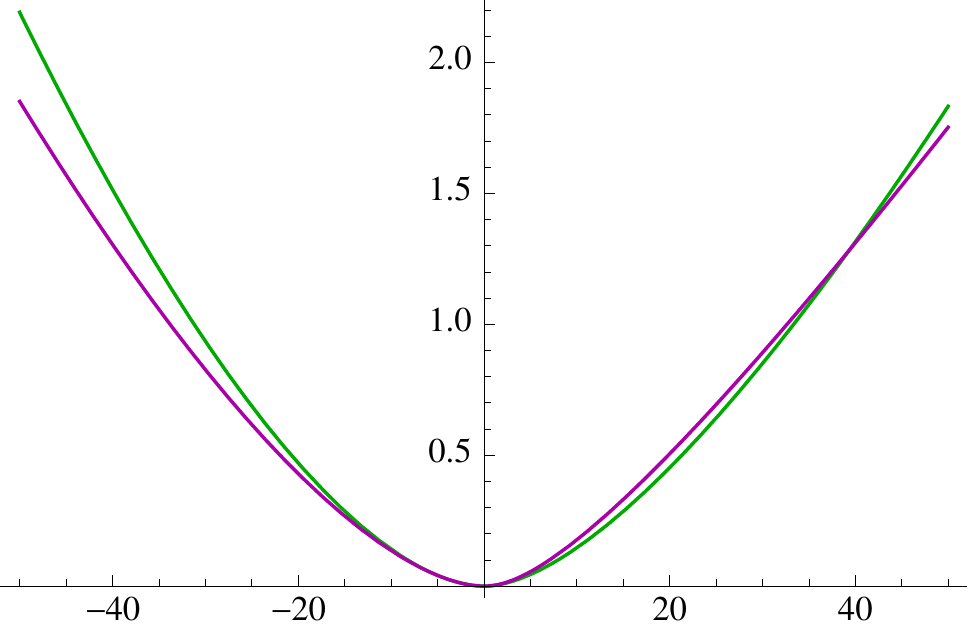}
\put(86,7.5){$Rm^{-2}$}
\put(53,60){$\overline{C}$}
\end{overpic}
\caption{\hphantom{A}}
\label{fig:CvsRplotsZoom}
\end{subfigure}%
\caption{\textbf{(a)-(b):} $\overline{C}(R m^{-2})$ for QFT$_1$ (green) with $\Delta=3.10$, $\Delta^{\textsc{ir}}=4.74$, $\tilde{a}_{\textsc{uv}} / \tilde{a}_{\textsc{ir}}=2$  and QFT$_2$ (magenta) with $\Delta=2.60$, $\Delta^{\textsc{ir}}=4.93$, $\tilde{a}_{\textsc{uv}} / \tilde{a}_{\textsc{ir}}=2$. The plot in (b) is the magnified version of the region within the black box in (a).}
\label{fig:CvsRplots}
\end{figure}

For a given QFT (i.e.~fixed $\Delta$, $\Delta^{\textsc{ir}}$, $\tilde{a}_{\textsc{ir}} / \tilde{a}_{\textsc{uv}}$) the function $\overline{C}(R m^{-2})$ is determined as follows: Choosing a value for the background curvature $R$ and the UV coupling mass scale $m$ one calculates the corresponding value of the dimensionless vev $C$. Repeating this process for different choices of $R$ and $m$ so that the dimensionless combination $R m^{-2}$ scans over the desired domain, one can build up the function $C(R m^{-2})$. From this, we can obtain $\overline{C}(R m^{-2})$ using \eqref{SolRen4a}. In this work this is done in the gravity dual formulation and the calculational details can be found in section \ref{sec:holoRGreview}. In figure \ref{fig:CvsRplots} we display the function $\overline{C}(R m^{-2})$ for two example theories QFT$_1$ and QFT$_2$ introduced above.

For $|Rm^{-2}| \rightarrow 0$ and $|Rm^{-2}| \rightarrow \infty$ we can derive analytic expressions for $\overline{C}(Rm^{-2})$. How this is done is shown in appendix \ref{app:largesmallR} and here we just quote the results:\footnote{For QFT$_2$ and $Rm^{-2} \rightarrow -\infty$ the expression in \eqref{eq:CbarlargeR} has to be replaced by $\overline{C}(Rm^{-2}) \sim  \mathcal{O} \big( (Rm^{-2})^{\Delta /2} \big)$. This difference will have no consequences.}
\begin{align}
\label{eq:CbarlargeR} \overline{C}(Rm^{-2}) &\overset{|Rm^{-2}| \rightarrow \infty}{=} \mathcal{O} \big( (Rm^{-2})^{\Delta -2}\big) \, , \\
\label{eq:CbarsmallR} \overline{C}(Rm^{-2}) &\overset{|Rm^{-2}| \rightarrow 0}{=} \frac{1}{192} \bigg(1 - \frac{\tilde{a}_{\textsc{ir}}}{\tilde{a}_{\textsc{uv}}} \bigg) R^2 m^{-4} + \mathcal{O} \big(R^3 m^{-6} \big) + \mathcal{O} \big( (Rm^{-2})^{\Delta_-^{\textrm{IR}}-2}\big) \, .
\end{align}
With this, we will be able to solve for constant-curvature solutions analytically in appropriate parametric regimes which we will do presently.

We begin by considering the parametric regime
\begin{align}
\label{eq:parametricmsmall} \frac{m}{M_\textrm{ren}} \rightarrow 0 \, , \qquad \frac{m^2}{\lambda_\textrm{ren}} \rightarrow 0 \, ,
\end{align}
i.e.~the QFT scale $m$ is vanishingly small compared to the other scales in the theory. We then expect that the backreaction effect due to the QFT is effectively that of the UV CFT (as for $m=0$ the QFT reduces to the UV CFT). This is indeed what we observe. As we will show, in the parametric regime \eqref{eq:parametricmsmall} we can self-consistently solve for constant-curvature solutions with $|R m^{-2}| \rightarrow \infty$. To this end we insert \eqref{eq:CbarlargeR} into \eqref{SolRen4b} to find that the equation for constant-curvature solution becomes:
\begin{align}
\label{eq:cceqsmallm} M_{\textrm{ren}}^{2} R - 4 M_{\textrm{ren}}^{2} \lambda_{\textrm{ren}} - \tfrac{1}{48} \, \tilde{a}_{\textsc{uv}} \, R^{2} + \mathcal{O} \Big( \tilde{a}_{\textsc{uv}} R^{\Delta-2} m^{8-2 \Delta} \Big)  = 0 \, .
\end{align}
In the parametric regime \eqref{eq:parametricmsmall} and assuming $|R m^{-2}| \rightarrow \infty$, the last term on the LHS of \eqref{eq:cceqsmallm} is subleading, while the leading part of the equation reduces to that obtained for backreaction due to a CFT, \eqref{SolRen9a}, but with anomaly parameter $\tilde{a}=\tilde{a}_{\textsc{uv}}$. Thus the leading backreaction effect is that of the UV CFT as expected. The constant-curvature solutions are hence just given by \eqref{SolRen11} with $\tilde{a}= \tilde{a}_{\textsc{uv}}$. These scale as
\begin{align}
R \sim \frac{M_{\textrm{ren}}^2}{\tilde{a}_{\textsc{uv}}} \, , \quad \textrm{or} \quad R \sim \lambda_{\textrm{ren}} \, ,
\end{align}
depending on the hierarchy between $M_{\textrm{ren}}$ and $\lambda_{\textrm{ren}}$. In any case, in the parametric regime \eqref{eq:parametricmsmall} this implies that $|R m^{-2}| \rightarrow \infty$ so that our use of \eqref{eq:CbarlargeR} was justified and the analysis is self-consistent.

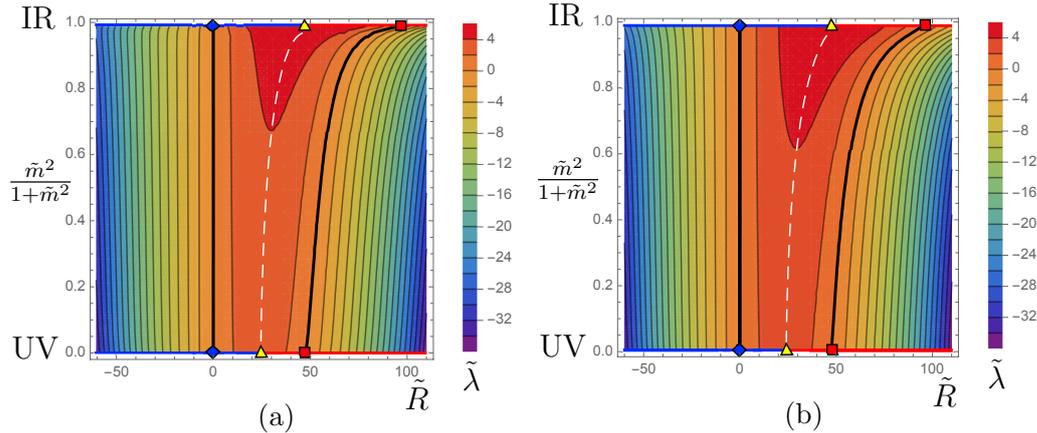
\begin{figure}[t]
\centering
\begin{subfigure}{.4\textwidth}
\centering
   \begin{overpic}
[width=1.0\textwidth]{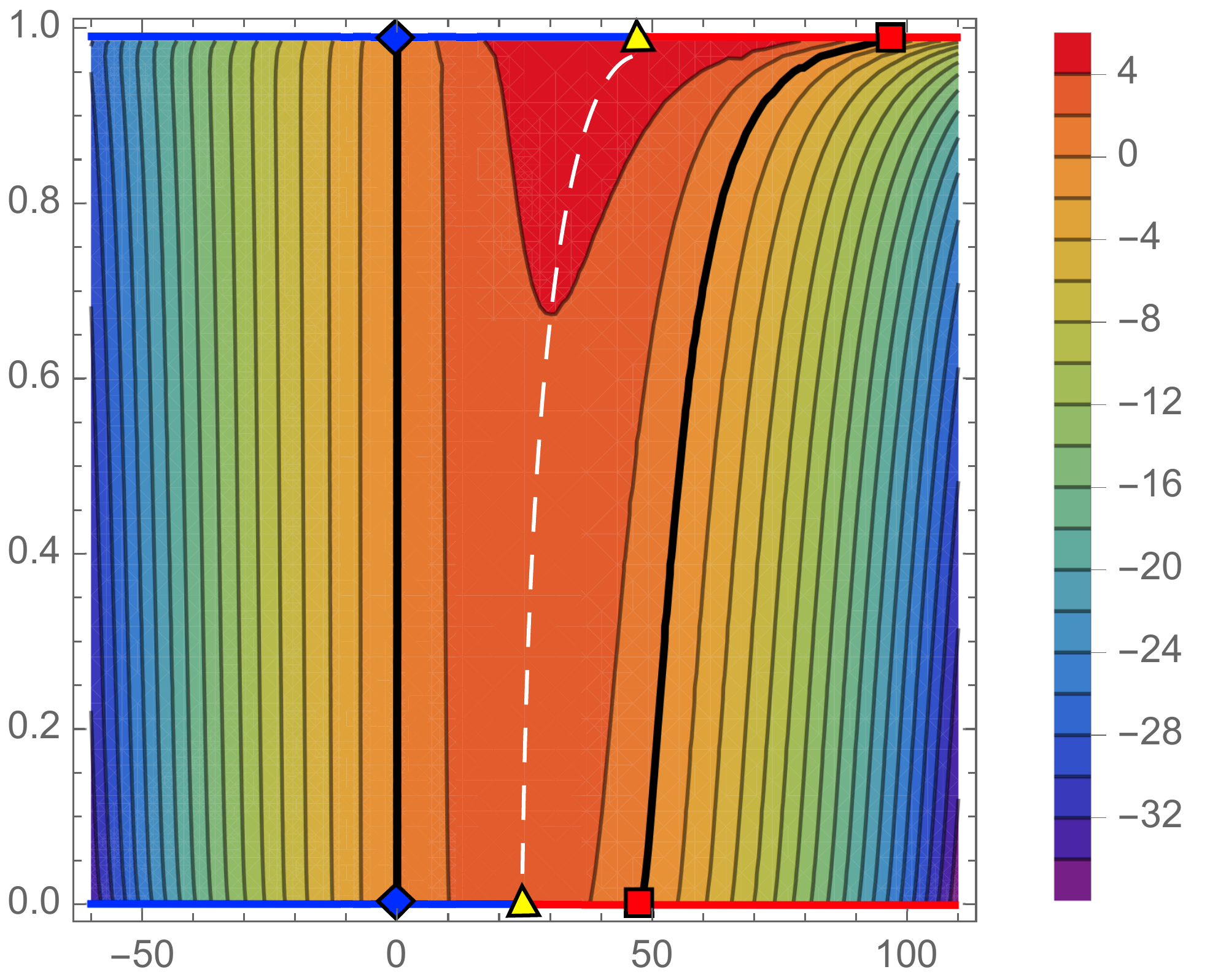}
\put(-12,42){$\frac{\tilde{m}^2}{1 + \tilde{m}^2}$}
\put(-9,77.5){IR}
\put(-11,5){UV}
\put(74.5,-6){$\tilde{R}$}
\put(87,-2){$\tilde{\lambda}$}
\end{overpic}
\caption{\hphantom{A}}
\label{fig:Lambda_Delta_0p9LIR0p5}
\end{subfigure}%
\begin{subfigure}{.05\textwidth}
\hfill
\end{subfigure}
\begin{subfigure}{.4\textwidth}
\hfill
   \begin{overpic}
[width=1.0\textwidth]{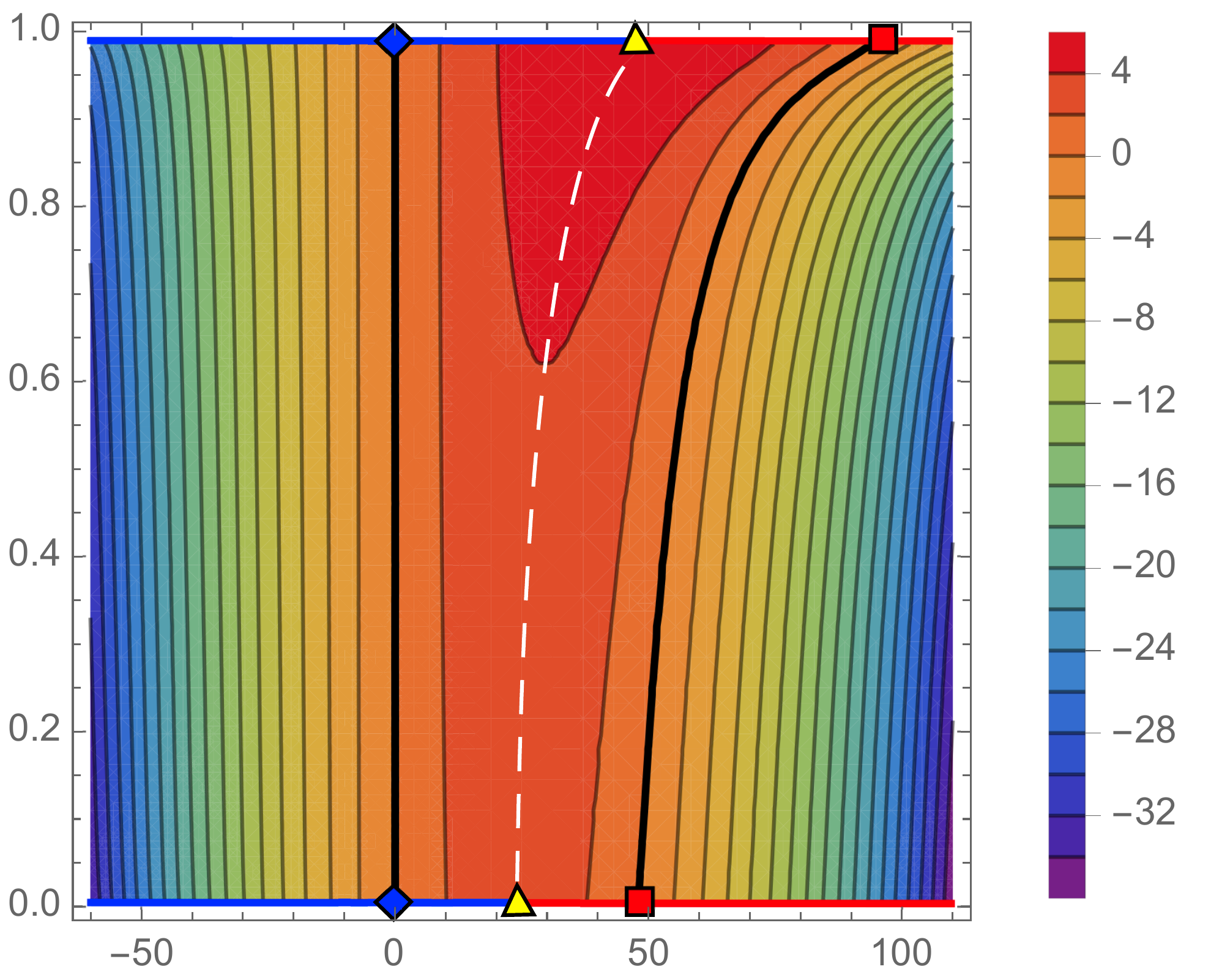}
\put(-12,42){$\frac{\tilde{m}^2}{1 + \tilde{m}^2}$}
\put(-9,77.5){IR}
\put(-11,5){UV}
\put(74.5,-6){$\tilde{R}$}
\put(86.5,-2){$\tilde{\lambda}$}
\end{overpic}
\caption{\hphantom{A}}
\label{fig:Lambda_Delta_1p4LIR0p5}
\end{subfigure}%
\caption{Solution to \protect\eqref{SolRen6} as contours of fixed $\tilde{\lambda}$ on the $\big( \tilde{R}, \, \tfrac{\tilde{m}^2}{1+\tilde{m}^2} \big)$-plane for backreaction due to \textbf{(a)}  QFT$_1$ ($\Delta=3.10$, $\Delta^{\textsc{ir}}=4.74$ and $\tilde{a}_{\textsc{uv}} / \tilde{a}_{\textsc{ir}}=2$), \textbf{(b)}  QFT$_2$ ($\Delta=2.60$, $\Delta^{\textsc{ir}}=4.93$ and $\tilde{a}_{\textsc{uv}} / \tilde{a}_{\textsc{ir}}=2$). For $\tfrac{\tilde{m}^2}{1+\tilde{m}^2} =0$ and $\tfrac{\tilde{m}^2}{1+\tilde{m}^2} =1$ we recover the results for backreaction due to the UV and IR CFTs respectively which can also be read off from fig.~\protect\ref{fig:RvsLambdaCFT}. For easier comparison we use the same colour-coding as in fig.~\protect\ref{fig:RvsLambdaCFT}, i.e.~the red lines correspond to the $(+)$-branch in fig.~\protect\ref{fig:RvsLambdaCFT}, while the blue lines denote the $(-)$-branch. We also use the same coloured markers to indicate the special solutions previously highlighted in fig.~\protect\ref{fig:RvsLambdaCFT}. The dotted white line (not a contour) marks solutions with the maximal value of $\tilde{\lambda}$ on a given slice of fixed $\tilde{m}$.}
\label{fig:Lambda_Delta_0p9and1p4LIR0p5}
\end{figure}

We now turn to the ``opposite'' choice by instead considering the parametric regime
\begin{align}
\label{eq:parametricmlarge} \frac{m}{M_\textrm{ren}} \rightarrow \infty \, , \qquad \frac{m^2}{\lambda_\textrm{ren}} \rightarrow \infty \, ,
\end{align}
i.e.~the QFT scale $m$ is much larger than all other scales in the theory. We will find that this will be consistent with constant-curvature solutions with $|Rm^{-2}| \rightarrow 0$, which we will confirm a posteriori. Hence we insert \eqref{eq:CbarsmallR} into \eqref{SolRen4b} to obtain:
\begin{align}
\label{eq:cceqlargem} M_{\textrm{ren}}^{2} R - 4 M_{\textrm{ren}}^{2} \lambda_{\textrm{ren}} - \tfrac{1}{48} \, \tilde{a}_{\textsc{ir}} \, R^{2} + \mathcal{O} \Big( \tilde{a}_{\textsc{uv}} R^{\Delta_-^{\textrm{IR}}-2} m^{8-2 \Delta^{\textrm{IR}}} \Big)  + \mathcal{O} \Big(\tilde{a}_{\textsc{uv}} R^3 m^{-2} \Big) = 0 \, .
\end{align}
In the parametric regime \eqref{eq:parametricmlarge} and assuming $|R m^{-2}| \rightarrow 0$, the last two terms on the LHS of \eqref{eq:cceqsmallm} are now subleading.\footnote{Recall that $\Delta^{\textrm{IR}}>4$ for both QFT$_1$ and QFT$_2$. In fact, for any 4d RG flow QFT whose IR CFT has a dual representation in terms of a AdS minimum of a scalar potential, the standard holographic dictionary states that $\Delta^{\textrm{IR}} = 2+2\sqrt{1+3 V''(\f_{\textrm{min}}) / |V(\f_{\textrm{min}})|} > 4$.} The leading part of the equation can again be identified with the equation obtained for backreaction due to a CFT, \eqref{SolRen9a}, but now it is the IR CFT as the anomaly coefficient is given by $\tilde{a}=\tilde{a}_{\textsc{ir}}$. The constant-curvature solutions are once more just given by \eqref{SolRen11}, but with $\tilde{a}= \tilde{a}_{\textsc{ir}}$. These again scale as
\begin{align}
R \sim \frac{M_{\textrm{ren}}^2}{\tilde{a}_{\textsc{ir}}} \, , \quad \textrm{or} \quad R \sim \lambda_{\textrm{ren}} \, .
\end{align}
Thus, in the parametric regime \eqref{eq:parametricmlarge} these satisfy $|R m^{-2}| \rightarrow 0$ which justifies our assumption that went into deriving \eqref{eq:cceqlargem}.

For intermediate values of $m M_{\textrm{ren}}^{-1}$ we need to turn to numerical methods to extract the constant-curvature solutions. To this end it will be convenient to use the rescaled variables \eqref{SolRen5b}, so that the equation for constant curvature solutions simplifies to \eqref{SolRen6b}. To present the numerical results, it will be convenient to rewrite \eqref{SolRen6b} as:
\begin{align}
\label{SolRen14} \frac{\tilde{R}}{4} - \frac{\tilde{R}^2 }{192} + \tilde{m}^4 \, C \big( \tilde{R} \tilde{m}^{-2} \big) = \tilde{\lambda} \, .
\end{align}
i.e.~\eqref{SolRen14} gives $\tilde{\lambda}$ as a function of $(\tilde{R}, \, \tilde{m}^2)$. This implies that we can display solutions to this equation as contours of constant $\tilde{\lambda}$ on the $(\tilde{R}, \, \tilde{m}^2)$-plane. This is  what we will do here and the corresponding plots for constant-curvature solutions are shown in figure \ref{fig:Lambda_Delta_0p9and1p4LIR0p5}.

More precisely, while on one of the axes in fig.~\ref{fig:Lambda_Delta_0p9and1p4LIR0p5} we display the value of $\tilde{R}$, the other axis gives the value of $\tfrac{\tilde{m}^2}{1+\tilde{m}^2}$. The advantage is that for
\begin{align}
\nonumber \tilde{m} \rightarrow 0 \quad &\Rightarrow \quad \tfrac{\tilde{m}^2}{1+\tilde{m}^2} \rightarrow 0 \, , \\
\tilde{m} \rightarrow \infty \quad &\Rightarrow \quad \tfrac{\tilde{m}^2}{1+\tilde{m}^2} \rightarrow 1 \, ,
\end{align}
so that we can display the full range $0 < \tilde{m} < \infty$ on the interval $[0,1]$ for the variable $\tfrac{\tilde{m}^2}{1+\tilde{m}^2}$. In figures \ref{fig:Lambda_Delta_0p9LIR0p5} and \ref{fig:Lambda_Delta_1p4LIR0p5} we hence plot contours of constant $\tilde{\lambda}$ on the $\big(\tilde{R}, \,\tfrac{\tilde{m}^2}{1+\tilde{m}^2} \big)$-plane for results from the backreaction of QFT$_1$ and QFT$_2$, respectively. We can make the following observations:
\begin{itemize}
\item As shown above, for $\tilde{m} \rightarrow 0$ or, equivalently, $\tfrac{\tilde{m}^2}{1+\tilde{m}^2} \rightarrow 0$ the effect of backreaction on constant-curvature solutions is that of the UV CFT. For $\tilde{m} \rightarrow \infty$ or, equivalently, $\tfrac{\tilde{m}^2}{1+\tilde{m}^2} \rightarrow 1$ the effect of backreaction on constant-curvature solutions is that of the IR CFT. Backreaction due to a CFT has been discussed in section \ref{sec:ResultsQFTrenCFT} with the main results plotted in fig.~\ref{fig:RvsLambdaCFT}. To be able to compare between fig.~\ref{fig:RvsLambdaCFT} and the corresponding results in fig.~\ref{fig:Lambda_Delta_0p9and1p4LIR0p5} we employ the same colour-coding. In particular, we again marked the $(-)$-branch of solutions in fig.~\ref{fig:RvsLambdaCFT} by a blue line, while the $(+)$-branch is highlighted by the red line. Similarly, the markers (blue diamond, yellow triangle, red square) denoting special points in fig.~\ref{fig:RvsLambdaCFT} are also reproduced here in fig.~\ref{fig:Lambda_Delta_0p9and1p4LIR0p5}. The precise numerical values for these special solutions can be calculated from the general result for CFT backreaction in \eqref{SolRen11}, the definition of the rescaled variables in \eqref{SolRen5b} and the fact that for the RG flows considered we have $\tilde{a}_{\textsc{uv}} / \tilde{a}_{\textsc{ir}} =2$.
\item The thick black lines correspond to contours with $\lambda_{\textrm{ren}}=0$ or, equivalently, $\tilde{\lambda}=0$. For backreaction due to a CFT there exist two solutions for $\lambda_{\textrm{ren}}=0$, the `regular' solution $R=0$ (denoted by a blue diamond) and the `exotic' one with $R=48 / \tilde{a}$ (denoted by a red square). Hence, in both figures \ref{fig:Lambda_Delta_0p9LIR0p5} and \ref{fig:Lambda_Delta_1p4LIR0p5} there exist two thick black contours displaying how these two solutions for $\tilde{\lambda}=0$ evolve as $\tilde{m}$ is changed. For one, we see that the `regular' solution $R=0$ persists unaffected for all $\tilde{m}$. In contrast, the value of $R$ of the `exotic' solution grows monotonically as $\tilde{m}$ is increased, interpolating between the values $\tilde{R}=48$ for $\tilde{m} \rightarrow 0$ and $\tilde{R}=96$ for $\tilde{m} \rightarrow \infty$.
\item For every value of $\tilde{m}$ there exists a maximum value $\tilde{\lambda}_{\textrm{max}}(\tilde{m})$ above which no backreacted solutions exist. In figures \ref{fig:Lambda_Delta_0p9LIR0p5} and \ref{fig:Lambda_Delta_1p4LIR0p5} this is denoted by the dashed white line (not a contour of constant $\tilde{\lambda}$). For $\tilde{m} \rightarrow 0$ and $\tilde{m} \rightarrow \infty$, i.e.~when backreaction reduces to that of the UV or IR CFTs, this corresponds to the solutions denoted by a yellow triangle. As $\tilde{m}$ is increased from $0$ to $\infty$ this bound on $\tilde{\lambda}$ relaxes monotonically from $\tilde{\lambda}_{\textrm{max}}=3$ to $\tilde{\lambda}_{\textrm{max}}=6$, with corresponding values $\tilde{R}=24$ and $\tilde{R}=48$, respectively. The overall increase is a direct consequence of the fact that the IR CFT has fewer degrees of freedom than the UV CFT (i.e.~$\tilde{a}_{\textsc{ir}} < \tilde{a}_{\textsc{uv}}$). The fact that the increase is monotonic in $\tilde{m}$ is however non-trivial and reminiscent of monotonicity properties of QFTs under RG flow.
\item For any value $\tilde{\lambda} < \tilde{\lambda}_{\textrm{max}}(\tilde{m})$ there exist two solutions, one to the left of the white dashed line and one to the right. On the left branch the sign of $\tilde{R}$ is correlated to that of $\tilde{\lambda}$, while on the right branch one finds $\tilde{R}>0$ regardless of the sign of $\tilde{\lambda}$. This is just like for the $(-)$- and $(+)$-branches observed in the case of CFT-backreaction.
\item Another important observation is that while figures \ref{fig:Lambda_Delta_0p9LIR0p5} and \ref{fig:Lambda_Delta_1p4LIR0p5} differ in quantitative aspects, they exhibit the same qualitative features, in particular those discussed above. As both plots display results from backreacting RG flow QFTs, this implies that the observed qualitative features appear representative for backreaction effects from this class of theories. The quantitative differences are due QFT$_1$ and QFT$_2$ differing in the dimensions of the operators perturbing the UV and IR CFTs.
\end{itemize}
To summarise, backreaction on constant-curvature solutions due to RG flow QFTs is qualitatively very similar to that observed for CFTs: For sufficiently small values values of $\lambda_{\textrm{ren}}$ there generically exist two branches of solutions, one `regular' branch where the signs of $R$ and $\lambda_{\textrm{ren}}$ coincide, and an `exotic' branch where all solutions are de Sitter solutions regardless of the sign of $\lambda_{\textrm{ren}}$. Further, when dialing the QFT scale $m M_{\textrm{ren}}^{-1}$ from $0$ to $\infty$, backreaction effects from to the RG flow QFT interpolate between those of the UV and IR CFTs. At the same time the maximally permitted value $\tilde{\lambda}_{\textrm{max}}$ for a solution to exist is relaxed as $m$ is increased.

\section{Constant-curvature solutions: Backreacted 4d QFT with a UV cutoff}
\label{sec:ResultsQFTcutoff}
Here we present constant curvature solutions when coupling a QFT with a UV (energy) cutoff $\Lambda$ to the gravitational theory described by $S_0$ in \eqref{eq:S0def}. Once more, we restrict our analysis to $d=4$ as this is the phenomenologically most relevant case. As in section \ref{sec:ResultsQFTren}, the discussion presented here does not rely on the concepts and the notation introduced in section \ref{sec:holo}. Hence it can be read without having read section \ref{sec:holo} at the expense that certain expressions have to be taken for granted.

The action for the combined system was given in \eqref{eq:Stotcutoff}, which for $d=4$ becomes
\begin{gather}
\label{eq:CutoffResult1} S_{\textrm{tot}} = \int d^4x \sqrt{|g|} \, f(R) \, , \\
\nonumber \textrm{with} \quad f(R) = \tfrac{1}{2} M_0^2 R - M_0^2 \lambda_0 + \tfrac{1}{2} a R^2 + \tilde{a}_{\textsc{uv}} \Lambda^4 \mathcal{F} \Big(\tfrac{m}{\Lambda}, \tfrac{R}{\Lambda^2} \Big) \, ,
\end{gather}
where we also dropped the subscript on $a$. The equation for constant curvature solutions is given by \eqref{eq:fofR4}, which for $d=4$ can be written as
\begin{align}
\label{eq:CutoffResult2} \Big( 2 - R \frac{\partial}{\partial R} \Big) f(R) =0 \, .
\end{align}
The non-trivial contribution due to the backreacted QFT is given by the term $\sim \tilde{a}_{\textsc{uv}} \Lambda^4 \mathcal{F}$. For a CFT $\mathcal{F}$ is a function of $R \Lambda^{-2}$ only. Furthermore, we shall be able to give an explicit analytical expression. For a non-conformal QFT we only have numerical expressions for $\mathcal{F}$. In both cases the computation of these expressions crucially relies on the gravity dual and thus here we only state the results.

In the following, it will be useful to define dimensionless quantities with respect to a reference scale, which here we choose to be $M_0$, i.e.
\begin{align}
\label{eq:CutoffResult6} \hat{R} \equiv \frac{R}{M_{0}^2} \, , \qquad \hat{\lambda} \equiv \frac{\lambda_{0}}{M_{0}^2} \, , \qquad \hat{\Lambda} \equiv \frac{\Lambda}{M_{0}} \, , \qquad \hat{m} \equiv \frac{m}{M_{0}} \, .
\end{align}
In addition, we also use the rescaled quantities
\begin{align}
\label{eq:CutoffResult8} \tilde{R} \equiv \tilde{a}_{\textsc{uv}} \hat{R} \, , \qquad \tilde{\lambda} \equiv \tilde{a}_{\textsc{uv}} \hat{\lambda} \, , \qquad \tilde{\Lambda}^2 \equiv \tilde{a}_{\textsc{uv}} \hat{\Lambda}^2 \, , \qquad \tilde{m}^2 \equiv \tilde{a}_{\textsc{uv}} \hat{m}^2 \, .
\end{align}

\subsection{Backreacted 4d CFT with a UV cutoff}
\label{sec:ResultsQFTcutoffCFT}
To consider backreaction due to a CFT, here we set $m=0$, i.e.~we switch off the UV source for the relevant operator, in which case the QFT reduces to what we have been referring to as the UV CFT. This is the CFT with anomaly coefficient $\tilde{a}_{\textsc{uv}}$ governing the dynamics at the UV fixed point of the flow. However, as we shall be able to display results for arbitrary values of $\tilde{a}_{\textsc{uv}}$, the findings in this section will be valid for any CFT with a gravity dual.

For a CFT the function $\mathcal{F}$ only depends on the variable $R \Lambda^{-2}$. {In the following it will be useful to employ the fact that the contribution $\mathcal{F}$ from the QFT can be split into two parts. The first part can be identified as  the energy of the QFT in the static patch of de Sitter space, denoted by $\mathcal{W}$. The second part is related to the entropy of the QFT in the static patch, denoted by $\mathcal{U}$.\footnote{This same entropy can be shown, for any holographic QFT, to be equal to the entanglement entropy of the two hemispheres of the spacial $S^3$ in de Sitter space in global coordinates, generalizing the CFT result, \cite{F}. See \cite{Smolkin} for the same identification between thermal entropy and entanglement entropy for a non-holographic case.} More details are given in appendix \ref{app:thermo} upon using the dictionary in \ref{app:dictionary} and in \cite{F}. Therefore, we write the function $\mathcal{F}$ as
\begin{align}
\label{eq:CutoffResult3} \mathcal{F} \Big(\tfrac{R}{\Lambda^2} \Big) = \mathcal{W} \Big(\tfrac{R}{\Lambda^2} \Big) + \tfrac{R}{\Lambda^2} \, \mathcal{U} \Big(\tfrac{R}{\Lambda^2} \Big) \, .
\end{align}
For $d=4$ the functions $\mathcal{W}$ and $\mathcal{U}$ are given by
\begin{align}
\label{eq:CutoffResult10} \mathcal{W} (R \Lambda^{-2}) = 6 \, \sqrt{1+\frac{R \Lambda^{-2}}{12}} \, ,
\end{align}
\begin{align}
\label{eq:CutoffResult11}
\mathcal{U} (R \Lambda^{-2}) = \left\{
  \begin{array}{l l}
   \tfrac{1}{48} \bigg[ R \Lambda^{-2} \, \log \bigg( \sqrt{1 +  \frac{12}{R \Lambda^{-2}} } - \sqrt{ \frac{12}{R \Lambda^{-2}} } \bigg) + 12 \sqrt{1 + \frac{R \Lambda^{-2}}{12}} \bigg] \, , &  \quad R>0 \, , \\
& \\
 \tfrac{1}{48} \bigg[-R \Lambda^{-2} \, \log \bigg(\sqrt{ -\frac{12}{R \Lambda^{-2}} } + \sqrt{-  \frac{12}{R \Lambda^{-2}} -1} \bigg) + 12 \sqrt{1 + \frac{R \Lambda^{-2}}{12}} \bigg] \, , & \quad R<0 \, . \\
  \end{array} \right.
\end{align}
For the origin of these results from the holographic construction see section \ref{sec:WU-CFT} and the dictionary in appendix \ref{app:dictionary}.

\begin{figure}[t]
\centering
\begin{overpic}
[width=0.85\textwidth]{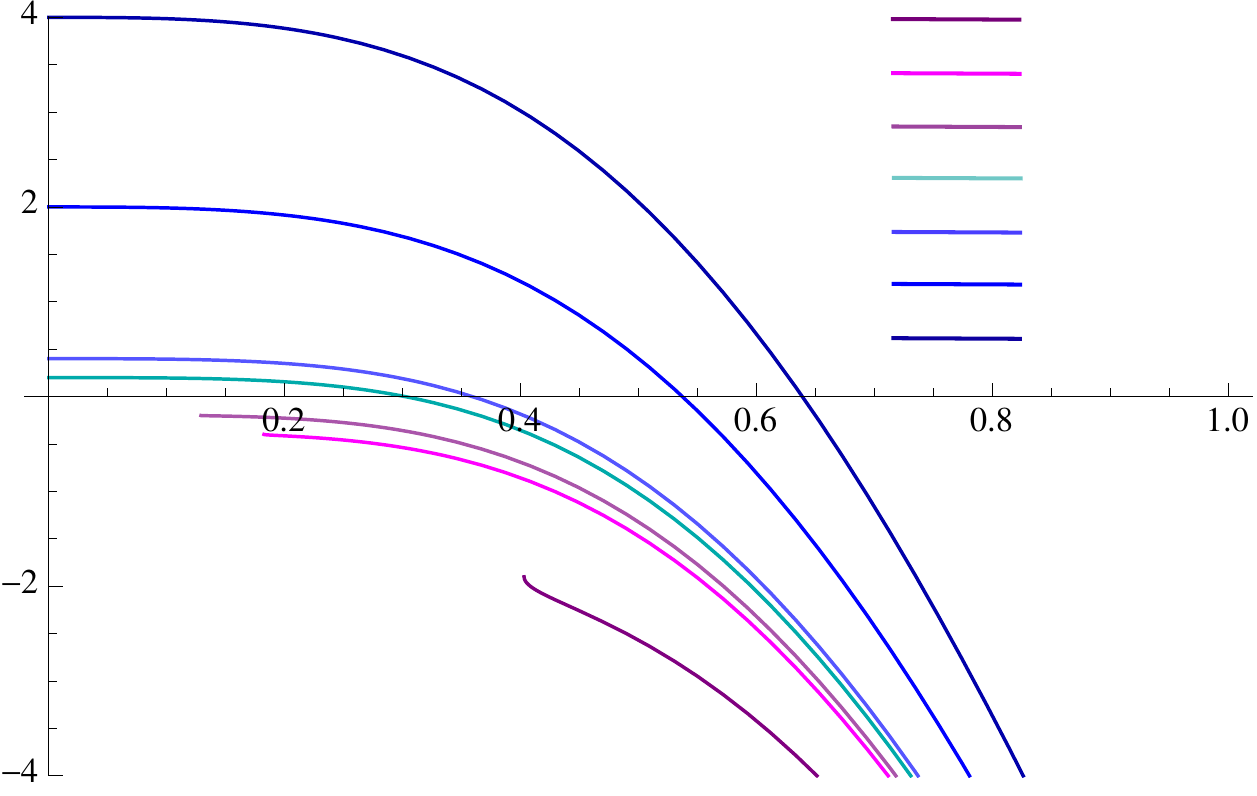}
\put(96,23){$\tilde{\Lambda}$}
\put(6,63){$\tilde{R}$}
\put(83,60.5){$\tilde{\lambda} = -0.50$}
\put(83,56.5){$\tilde{\lambda} = -0.10$}
\put(83,52.5){$\tilde{\lambda} = -0.05$}
\put(83,48){$\tilde{\lambda} = 0.05$}
\put(83,43.8){$\tilde{\lambda} = 0.10$}
\put(83,39.8){$\tilde{\lambda} = 0.50$}
\put(83,35.2){$\tilde{\lambda} = 1.00$}
\end{overpic}
\caption{$\tilde{R}$ vs.~$\tilde{\Lambda}$ for a backreacted CFT for various values of $\tilde{\lambda}$ as given in \protect\eqref{eq:CutoffResult13}. Note that for $\tilde{\lambda} <0$ there is a lower bound on $\tilde{\L}$ for a solution to exist.}
\label{fig:RvsLambda_CFTcutoff}
\end{figure}
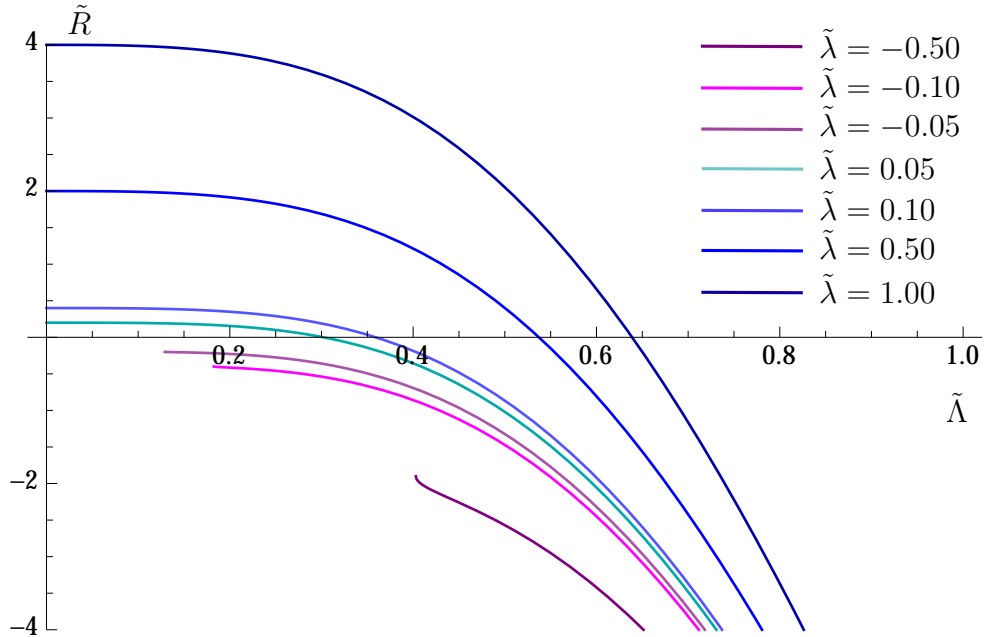

This parameterisation of $\mathcal{F}$ is convenient for the following reason. {Thermodynamics of dS space implies that the functions $\mathcal{W}$ and $\mathcal{U}$ satisfy the identity (see once more appendix \ref{app:thermo} upon using the dictionary in \ref{app:dictionary}):}
\begin{align}
\label{eq:CutoffResult4} \frac{\partial \mathcal{W}}{\partial R} = \Lambda^{-2} \, \bigg( \mathcal{U} - R \frac{\partial \mathcal{U}}{\partial R} \bigg) \, .
\end{align}
{This can be checked explicitly using the expressions in \eqref{eq:CutoffResult10} and \eqref{eq:CutoffResult11} and can be confirmed to hold also for the case of AdS backgrounds.}
As a result, when applying the differential operator in \eqref{eq:CutoffResult2} to $\mathcal{F}$, only the $\mathcal{W}$-part of $\mathcal{F}$ will contribute, while any dependence on $\mathcal{U}$ is eliminated. In particular, the equation for constant curvature solutions \eqref{eq:CutoffResult2} becomes
\begin{align}
\label{eq:CutoffResult5} M_0^2 R -4 M_0^2 \lambda_0 + 4 \tilde{a} \Lambda^4 \mathcal{W} \big( R \Lambda^{-2} \big) =0 \, .
\end{align}
In equation (\ref{eq:CutoffResult5}) only the ``energy" part of the QFT partition function contributes and backreacts on the bare cosmological constant. Moreover, as shown in \cite{F}, all holographic theories 
have manifestly $\mathcal{W}>0$. Therefore, all holographic theories contribute a negative correction to the bare cosmological constant. Moreover, such a correction scales as ${\cal O}(N^2)$ and, for large enough $N$, can overcome any finite cosmological constant and render it negative.

This result is very different from the case of weakly coupled theories. In such theories, at finite cutoff, the corrections to the cosmological constant 
can have either sign. Bosons contribute negatively while fermions contribute positively. Depending on the boson-fermion balance, at weak coupling the cosmological constant correction can have either sign.
Our finding seems to imply that at large $N$ and strong coupling all theories give negative contributions to the cosmological constant. Especially since this contribution has a thermodynamic interpretation, \cite{F}, this suggest that the dichotomy is similar to microscopic dynamics vs.~thermodynamics. For example, entropy increase can be violated microscopically but is always valid in the thermodynamic limit.

It is important to have a finite cutoff for the arguments above. Proceeding with renormalization hides some of these contributions into the renormalized cosmological constant and Planck mass and monotonicity is not further visible. 

To proceed with our calculation, we again suppress the subscript on $\tilde{a}$ to highlight that this expression is valid for any CFT.
Inserting the expression (\ref{eq:CutoffResult10}) for $\mathcal{W}$
and using the rescaled variables defined in \eqref{eq:CutoffResult8} this becomes
\begin{align}
\label{eq:CutoffResult12} \tilde{R} -4 \tilde{\lambda} + 24 \tilde{\Lambda}^4 \sqrt{1+\frac{\tilde{R} \tilde{\Lambda}^{-2}}{12}} =0 \, .
\end{align}
Taking the term involving the square root to the RHS and squaring this becomes a quadratic equation for $\tilde{R}$ which admits two branches of solutions. However, only one of them is also a solution to the original equation \eqref{eq:CutoffResult12}. On this branch we find:
\begin{align}
\label{eq:CutoffResult13} \tilde{R} = 4 \tilde{\lambda} - 24 \, \tilde{\Lambda}^6 \bigg( \sqrt{1+ \frac{1}{3} \tilde{\lambda} \tilde{\Lambda}^{-6} + \tilde{\Lambda}^{-4}} -1 \bigg)  \, .
\end{align}
Note that at large $N$, taking into account the $N$-dependent rescaling of our variables\footnote{$N$ is hidden in $\tilde{a}$ as $\tilde{a}=\tilde{a}_0 N^2$ with $\tilde{a}_{0}$ an $\mathcal{O}(1)$ number.} the correction to the cosmological constant above scales as $O(N^2)$. Therefore, for any finite original cosmological constant, at large enough $N$ it is overwhelmed by the holographic contribution and one is driven to AdS solutions.

This is our general result for constant-curvature solutions from backreacting a (holographic) CFT with UV cutoff $\tilde{\L}$ on a system with cosmological constant $\tilde{\lambda}$. In fig.~\ref{fig:RvsLambda_CFTcutoff} we plot $\tilde{R}$ for this solution vs.~$\tilde{\Lambda}$ for various values of $\tilde{\lambda}$. We make the following observations:
\begin{itemize}
\item As $\tilde{\Lambda}$ is increased for fixed $\tilde{\lambda}$ the corresponding value $\tilde{R}$ always decreases, i.e.~as `more' of the CFT is integrated out, the effect is to reduce $\tilde{R}$.
\item This monotonic decrease implies that even if $\tilde{R} >0$ for $\tilde{\Lambda}=0$, for sufficiently large $\tilde{\L}$ backreaction always leads to $\tilde{R} <0$. The value of $\tilde{\L}_0$ where backreaction gives $\tilde{R}=0$ depends on $\tilde{\lambda}$ and is given by:
\begin{align}
\tilde{\L}_0^4 = \frac{\tilde{\lambda}}{6} \, .
\end{align}
\item Note that for $\tilde{\lambda} <0$ there is a lower bound on $\tilde{\L}$ for a real solution to exist. This condition can be traced back to the fact that for \eqref{eq:CutoffResult10} and \eqref{eq:CutoffResult11} to be real-valued we require $\tilde{\L}^2 \geq - \tilde{R} /12$.
\end{itemize}

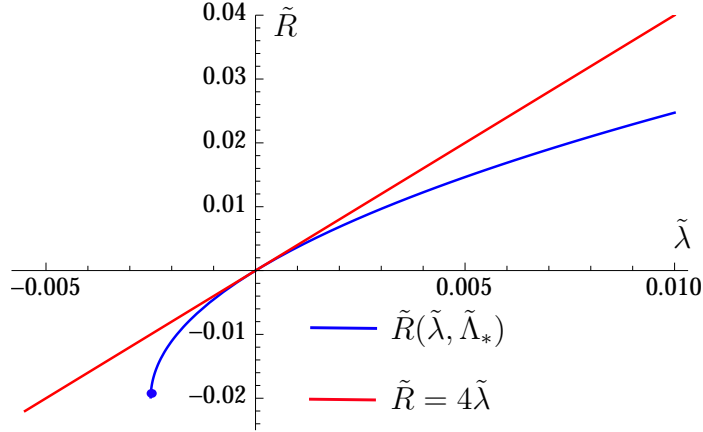
\begin{figure}[t]
\centering
\begin{overpic}
[width=0.6\textwidth]{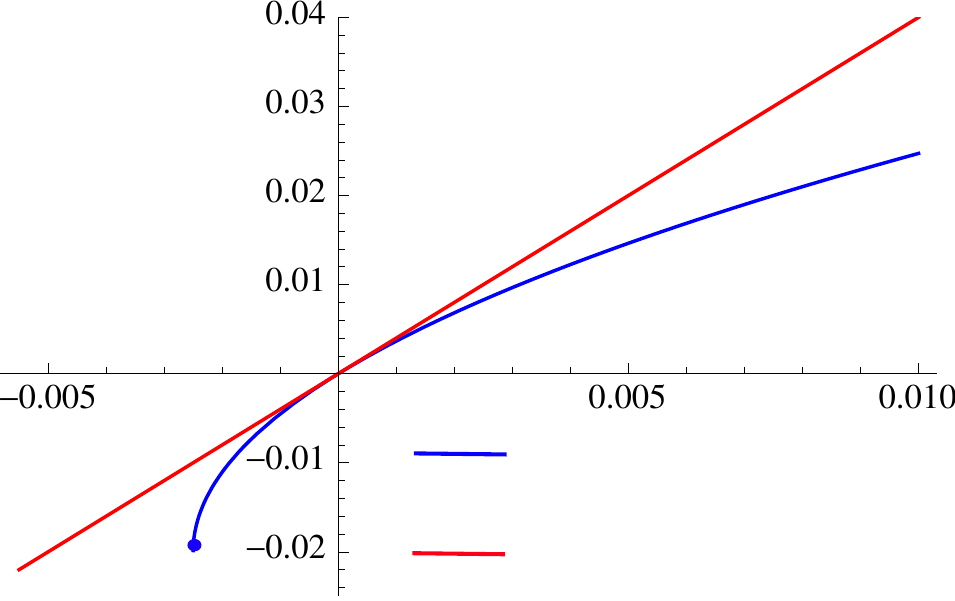}
\put(96,27){$\tilde{\lambda}$}
\put(38,58){$\tilde{R}$}
\put(55,13.75){$\tilde{R}(\tilde{\lambda}, \tilde{\Lambda}_*)$}
\put(55,3.5){$\tilde{R}=4 \tilde{\lambda}$}
\end{overpic}
\caption{\textbf{Blue:} $\tilde{R}$ vs.~$\tilde{\lambda}$ as given in \protect\eqref{eq:superhorizonresult} for backreaction due to the `superhorizon modes' of a 4d CFT, i.e.~for a CFT with UV cutoff $\tilde{\Lambda}_*$ satisfying \protect\eqref{eq:Lambdatildestareq}. The plot ends at the blue dot. \textbf{Red:} $\tilde{R} =4 \tilde{\lambda}$, i.e.~the solution for $\tilde{R}$ without backreaction ($\tilde{\Lambda}=0$).}
\label{fig:RvslambdaCFTcutoffself}
\end{figure}

While the general result for backreaction due to a cutoff CFT is already given in \eqref{eq:CutoffResult13}, it will be instructive to consider a special case next. In particular, we will examine the effect of backreaction due to `superhorizon modes' only. As we do not really employ a mode expansion here, the term `superhorizon modes' is not strictly applicable, but we use it to refer to the case where the cutoff is given by the backreacted curvature of the background. That is we will consider a CFT with cutoff $\tilde{\Lambda}_*$, which is self-consistently determined as
\be
\label{eq:Lambdatildestareq} \tilde{\Lambda}_*^2 = \tilde{R}(\tilde{\lambda}, \tilde{\Lambda}_*) \, .
\ee
If $\tilde{\Lambda}_*$ was a momentum-cutoff then the above relation would state that only the effect of superhorizon modes is included. Using \eqref{eq:CutoffResult13} in \eqref{eq:Lambdatildestareq} we can solve for $\tilde{\Lambda}_*$ and hence $\tilde{R}(\tilde{\lambda}, \tilde{\Lambda}_*)$ as
\begin{align}
\label{eq:superhorizonresult} \tilde{R}(\tilde{\lambda}, \tilde{\Lambda}_*) = \tilde{\Lambda}_*^2 = \frac{1}{8 \sqrt{39}} \bigg( \sqrt{1 + 64 \sqrt{39} \, \tilde{\lambda}} -1 \bigg) \, .
\end{align}
In figure \ref{fig:RvslambdaCFTcutoffself} we plot $\tilde{R}(\tilde{\lambda}, \tilde{\Lambda}_*)$ vs.~$\tilde{\lambda}$ (blue line). This can then be compared with the solution $\tilde{R} = 4 \tilde{\lambda}$ in absence of a backreacting CFT (red line). We make the following observations:
\begin{itemize}
\item Most importantly, backreaction always reduces the value of $\tilde{R}$ compared to the case without backreaction.
\item For $\tilde{\lambda} < - \tfrac{1}{64 \sqrt{39}}$ there is no backreacted solution $\tilde{R}(\tilde{\lambda}, \tilde{\Lambda}_*)$ as the condition \eqref{eq:Lambdatildestareq} cannot be satisfied.
\item For $\tilde{\lambda} \rightarrow \infty$ the backreacted solution behaves as
\begin{align}
\tilde{R}(\tilde{\lambda}, \tilde{\Lambda}_*) \underset{\tilde{\lambda} \rightarrow \infty}{=} \sqrt{\tilde{\lambda} / \sqrt{39}} \, ,
\end{align}
i.e.~it grows with increasing $\tilde{\lambda}$, but is parametrically suppressed by a power of $1 / \sqrt{\tilde{\lambda}}$ compared to the solution without backreaction.
\end{itemize}

\subsection{Backreacted 4d QFT with a UV cutoff}
\label{sec:ResultsQFTcutoffQFT}
Here we consider backreaction of a QFT as described in section \ref{sec:QFTdef} with a UV energy cutoff $\Lambda$. Once we consider backreaction of a non-conformal QFT we are not able to give an analytic expressions for $\mathcal{F}(m \Lambda^{-1}, R\Lambda^{-2})$, except in particular corners of parameter space. Thus here we present results based on numerical expressions for $\mathcal{F}(m \Lambda^{-1}, R\Lambda^{-2})$. For details on how these are obtained see section \ref{sec:onshell}.

Explicit numerical results will be displayed for two example QFTs. Here we again choose the theories labelled QFT$_1$ and QFT$_2$ before, whose backreaction we studied in section \ref{sec:ResultsQFTrenQFT} in the case of UV-completeness. For a specification of QFT$_{1,2}$ in terms of the choice for the parameters $\Delta$, $\Delta^{\textsc{ir}}$ and $\tilde{a}_{\textsc{uv}} / \tilde{a}_{\textsc{ir}}$ see the definition below equation \eqref{eq:modelparameters}.

\begin{figure}[t]
\centering
\begin{overpic}
[width=0.65\textwidth]{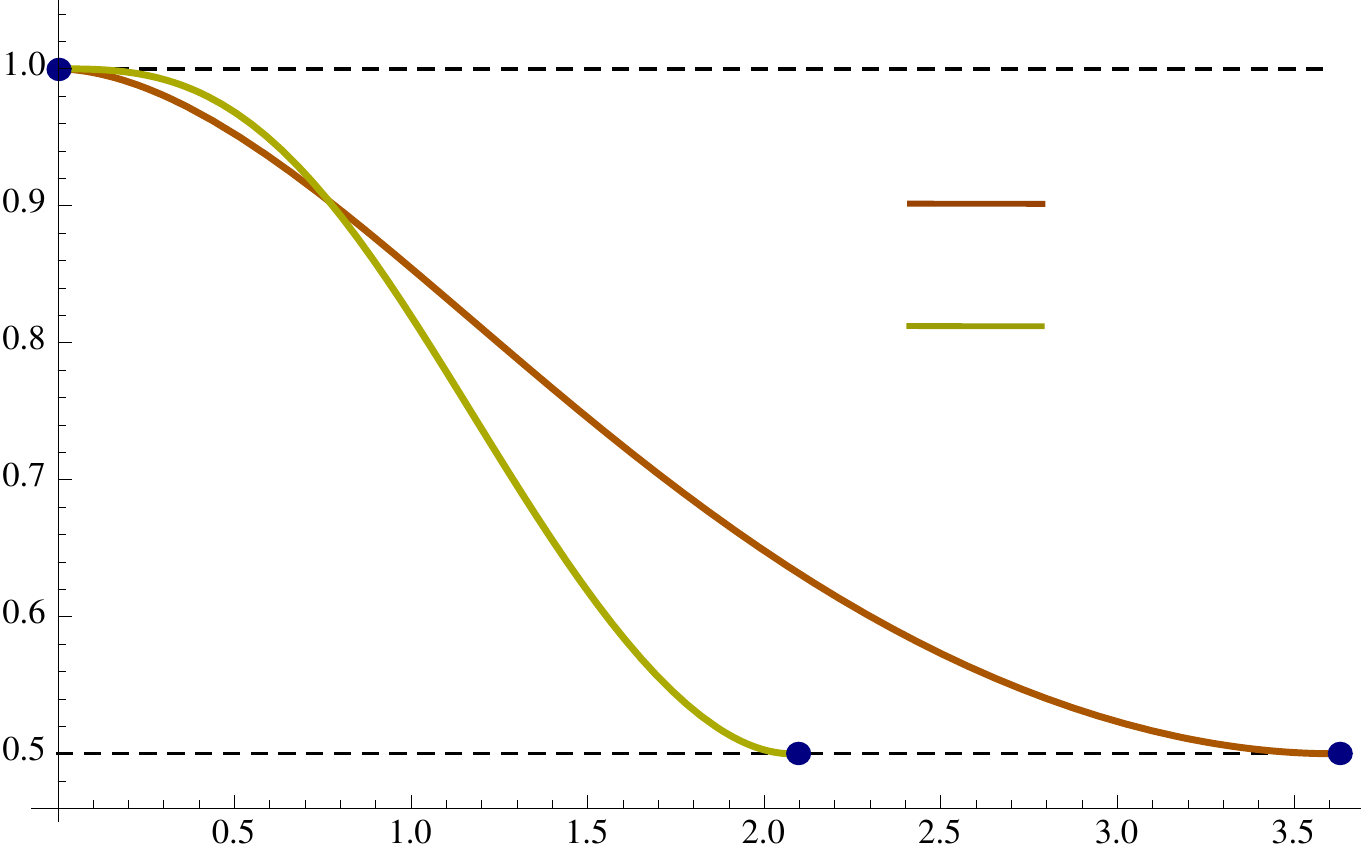}
\put(102,3){$m \Lambda^{-1}$}
\put(0,65.5){$\tilde{a} / \tilde{a}_{\textsc{uv}}$}
\put(7,10){$\tilde{a}_{\textsc{ir}} / \tilde{a}_{\textsc{uv}}$}
\put(7,60){${\textsc{uv}}$}
\put(57,10){${\textsc{ir}}$}
\put(96,10){${\textsc{ir}}$}
\put(79,47){$\Delta=3.1$}
\put(79,38){$\Delta=2.6$}
\end{overpic}
\caption{Plot of $\tilde{a}(m\Lambda^{-1}) / \tilde{a}_{\textsc{uv}}$ vs.~$m\Lambda^{-1}$ for QFT$_1$ and QFT$_2$. The two QFTs differ in the dimension of the operator perturbing the UV QFT and in the maximal value of $m \Lambda^{-1}$. For QFT$_1$ we have $\Delta=3.1$ and $m \Lambda^{-1} |_{\textrm{max}}= 3.61$ (orange plot), for QFT$_2$ we have $\Delta=2.6$ and $m \Lambda^{-1} |_{\textrm{max}}= 2.08$ (yellow plot). In both cases we find that $\tilde{a}(m\Lambda^{-1})$ interpolates monotonically between $\tilde{a}_{\textsc{uv}}$ and $\tilde{a}_{\textsc{ir}}$ as $m\Lambda^{-1}$ is varied from $0$ to $m \Lambda^{-1} |_{\textrm{max}}$.}
\label{fig:atildePlots}
\end{figure}
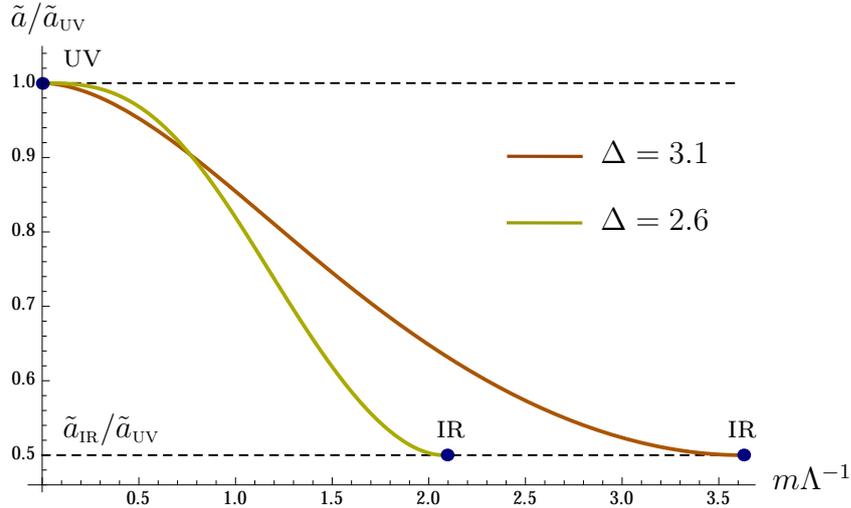

To study constant-curvature solutions, our focus is not primarily on $\mathcal{F}$, but how it contributes to the equation for constant-curvature solutions, \eqref{eq:CutoffResult2}. Here we find it useful to write this contribution as
\begin{align}
\label{eq:CutoffResult17} \Big( 2 - R \frac{\partial}{\partial R} \Big) \, \tilde{a}_{\textsc{uv}} \mathcal{F} \Big( \tfrac{m}{\Lambda},  \tfrac{R}{\Lambda^2} \Big)  = 2 \Lambda^4 \, \tilde{a} \Big( \tfrac{m}{\Lambda} \Big) \, \mathcal{W} \Big(  \tfrac{m}{\Lambda} , \tfrac{R}{\Lambda^2} \Big) \, ,
\end{align}
where the functions $\tilde{a}(m \Lambda^{-1})$ and $\mathcal{W}(m \Lambda^{-1}, R\Lambda^{-2})$ will be described presently. Using this in \eqref{eq:CutoffResult2} the equation for constant-curvature solutions becomes
\begin{align}
\label{eq:CutoffResult18} M_0^2 R -4 M_0^2 \lambda_0 + 4 \tilde{a} \Big( \tfrac{m}{\Lambda} \Big) \, \Lambda^4 \, \mathcal{W} \Big( \tfrac{m}{\Lambda} , \tfrac{R}{\Lambda^2} \Big) =0 \, .
\end{align}
With our choice in \eqref{eq:CutoffResult17} this is superficially similar in form to the equation obtained in the case of the UV CFT in \eqref{eq:CutoffResult5}, but with $\tilde{a}_{\textsc{uv}}$ replaced by $\tilde{a}(m \Lambda^{-1})$ and $\mathcal{W}$ now a function of both $m \Lambda^{-1}$ and $R \Lambda^{-2}$. We  find that this is helpful when discussing how backreaction due to non-conformal QFTs differs from that of CFTs.

One reason why the parameterisation in \eqref{eq:CutoffResult17} is useful is that the function $\tilde{a} (m \Lambda^{-1})$ interpolates between the anomaly coefficient $\tilde{a}_{\textsc{uv}}$ of the UV CFT and the anomaly coefficient $\tilde{a}_{\textsc{ir}}$ of the IR CFT as $m \Lambda^{-1}$ is varied from $0$ to its maximal value:
\begin{align}
\label{eq:CutoffResult19} \tilde{a} \Big( \tfrac{m}{\Lambda}  \rightarrow 0 \Big) & \longrightarrow \tilde{a}_{\textsc{uv}} \, , \\
\label{eq:CutoffResult20} \tilde{a} \Big( \tfrac{m}{\Lambda} \rightarrow {\left.\tfrac{m}{\Lambda} \right|}_{\textrm{max}} \Big) & \longrightarrow \tilde{a}_{\textsc{ir}} \, ,
\end{align}
where ${\left|\tfrac{m}{\Lambda} \right|}_{\textrm{max}}$ is the maximal value this quantity can take, which is a fixed number for a given QFT considered here. Further, for all examples considered here the function $\tilde{a}$ decreases monotonically with $m \Lambda^{-1}$, i.e.
\begin{align}
\label{eq:CutoffResult21} \frac{\partial \tilde{a}}{\partial (\tfrac{m}{\Lambda})} \leq 0 \, .
\end{align}
The function $\tilde{a}$ can hence be seen as a running anomaly parameter. In virtue of \eqref{eq:CutoffResult21} the parameter $\tilde{a}$ evolves monotonically with $\tfrac{m}{\Lambda}$, thus realising the $a$-theorem in 4 dimensions. For a definition of $\tilde{a}(m \Lambda^{-1})$ in terms of quantities in the holographic dual see section \ref{sec:QFTcontrib} or appendix \ref{app:dictionary}.

To illustrate this, in figure \ref{fig:atildePlots} we show $\tilde{a}$ as a function of $m \Lambda^{-1}$ for both QFT$_1$ and QFT$_2$. In both cases $\tilde{a}$ interpolates monotonically between the values of the UV and IR anomaly coefficients as $m\Lambda^{-1}$ is varied from $0$ to ${m \L^{-1} |}_{\textrm{max}}$, as asserted above.

\begin{figure}[t]
\centering
\begin{subfigure}{.5\textwidth}
 \centering
   \begin{overpic}
[width=1.0\textwidth]{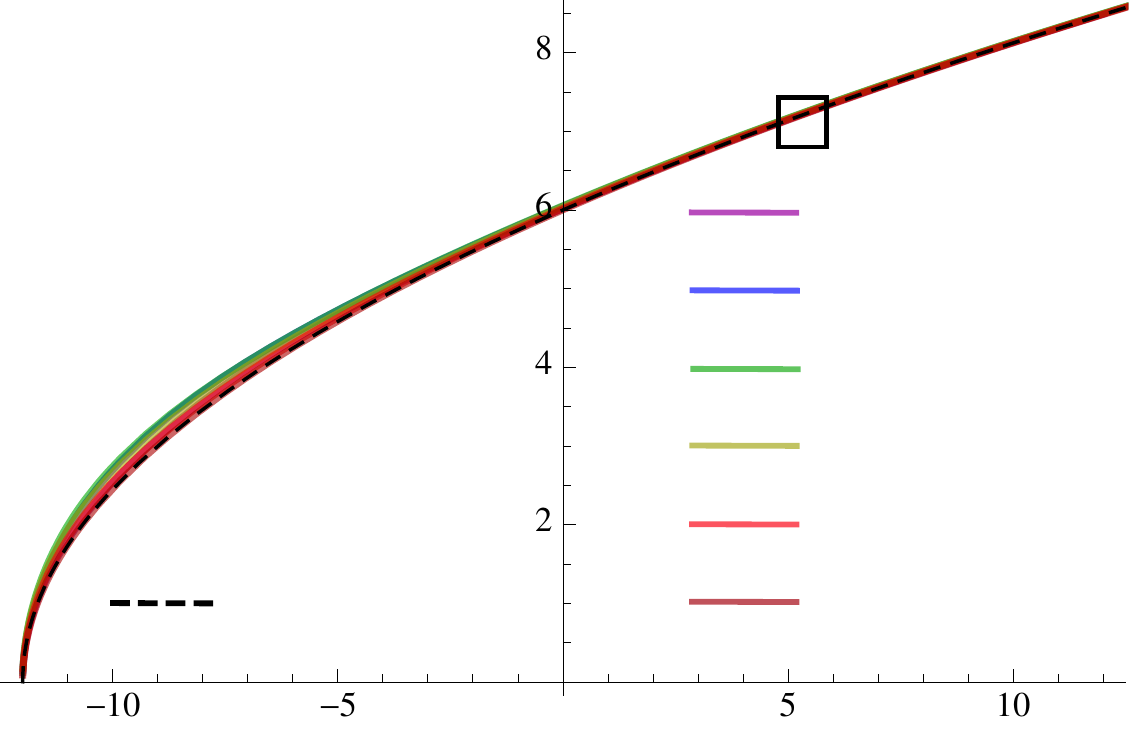}
\put(86,-5){$R \L^{-2}$}
\put(53,61){$\mathcal{W}$}
\put(74.5,44.5){${\scriptstyle m \L^{-1} = 0.46}$}
\put(74.5,37.6){${\scriptstyle m \L^{-1} = 1.00}$}
\put(74.5,30.7){${\scriptstyle m \L^{-1} = 1.57}$}
\put(74.5,23.8){${\scriptstyle m \L^{-1} = 2.16}$}
\put(74.5,16.9){${\scriptstyle m \L^{-1} = 2.77}$}
\put(74.5,10){${\scriptstyle m \L^{-1} = 3.39}$}
\put(22,10){${\scriptstyle m \L^{-1} = 0}$}
\end{overpic}
\caption{\hphantom{A}}
\label{fig:WvsRLm2Delta0p9LIR0p5large}
\end{subfigure}%
\begin{subfigure}{.5\textwidth}
 \centering
   \begin{overpic}
[width=1.0\textwidth]{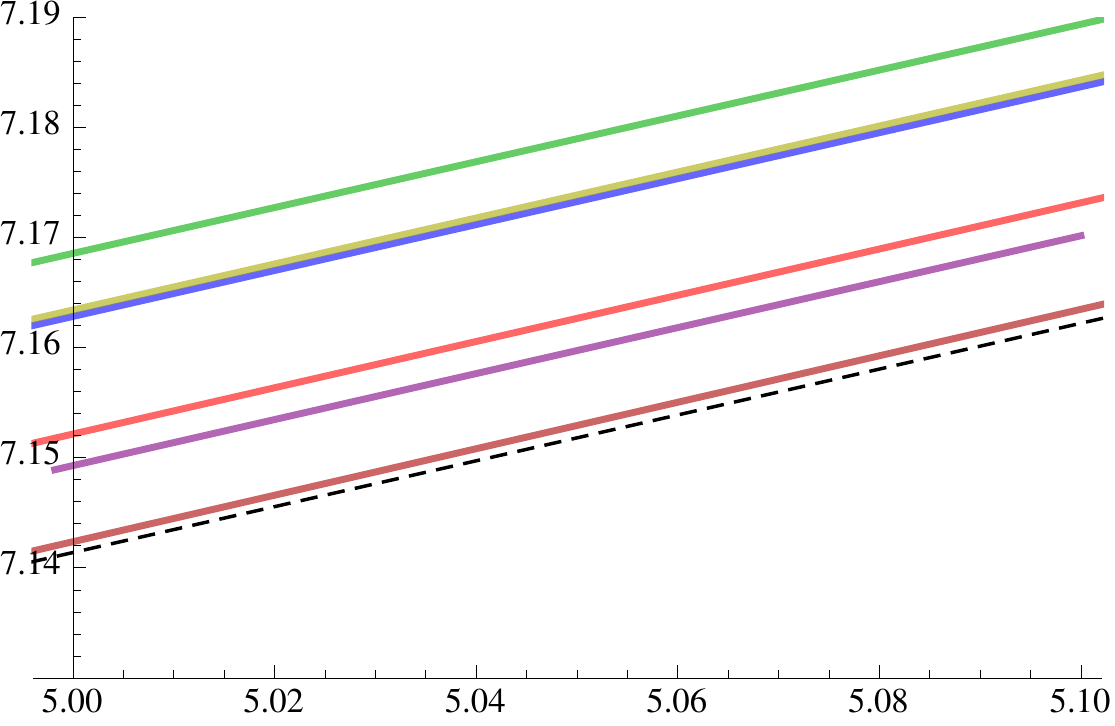}
\put(86,-5){$R \L^{-2}$}
\put(10,61){$\mathcal{W}$}
\end{overpic}
\caption{\hphantom{A}}
\label{fig:WvsRLm2Delta0p9LIR0p5zoom}
\end{subfigure}%
\caption{\textbf{(a)-(b):} $\mathcal{W}$ vs.~$R \Lambda^{-2}$ for several values of $m \Lambda^{-1}$ for QFT$_1$ (i.e.~$\Delta=3.10$, $\Delta^{\textsc{ir}}=4.74$, $\tilde{a}_{\textsc{uv}} / \tilde{a}_{\textsc{ir}}=2$ and $m \Lambda^{-1} |_{\textrm{max}}= 3.61$). The plot in (b) is the magnified version of the region within the black box in (a). The black dashed line corresponds to the function $\mathcal{W}_{\textsc{CFT}} (R \L^{-2}) = 6 \, \sqrt{1 + R \L^{-2} / 12}$.}
\label{fig:WvsRLm2Delta0p9LIR0p5}
\end{figure}

\begin{figure}[t]
\centering
\begin{subfigure}{.5\textwidth}
 \centering
   \begin{overpic}
[width=1.0\textwidth]{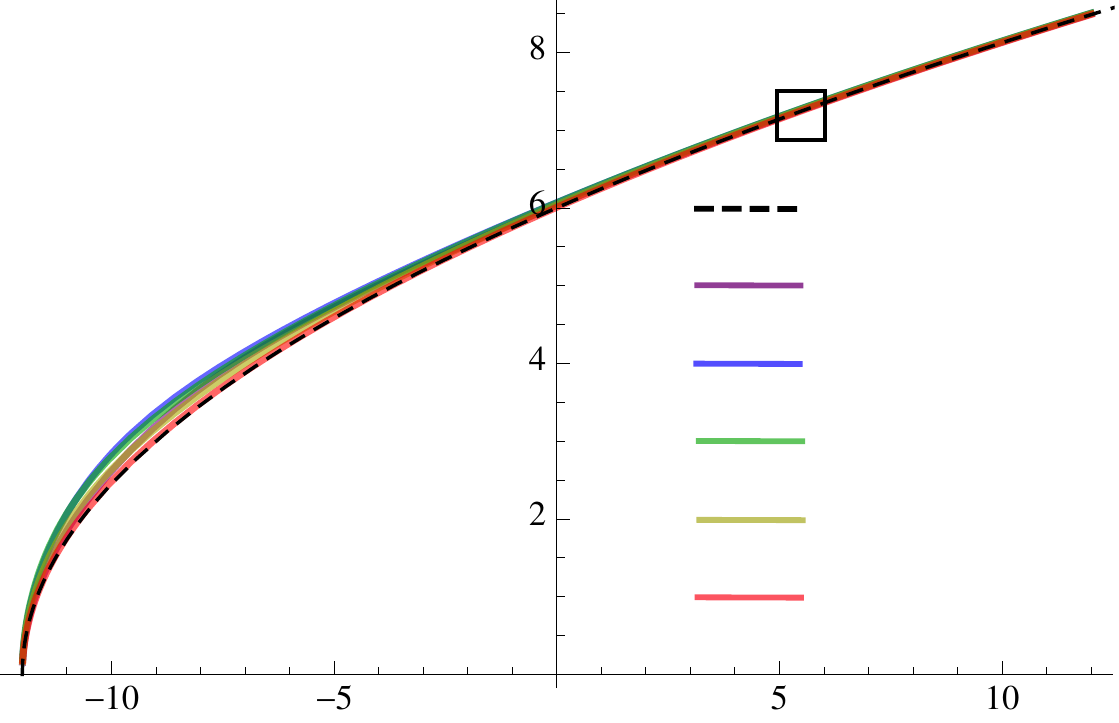}
\put(86,-5){$R \L^{-2}$}
\put(53,60){$\mathcal{W}$}
\put(75,45){${\scriptstyle m \L^{-1} = 0}$}
\put(75,38){${\scriptstyle m \L^{-1} = 0.61}$}
\put(75,31){${\scriptstyle m \L^{-1} = 1.00}$}
\put(75,24){${\scriptstyle m \L^{-1} = 1.34}$}
\put(75,17){${\scriptstyle m \L^{-1} = 1.64}$}
\put(75,10){${\scriptstyle m \L^{-1} = 1.92}$}
\end{overpic}
\caption{\hphantom{A}}
\label{fig:WvsRLm2Delta1p4LIR0p5large}
\end{subfigure}%
\begin{subfigure}{.5\textwidth}
 \centering
   \begin{overpic}
[width=1.0\textwidth]{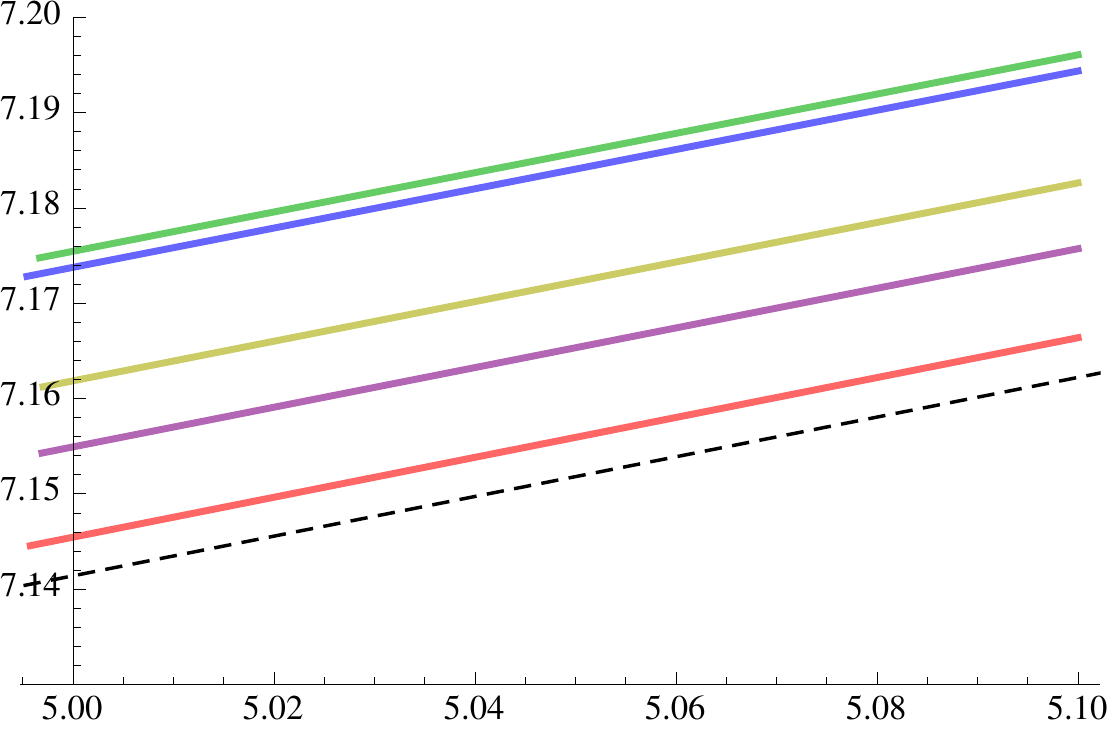}
\put(86,-5){$R \L^{-2}$}
\put(10,61){$\mathcal{W}$}
\end{overpic}
\caption{\hphantom{A}}
\label{fig:WvsRLm2Delta1p4LIR0p5zoom}
\end{subfigure}%
\caption{\textbf{(a)-(b):} $\mathcal{W}$ vs.~$R \Lambda^{-2}$ for several values of $m \Lambda^{-1}$ for QFT$_2$ (i.e.~$\Delta=2.60$, $\Delta^{\textsc{ir}}=4.93$, $\tilde{a}_{\textsc{uv}} / \tilde{a}_{\textsc{ir}}=2$ and $m \Lambda^{-1} |_{\textrm{max}}= 2.08$). The plot in (b) is the magnified version of the region within the black box in (a). The black dashed line corresponds to the function $\mathcal{W}_{\textsc{CFT}} (R \L^{-2}) = 6 \, \sqrt{1 + R \L^{-2} / 12}$.}
\label{fig:WvsRLm2Delta1p4LIR0p5}
\end{figure}

We now turn to the function $\mathcal{W}(m \L^{-1}, R \L^{-2})$ appearing in \eqref{eq:CutoffResult17} and \eqref{eq:CutoffResult18}. For $m \Lambda^{-1} \rightarrow 0$ this asymptotes to the function $\mathcal{W}_{\textsc{CFT}} (R \L^{-2})$ encountered in the case of a CFT and given in \eqref{eq:CutoffResult10}. This is as expected and can be understood analytically (see e.g.~section \ref{sec:onshell}). Then, from numerical observation we find that for $m \L^{-1} \rightarrow {m \L^{-1} |}_{\textrm{max}}$ the function $\mathcal{W}(m \L^{-1}, R \L^{-2})$ once more asymptotes to $\mathcal{W}_{\textsc{CFT}} (R \L^{-2})$ as given in \eqref{eq:CutoffResult10}. To, summarise, we have that
\begin{align}
\label{eq:CutoffResult22} \mathcal{W} \Big( \tfrac{m}{\Lambda}  \rightarrow 0 \, , \tfrac{R}{\Lambda^2} \Big) & \longrightarrow \mathcal{W}_{\textsc{CFT}} \Big( \tfrac{R}{\L^2} \Big) = 6 \, \sqrt{1 + \tfrac{R \L^{-2}}{12}} \, , \\
\label{eq:CutoffResult23} \mathcal{W} \Big( \tfrac{m}{\Lambda} \rightarrow {\left. \tfrac{m}{\Lambda} \right|}_{\textrm{max}} \, , \tfrac{R}{\Lambda^2} \Big) & \longrightarrow \mathcal{W}_{\textsc{CFT}} \Big( \tfrac{R}{\L^2} \Big) = 6 \, \sqrt{1 + \tfrac{R \L^{-2}}{12}} \, .
\end{align}
Overall, we observe that the dependence of $\mathcal{W}$ on $m \L^{-1}$ is rather `weak', in the sense that the value of $\mathcal{W}$ for fixed $R \L^{-2}$ barely changes when $m \L^{-1}$ is modified. This is best illustrated by explicit plots. In figures \ref{fig:WvsRLm2Delta0p9LIR0p5} and \ref{fig:WvsRLm2Delta1p4LIR0p5} we plot $\mathcal{W}$ vs.~$R \L^{-2}$ for various values of $m \L^{-1}$ for two example QFTs. Here we consider the same two QFTs whose functions $\tilde{a}(m \L^{-1})$ were displayed in figure \ref{fig:atildePlots}. The main observation from figures \ref{fig:WvsRLm2Delta0p9LIR0p5} and \ref{fig:WvsRLm2Delta1p4LIR0p5} is that for any fixed value of $m \L^{-1}$ the function $\mathcal{W}$ very closely matches $\mathcal{W}_{\textsc{CFT}} (R \L^{-2})$ over the whole range of $R \Lambda^{-2}$. That is, we find that any value for $m \L^{-1}$ in general only leads to a small perturbation of $\mathcal{W}$ compared to the CFT case, i.e.
\begin{align}
\label{eq:CutoffResult24} \mathcal{W} \Big( \tfrac{m}{\Lambda} \, , \tfrac{R}{\Lambda^2} \Big) = \mathcal{W}_{\textsc{CFT}} \Big( \tfrac{R}{\L^2} \Big) + \omega \Big( \tfrac{m}{\Lambda} \, , \tfrac{R}{\Lambda^2} \Big) \, , \quad \textrm{with} \quad \frac{\big|\omega \big( \tfrac{m}{\Lambda} \, , \tfrac{R}{\Lambda^2} \big) \big|}{\big|\mathcal{W}_{\textrm{CFT}} \big( \tfrac{R}{\Lambda^2} \big) \big|} \ll 1 \quad \forall \quad \tfrac{m}{\L} \, , \tfrac{R}{\L^2} \, .
\end{align}
The plots in figures \ref{fig:WvsRLm2Delta0p9LIR0p5} and \ref{fig:WvsRLm2Delta1p4LIR0p5} exhibit further details regarding the behaviour of $\mathcal{W}$ as $m \L^{-1}$ is varied. In particular, one finds that when $m \L^{-1}$ is initially increased from 0, the value for $\mathcal{W}$ (at fixed $R \L^{-2}$) departs more and more from the corresponding value $\mathcal{W}_{\textrm{CFT}}$. This is best seen in figures \ref{fig:WvsRLm2Delta0p9LIR0p5zoom} and \ref{fig:WvsRLm2Delta1p4LIR0p5zoom}. Then, for some intermediate value of $m \L^{-1}$ the departure of $\mathcal{W}$ from $\mathcal{W}_{\textrm{CFT}}$ reaches some maximal value and starts to shrink again as $m \L^{-1}$ is increased further. Finally, for $m \L^{-1} \rightarrow {m \L^{-1} |}_{\textrm{max}}$ the value of $\mathcal{W}$ coincides once more with $\mathcal{W}_{\textrm{CFT}}$.

All these observations have the following implications for constant-curvature solutions as determined by \eqref{eq:CutoffResult18}. The above discussion implies that for a first approximation we can ignore the $m \L^{-1}$-dependence of $\mathcal{W}$ and make the replacement
\begin{align}
\label{eq:CutoffResult25}  \tilde{a} \Big( \tfrac{m}{\Lambda} \Big) \, \mathcal{W} \Big(  \tfrac{m}{\Lambda} , \tfrac{R}{\Lambda^2} \Big) \longrightarrow \tilde{a} \Big( \tfrac{m}{\Lambda} \Big) \, \mathcal{W}_{\textrm{CFT}} \Big( \tfrac{R}{\Lambda^2} \Big) \, ,
\end{align}
in \eqref{eq:CutoffResult18}, which gives
\begin{align}
\label{eq:CutoffResult26} M_0^2 R -4 M_0^2 \lambda_0 + 4 \tilde{a} \Big( \tfrac{m}{\Lambda} \Big) \, \Lambda^4 \, \mathcal{W}_{\textrm{CFT}} \Big(\tfrac{R}{\Lambda^2} \Big) =0 \, .
\end{align}
That is, backreaction on constant-curvature solutions due to a QFT with finite $m \L^{-1}$ is expected to be approximately like that of a CFT with anomaly coefficient $\tilde{a}(m \L^{-1})$. Since $\tilde{a}(m \L^{-1})$ interpolates monotonically between $\tilde{a}_{\textsc{uv}}$ and $\tilde{a}_{\textsc{ir}}$ as $m \L^{-1}$ is varied over its whole range, the corresponding backreaction interpolates between that due to the UV and IR CFTs. The fact that $\mathcal{W}$ does not exactly coincide with $\mathcal{W}_{\textrm{CFT}}$ for finite $m$ is expected to give rise to small quantitative differences in the backreaction effects between a QFT and a CFT, but not to change the picture qualitatively.

\begin{figure}[t]
\centering
\begin{subfigure}{.5\textwidth}
 \centering
   \begin{overpic}
[width=1.0\textwidth]{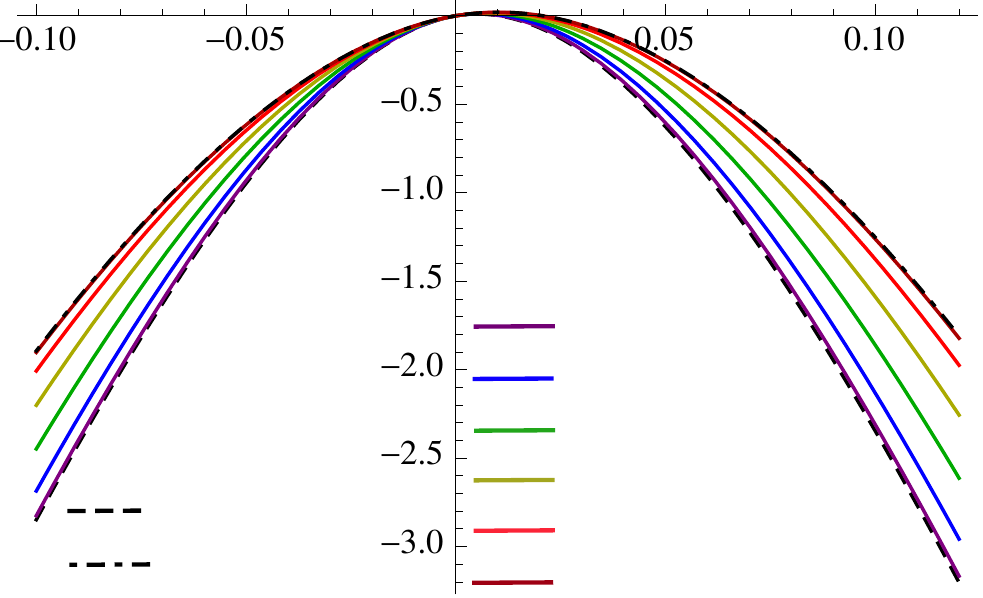}
\put(89,62){$\tilde{a}_{\textsc{uv}}  \hat{\lambda}$}
\put(43,65){$\tilde{a}_{\textsc{uv}} \hat{R}$}
\put(17.2,7.5){{\small UV CFT}}
\put(17.2,1.5){{\small IR CFT}}
\put(59,27){${\scriptstyle m \L^{-1} = 0.46}$}
\put(59,21.75){${\scriptstyle m \L^{-1} = 1.00}$}
\put(59,16.5){${\scriptstyle m \L^{-1} = 1.57}$}
\put(59,11.25){${\scriptstyle m \L^{-1} = 2.16}$}
\put(59,6){${\scriptstyle m \L^{-1} = 2.77}$}
\put(59,0.75){${\scriptstyle m \L^{-1} = 3.39}$}
\end{overpic}
\caption{\hphantom{A}}
\label{fig:RvslambdaQFT0p9LIR0p5large}
\end{subfigure}%
\begin{subfigure}{.5\textwidth}
 \centering
   \begin{overpic}
[width=1.0\textwidth]{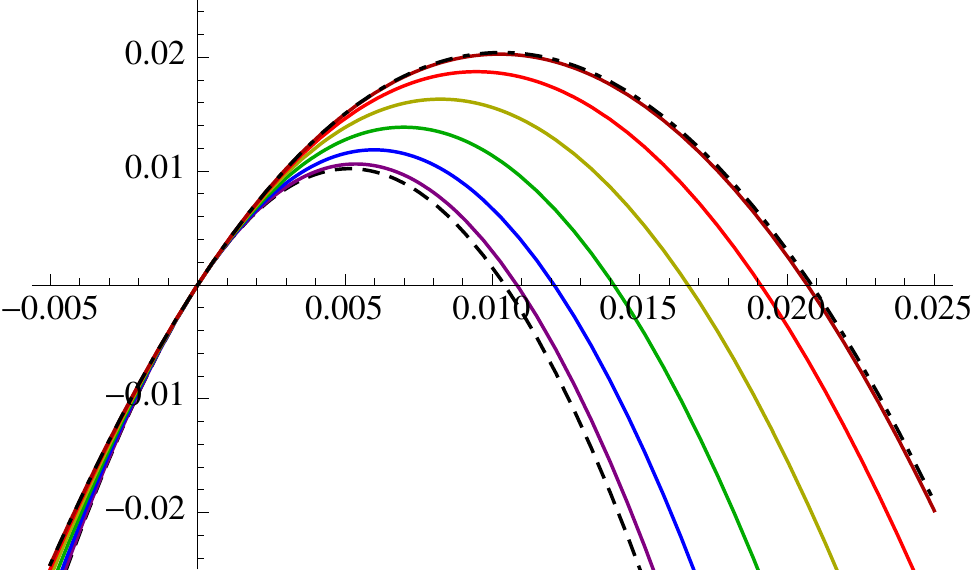}
\put(89,33){$\tilde{a}_{\textsc{uv}} \hat{\lambda}$}
\put(30,57){$\tilde{a}_{\textsc{uv}} \hat{R}$}
\end{overpic}
\caption{\hphantom{A}}
\label{fig:RvslambdaQFT0p9LIR0p5small}
\end{subfigure}%
\caption{\textbf{(a)-(b):} $\hat{R}$ vs.~$\hat{\lambda}$ for backreaction due to QFT$_1$ and cutoff $\hat{\Lambda}^2=4 |\hat{\lambda}|$. The various coloured plots correspond to different choices of $m \Lambda^{-1}$. For comparison, $\hat{R}$ vs.~$\hat{\lambda}$ for backreaction due to the UV CFT (dashed) and the IR CFT (dot-dashed) is shown, again with cutoff $\hat{\Lambda}^2=4 |\hat{\lambda}|$. Here the anomaly coefficients of the two CFTs are related as $\tilde{a}_\textsc{uv} = 2 \tilde{a}_{\textsc{ir}}$. Plot (b) is a zoomed-in version of the plot in (a). As $m \Lambda^{-1}$ is increased the solution for $\hat{R}$ interpolates between the values obtained for backreaction from the UV and IR CFTs.}
\label{fig:RvslambdaQFT0p9LIR0p5}
\end{figure}

\begin{figure}[t]
\centering
\begin{subfigure}{.5\textwidth}
 \centering
   \begin{overpic}
[width=1.0\textwidth]{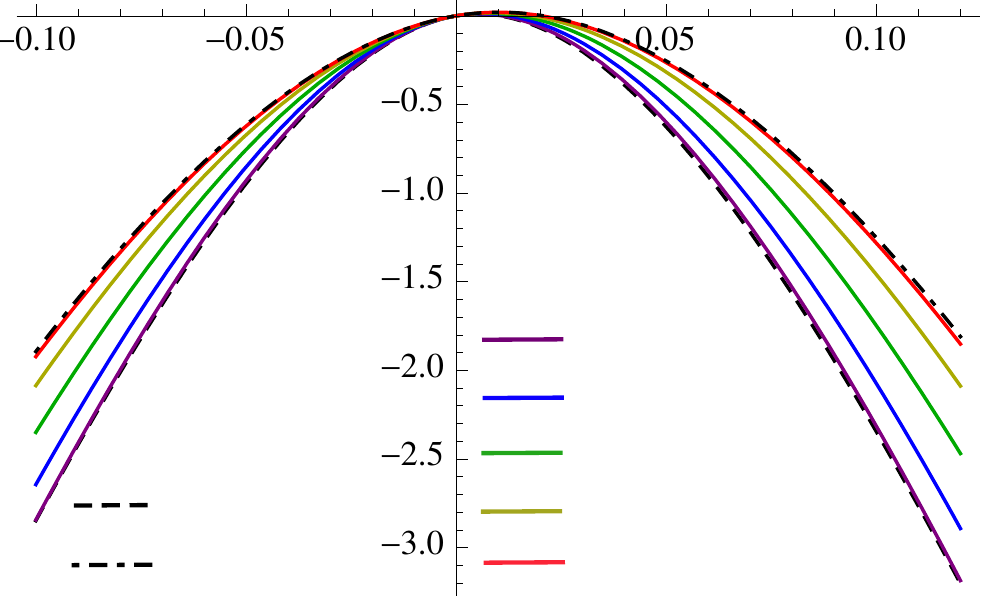}
\put(89,62){$\tilde{a}_{\textsc{uv}}  \hat{\lambda}$}
\put(43,65){$\tilde{a}_{\textsc{uv}} \hat{R}$}
\put(17.2,8.5){{\small UV CFT}}
\put(17.2,2.5){{\small IR CFT}}
\put(59,26){${\scriptstyle m \L^{-1} = 0.61}$}
\put(59,20){${\scriptstyle m \L^{-1} = 1.00}$}
\put(59,14.25){${\scriptstyle m \L^{-1} = 1.34}$}
\put(59,8.5){${\scriptstyle m \L^{-1} = 1.64}$}
\put(59,3){${\scriptstyle m \L^{-1} = 1.92}$}
\end{overpic}
\caption{\hphantom{A}}
\label{fig:RvslambdaQFT1p4LIR0p5large}
\end{subfigure}%
\begin{subfigure}{.5\textwidth}
 \centering
   \begin{overpic}
[width=1.0\textwidth]{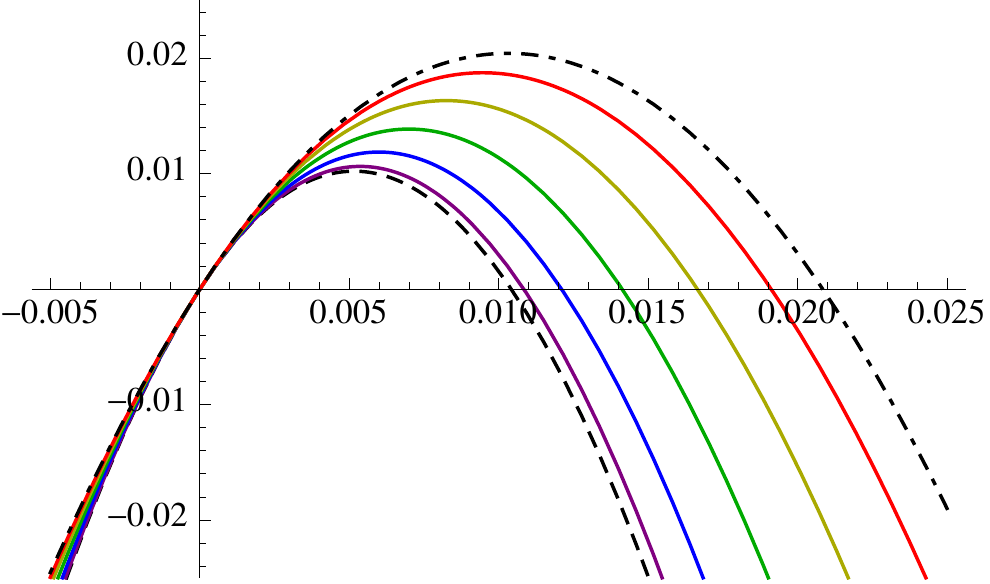}
\put(89,33){$\tilde{a}_{\textsc{uv}} \hat{\lambda}$}
\put(30,57){$\tilde{a}_{\textsc{uv}} \hat{R}$}
\end{overpic}
\caption{\hphantom{A}}
\label{fig:RvslambdaQFT1p4LIR0p5small}
\end{subfigure}%
\caption{\textbf{(a)-(b):} $\hat{R}$ vs.~$\hat{\lambda}$ for backreaction due to QFT$_2$ and cutoff $\hat{\Lambda}^2=4 |\hat{\lambda}|$. The various coloured plots correspond to different choices of $m \Lambda^{-1}$. For comparison, $\hat{R}$ vs.~$\hat{\lambda}$ for backreaction due to the UV CFT (dashed) and the IR CFT (dot-dashed) is shown, again with cutoff $\hat{\Lambda}^2=4 |\hat{\lambda}|$. Here the anomaly coefficients of the two CFTs are related as $\tilde{a}_\textsc{uv} = 2 \tilde{a}_{\textsc{ir}}$. Plot (b) is a zoomed-in version of the plot in (a). As $m \Lambda^{-1}$ is increased the solution for $\hat{R}$ interpolates between the values obtained for backreaction from the UV and IR CFTs.}
\label{fig:RvslambdaQFT1p4LIR0p5}
\end{figure}
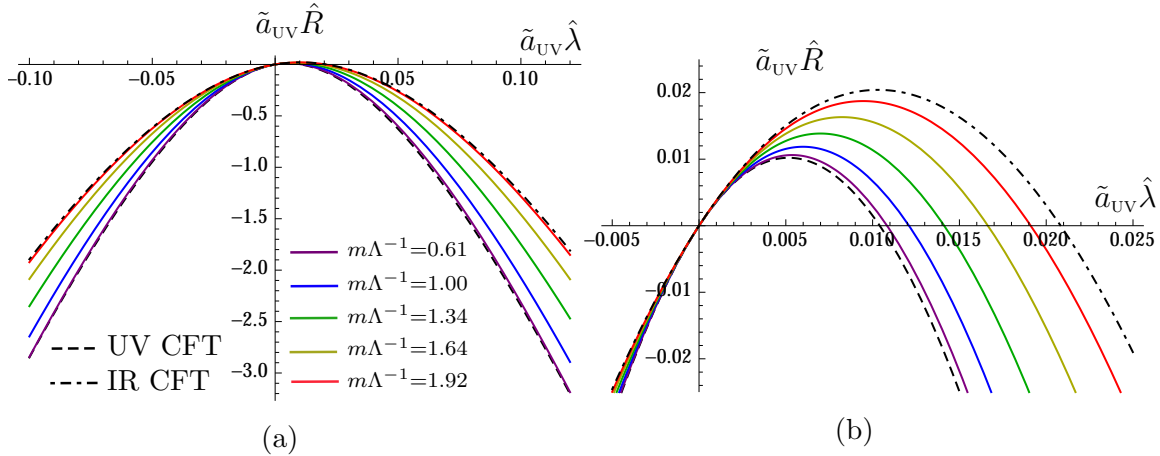

This view is confirmed by the full numerical results. In figures \ref{fig:RvslambdaQFT0p9LIR0p5} and \ref{fig:RvslambdaQFT1p4LIR0p5} we plot $\hat{R}$ vs.~$\hat{\lambda}$ for backreacted solutions with UV cutoff $\hat{\Lambda}^2 = 4 |\hat{\lambda}|$. Figure \ref{fig:RvslambdaQFT0p9LIR0p5} contains results for backreaction due to QFT$_1$ whereas figure \ref{fig:RvslambdaQFT1p4LIR0p5} contains results from backreacting QFT$_2$. The model parameters of QFT$_{1,2}$ were recorded below equation \eqref{eq:modelparameters}. In both figures \ref{fig:RvslambdaQFT0p9LIR0p5} and \ref{fig:RvslambdaQFT1p4LIR0p5} we display $\hat{R}$ vs.~$\hat{\lambda}$ for various values of $m \Lambda^{-1}$, as well as the results from backreacting the UV and IR CFTs governing the dynamics of the UV and IR fixed points. As can be observed from \ref{fig:RvslambdaQFT0p9LIR0p5} and \ref{fig:RvslambdaQFT1p4LIR0p5}, incrementally increasing $m \Lambda^{-1}$ causes the value for $\hat{R}$ to interpolate (monotonically) between the value obtained for backreaction due to the UV CFT and that for backreaction due to the IR CFT, as expected.

\section*{Note added}\label{ACKNOWL}
\addcontentsline{toc}{section}{Note added}

While we were finalising this paper, a paper appeared,   \cite{2003.05501} which computes the backreaction due to $\mathcal{N}=4$ supersymmetric Yang-Mills (SYM). In this work we will go beyond conformal field theories like $\mathcal{N}=4$ SYM to address also generic holographic RG flows.
Reference \cite{2003.05501},  however, addresses perturbations around de Sitter in a complete manner.

\vskip 0.5cm

\section*{Acknowledgements}\label{ACKNOWL}
\addcontentsline{toc}{section}{Acknowledgements}

We would like to thank D. Anninos, V. Gorbenko, E. Mottola, J. Serreau, K. Skenderis, N. Tsamis, V. Vennin and R. Woodard for discussions.

\noindent This work was supported in part  by the Advanced ERC grant SM-grav, No 669288. LW also acknowledges support from the European Research Council under the European Union's Horizon 2020 research and innovation programme (grant agreement No 758792, project GEODESI). JKG acknowledges the postdoctoral program at ICTS for funding support through the Department of Atomic Energy, Government of India, under project no.~12-R\&D-TFR-5.10-1100.

\appendix
\renewcommand{\theequation}{\thesection.\arabic{equation}}
\addcontentsline{toc}{section}{Appendix\label{app}}
\section*{Appendix}

\section{$f(R)$-gravity as an Einstein-scalar theory}
\label{app:constRfromFofR}
Consider a $f(R)$ theory of gravity, i.e.~a gravitational theory with a Lagrangian density that depends on the metric exclusively through the scalar curvature $R$, i.e.
\be
\label{eq:FRaction} S = \int d^d x \, \sqrt{|g|} \, f(R) \, .
\ee
The equations of motion that emerge from this action are
\begin{align}
f_R(R)R_{\m\n}-\nabla _{\m}\nabla_{\n} f_{R}+g_{\m\n}\square f_R(R)-{1\over 2}f(R)g_{\m\n}=0 \, , \quad f_R\equiv \frac{\partial f}{\partial R} \, .
\label{eq:FReqmot}
\end{align}
For constant-curvature solutions this can be simplified to give
\begin{align}
\label{eq:Rsolcondition} d f - 2 R f_R =0 \, .
\end{align}

We will now rewrite the theory in \eqref{eq:FRaction} as a scalar-tensor theory in Einstein frame. To this end we introduce a new dynamical field denoted by $Q$ to rewrite the action \eqref{eq:FRaction} as:
\be
\label{eq:FRaction2} S = \int d^d x \, \sqrt{|g|} \, \Big(f_Q(Q) \big(R-Q\big) + f(Q) \Big) \, .
\ee
By varying \eqref{eq:FRaction2} with respect to $Q$ one obtains $f_{QQ}(R-Q)=0$. We take $f_{QQ} \neq 0$ and hence the e.o.m.~requires $Q=R$, such that \eqref{eq:FRaction2} is classically equivalent to \eqref{eq:FRaction}.

To be able to keep track of dimensions more easily, let us define a function $F(R)$ which is related to $f(R)$ as
\begin{align}
F(R) = 2 \kappa \, f(R) \, ,
\end{align}
where $\kappa$ is a parameter with energy dimensions $[\kappa] =-(d-2)$. Then $F(R)$ has the same energy dimensions as $R$ and $F_R(R)$ is dimensionless. In practice, it will often be convenient to choose
\begin{align}
\kappa = \frac{1}{2 f_R(R_0)} \, ,
\end{align}
for some reference value $R_0$. Expression \eqref{eq:FRaction2} hence becomes:
\be
\label{eq:FRaction2b} S = \frac{1}{2 \kappa} \int d^d x \, \sqrt{|g|} \, \Big(F_Q(Q) \big(R-Q\big) + F(Q) \Big) \, .
\ee
Using \eqref{eq:FRaction2b} as a starting point we now perform a conformal transformation to arrive at a scalar-tensor theory in Einstein frame. In particular, we define
\begin{align}
\tilde{g}_{\mu \nu} &\equiv \Omega^2 g_{\mu \nu} \, , \\
\Omega^{d-2} &\equiv F_Q(Q) \, , \\
\label{eq:FRphi} F_Q(Q) &\equiv \exp \Big( \sqrt{\frac{d-2}{d-1}} \, \sqrt{\kappa} \, \phi \Big) \, .
\end{align}
One obtains
\begin{align}
\label{eq:FRaction3} S = \int d^d x \, \sqrt{|\tilde{g}|} \, \left[ \frac{1}{2 \kappa} \tilde{R} - \frac{1}{2} \tilde{g}^{\mu \nu} \partial_{\mu} \phi \partial_{\nu} \phi - V(\phi) \right] \, ,
\end{align}
with an implicit expression for the potential $V(\varphi)$ given by
\begin{align}
\label{eq:VQ} V(\phi) = \frac{1}{2 \kappa} \left(\frac{F_Q(Q) Q - F(Q)}{F_Q(Q)^{\frac{d}{d-2}}} \right) \, .
\end{align}
While it was useful to have introduced the auxiliary field $Q$ for the above analysis, we shall suppress it in the following by setting $Q=R$. Hence, from now on we will write
\begin{align}
\label{eq:VR} V(\phi) = \frac{1}{2 \kappa} \left(\frac{F_R R - F}{F_R^{\frac{d}{d-2}}} \right) \, , \quad \textrm{with} \quad F_R(\phi) = \exp \Big( \sqrt{\frac{d-2}{d-1}} \, \sqrt{\kappa} \, \phi \Big) \, .
\end{align}

We can now show that the solutions to \eqref{eq:Rsolcondition} correspond to extremal points of the potential $V(\phi)$. The necessary condition for an extremum of $V(\phi)$ is:
\begin{align}
\frac{d V}{d \phi} &= 0  \, .
\end{align}
With the help of \eqref{eq:VR} this can be rephrased as an equation for $R$ as follows:
\begin{align}
\label{eq:extremeumcond} 0= \frac{d V}{d \phi} &= \frac{dF_R}{d \phi} \frac{dR}{dF_R} \frac{dV}{dR} = \frac{1}{2(d-2) \, \sqrt{\kappa}} \, \sqrt{\frac{d-2}{d-1}} \, \frac{d F - 2 R F_R}{F_R^{\frac{d}{d-2}}} \, .
\end{align}
This equation is satisfied if
\begin{align}
d F - 2 R F_R =0 \, ,
\end{align}
which we identify with the e.o.m.~given in \eqref{eq:Rsolcondition}. Alternatively, \eqref{eq:extremeumcond} can be satisfied for
\begin{align}
\label{eq:Rsolcondition2} |F_R| \rightarrow \infty \ \ \textrm{with} \ \ (d F - 2 R F_R) \ \ \textrm{finite} \, .
\end{align}
The latter condition \eqref{eq:Rsolcondition2} shall not be relevant in practice as $|F_R| \rightarrow \infty$ is only attained at the boundary of field space for $\phi \rightarrow \infty$, as can be seen from \eqref{eq:FRphi}.

Another useful quantity will be the curvature of the potential at an extremum. This can be computed as
\begin{align}
\label{eq:VRR} \left. \frac{d^2 V}{d \phi^2} \right|_{\textrm{ext}} &= \left. \frac{dF_R}{d \phi} \frac{dR}{dF_R} \frac{d}{dR} \frac{dV}{d \phi} \right|_{\textrm{ext}} = \frac{1}{2(d-1)}  \, \frac{(d-2) F_R -2 R F_{RR}}{F_R^{\frac{2}{d-2}} F_{RR}} \, ,
\end{align}
where we have used that at an extremum $dF=2 R F_R$.

\section{Relations from thermodynamics}
\label{app:thermo}

Here we  derive a relation between $W$, $T$ and $U$ from a thermodynamic identity relating an entanglement entropy for a theory on dS$_d$ and the free energy for a theory on $S^d$. This is possible as solutions for $W$, $S$, $T$ and $U$ are insensitive to the signature of the metric, i.e.~we find the same solutions for a theory on dS$_d$ and $S^d$ for identical UV data $R, \f_-^{\L}$.

First, consider the (static) entanglement entropy across a spherical surface for the field theory on dS$_d$. We showed in our earlier work \cite{F} that, using the Ryu-Takayanagi prescription, this can be written as
\begin{align}
\label{eq:SEETU} S_{\textrm{EE}} = M^{d-1} \tilde{\Omega}_d \big[T_\L^{-\frac{d}{2} + 1} U_\L \big] \, ,
\end{align}
with
\begin{align}
\tilde{\Omega}_d = \frac{2 \pi^{\frac{d+1}{2}}}{\Gamma(\tfrac{d+1}{2})} \, d^{\frac{d}{2}} (d-1)^{\frac{d}{2}} \, .
\end{align}
Now consider a theory on $S^d$ instead. The corresponding free energy has been calculated in \cite{F} and can be written as
\begin{align}
\label{eq:FTWTU} F = - M^{d-1} \tilde{\Omega}_d \big[T_\L^{-\frac{d}{2}} W_\L + T_\L^{-\frac{d}{2} + 1} U_\L \big] \, .
\end{align}
In \cite{F} we also confirmed that, identifying the entanglement entropy $S_{\textrm{EE}}$ with a thermal entropy, thermodynamics of de Sitter space implies that
\begin{align}
\label{eq:thermorelRapp} S_{\textrm{EE}} = - \Big( \frac{2}{d} \mathcal{R}_{\L} \frac{\partial}{\partial \mathcal{R}_{\L}} + 1 \Big) F \, ,
\end{align}
where $\mathcal{R}_{\L}  = R \, |\f_-^{\L}|^{-2 / \Delta_-}$. For constant $\f_-^{\L}$ and $\Lambda$ this can be rewritten as
\begin{align}
\label{eq:thermorelT} S_{\textrm{EE}} = - \Big( \frac{2}{d} T_\L \frac{\partial}{\partial T_\L} + 1 \Big) F \, .
\end{align}
Inserting \eqref{eq:SEETU} and \eqref{eq:FTWTU} into \eqref{eq:thermorelT} and rearranging one finds
\begin{align}
\frac{\partial W_\L}{\partial T_\L} = \left( \frac{d-2}{2} \, U_\L - T_\L \frac{\partial U_\L}{\partial T_\L} \right) \, .
\end{align}
This will be helpful for simplifying expressions in the main text.

Now we restrict attention to $d=4$. Using the near-boundary expansions in section \ref{app:nearboundary} we can identify the universal contributions to the entanglement entropy and free energy, i.e.~the terms without any explicit appearance of $\Lambda$. In These are given by:
\begin{align}
S_{\textrm{EE}}^{\textrm{univ}} (\mathcal{R}_\L)& = \hphantom{-} (M \ell)^{3} \tilde{\Omega}_4 \ \ \mathcal{R}_\L^{-1} \, B(\mathcal{R}_\L) \, , \\
\nonumber F^{\textrm{univ}} (\mathcal{R}_\L) &= - (M \ell)^{3} \tilde{\Omega}_4 \, \Big[\mathcal{R}_\L^{-2} \, C(\mathcal{R}_\L) + \mathcal{R}_\L^{-1} \ B(\mathcal{R}_\L) - \frac{1}{192} \Big] \, .
\end{align}
Then \eqref{eq:thermorelRapp} implies
\begin{align}
\label{eq:CBidapp} C'(\mathcal{R}) = B(\mathcal{R}) - \mathcal{R} B'(\mathcal{R}) + \frac{\mathcal{R}}{96} \, .
\end{align}
where we have also set $\Lambda \rightarrow \infty$. This can be verified numerically.

Now consider the renormalized entanglement entropy and free energy for $d=4$. These are given by
\begin{align}
S_{\textrm{EE}}^{\textrm{ren}} ( \tilde{B}_{ct} , \tilde{\mathcal{A}}_{ct} \, | \, \mathcal{R})& = \hphantom{-} (M \ell)^{3} \tilde{\Omega}_4 \, \Big[ \mathcal{R}^{-1} \big( B(\mathcal{R}) - \tilde{B}_{ct} \big) - \tilde{\mathcal{A}}_{ct}  \Big] \, , \\
\nonumber F^{\textrm{ren}} (C_{ct} , B_{ct}, \mathcal{A}_{ct} \, | \, \mathcal{R}) &= - (M \ell)^{3} \tilde{\Omega}_4 \, \Big[\mathcal{R}^{-2} \big( C(\mathcal{R}) -C_{ct} \big) + \mathcal{R}^{-1} \big( B(\mathcal{R}) -B_{ct} \big) \\
& \hphantom{AAAAAAAAAAAAAAAAAAAAAAAll} - \frac{1}{192} - \mathcal{A}_{ct} \Big] \, .
\end{align}
The identity \eqref{eq:CBidapp} then implies that
\begin{align}
S_{\textrm{EE}}^{\textrm{ren}} ( \tfrac{1}{2} {B}_{ct} , {\mathcal{A}}_{ct} \, | \, \mathcal{R}) = - \left(\frac{1}{2} \mathcal{R} \, \frac{\partial}{\partial \mathcal{R}} + 1 \right) \, F^{\textrm{ren}} (C_{ct} , B_{ct}, \mathcal{A}_{ct} \, | \, \mathcal{R}) \, .
\end{align}

\section{Near-boundary expansions}
\label{app:nearboundary}

Here we  show that in the vicinity of a UV fixed point at $\f=0$ we can write the functions $W(\f)$, $S(\f)$ and $U(\f)$ as a double expansion in $\f$ and $T(\f)$. In particular, we  show that we solve for $W(\f)$, $S(\f)$ and $U(\f)$ self-consistently with the ansatz
\begin{align}
\label{eq:WansatznearUV} W(\f) &= \sum_{n=0}^{\infty} W_n(\f) \, \big(\ell^2 T(\f) \big)^n \, , \\
\label{eq:SansatznearUV} S(\f) &= \sum_{n=0}^{\infty} S_n(\f) \, \big(\ell^2 T(\f) \big)^n \, , \\
\label{eq:UansatznearUV} U(\f) &= \sum_{n=0}^{\infty} U_n(\f) \, \big(\ell^2 T(\f) \big)^n \, ,
\end{align}
with $W_n(\f)$, $S_n(\f)$ and $U_n(\f)$ power series in $\f$.

Consider the following subset of equations that the functions $W(\f)$, $S(\f)$ and $U(\f)$ have to satisfy:
\begin{align}
\label{eq:EOMapp1} S^2 - S W' + \frac{2}{d} T &= 0 \, , \\
\label{eq:EOMapp2} \frac{d}{2(d-1)} W^2 - S^2 - 2 T + 2 V &=0 \, , \\
\label{eq:EOMapp3} S U' - \frac{(d-2)}{2(d-1)} WU &= - \frac{2}{d} \, .
\end{align}
In the vicinity of a UV fixed point we can also write $V(\f)$ without loss of generality as
\begin{align}
V(\f) = - \frac{d(d-1)}{\ell^2} - \frac{\Delta_-(d-\Delta_-)}{2 \ell} \, \f^2 + \mathcal{O}(\f^3) \, ,
\end{align}
with $0 < \Delta_- < \tfrac{d}{2}$ at a maximum and $\Delta_- < 0$ at a minimum. For a symmetric potential the subleading term is $\mathcal{O}(\f^4)$ instead.

In the following we take $\f=0$ to be a maximum of $V$. Inserting the ans\"atze \eqref{eq:WansatznearUV}--\eqref{eq:UansatznearUV} into \eqref{eq:EOMapp1}--\eqref{eq:EOMapp3} we can then solve order by order in $\ell^2 T$. An important identity in this context is
\begin{align}
T^{\, \prime} = \frac{\partial T}{\partial \f} = \frac{\partial u}{\partial \f} \, \frac{\partial (R \, e^{-2A})}{\partial u} = \frac{1}{S} (-2 \dot A) \, R \, e^{-2A} =  \frac{W T}{(d-1) S} \, .
\end{align}

\noindent \textbf{Order} $(\ell^2 T)^0$: At order $(\ell^2 T)^0$ we then obtain the equations
\begin{align}
S_0(S_0 - W_0') &= 0 \, , \\
\frac{d}{2(d-1)} W_0^2 - S_0^2 + 2 V &=0 \, , \\
\label{eq:U0eqapp} S_0 U_0' - \frac{(d-2)}{2(d-1)} W_0 U_0 &= - \frac{2}{d} \, ,
\end{align}
which can be solved to give\footnote{Here we do not consider solutions with $S_0=0$.}
\begin{align}
W_0 & = \frac{1}{\ell} \left[2(d-1) + \frac{\Delta_-}{2} \f^2 + \mathcal{O}(\f^3) + C \, |\f|^{\frac{d}{\Delta_-}} \Big(1 + \mathcal{O}(\f) + \mathcal{O}\big(C|\f|^{\frac{d}{\Delta_-}} \big) \Big) \right] \, , \\
S_0 &= W_0' \, , \\
\nonumber U_0 &= \ell \bigg[\frac{2}{d(d-2)} + \frac{\Delta_-}{2d(d-1)(2 \Delta_- +2 -d)} \, \f^2 + \mathcal{O}\big(C^2 |\f|^{\frac{2d}{\Delta_-}-2} \big) \\
\label{eq:U0solapp} & \hphantom{AAA} + B \, |\f|^{\frac{d-2}{\Delta_-}} \Big(1 + \mathcal{O}(\f) + \mathcal{O}\big(C|\f|^{\frac{d}{\Delta_-}-2}\big) + \mathcal{O}\big(B |\f|^{\frac{d-2}{\Delta_-}}\big) \Big) \bigg] \, ,
\end{align}
with $C$ and $B$ are integration constants.

From this result we can also calculate the leading $\f$-dependence of $T(\f)$. Following the analysis in appendix D.1 of \cite{R} one finds
\begin{align}
T(\f) = \mathcal{R} \, |\f|^{-2 / \Delta_-} + \ldots \, .
\end{align}

\noindent \textbf{Order} $(\ell^2 T)^1$: The equations at order $(\ell^2 T)^1$ take the form:
\begin{align}
\label{eq:OT1eq1} S_1 W_0' - W_0' W_1' - \frac{1}{(d-1)} W_0 W_1 &= - \frac{2}{d} \, , \\
\frac{d}{(d-1)} W_0 W_1 - 2 S_1 W_0' &= 2 \, \\
\label{eq:OT1eq3} U_1' W_0' + \frac{(d-4)}{2(d-1)} U_1 W_0 - \frac{(d-2)}{2(d-1)} U_0 W_1 + S_1 U_0' &= 0 \, ,
\end{align}
where we have used that $S_0 = W_0'$. From the first two equations we find
\begin{align}
S_1 = \frac{d}{d-2} W_1' \, .
\end{align}
With this we can eliminate $S_1$ in \eqref{eq:OT1eq1} to arrive at the following equation for $W_1$:
\begin{align}
\label{eq:W1eqapp} W_0' W_1' - \frac{(d-2)}{2(d-1)} W_0 W_1 &= - \frac{(d-2)}{d} \, .
\end{align}
Note that the LHS takes the same form as equation \eqref{eq:U0eqapp} for $U_0$. Thus we can re-use the result for $U_0$ given in \eqref{eq:U0solapp} to write:
\begin{align}
\nonumber W_1 &= \frac{1}{\ell} \bigg[\frac{1}{d} -\frac{(d-2) \Delta_-}{4d(d-1)(d-2-2\Delta_-)} \f^2 + \mathcal{O}(\f^3) + \mathcal{O}\big(C^2 |\f|^{\frac{2d}{\Delta_-}-2} \big) \bigg] \, .
\end{align}
Note that solving for $W_1$ naively introduces an additional integration constant. However, one can check that this combines with the constant $C$ introduced in $W_0$ and we do not need to include it separately. This is just a manifestation of the fact that $W$ should only contain one integration constant, and by  including $C$ in $W_0$ we have already accounted for this.

Given an explicit expression for $W_1$ and hence also for $S_1$ we can now use \eqref{eq:OT1eq3} to solve for $U_1$. Here we need to distinguish between the cases $d=4$ and $d \neq 4$. As we are mainly interested in the former case, we only present the results for $d=4$. One finds:
\begin{align}
U_1 \underset{d=4}{=} \ell \bigg[ \frac{1}{48 \Delta_-} \, \ln \f + \mathcal{O}(\f) + \mathcal{O}\big(C |\f|^{\frac{4}{\Delta_-}-2} \big)  + \mathcal{O}\big(B |\f|^{\frac{2}{\Delta_-}}\big) \bigg] \, .
\end{align}
Here we again neglected to explicitly include an integration constant as it combines with $B$ contained in $U_0$.

\noindent \textbf{Order} $(\ell^2 T)^2$: In this case we shall restrict attention to $W_2$, as this term contributes to the renormalized on-shell action in $d=4$ dimensions. However, for $d=4$ one can show that the equation for $W_2$ reduces to
\begin{align}
W_2' =0 \, ,
\end{align}
i.e.~$W_2$ is purely an integration constant. This is not surprising: For $d=4$ a term $T^2 \sim \mathcal{R}^2 \, |\f|^{4 / \Delta_-}$ and the term $C \, |\f|^{4 / \Delta_-}$ containing the integration constant $C$ come with the same power of $\f$. Hence, in a power expansion in $\f$, we cannot really distinguish between a contribution to $C$ and a constant term in $W_2$.

However, we can make the following observation. When dialing the source of the perturbing operator $\f_- \rightarrow 0$ the QFT reduces to a CFT and we expect to recover the expressions for $W$ and $U$ for a CFT, derived in section \ref{sec:WU-CFT}. There we found that in $d=4$ the function $W$ has the following expansion in powers of $\ell^2 T$:
\begin{align}
\label{eq:WCFT4dapp} \ell W = 6 + \frac{\ell^2 T}{4} - \frac{\ell^4 T^2}{192} + \mathcal{O}(\ell^6 T^3) \, ,
\end{align}

We combine this with another observation. In the dual field theory the parameter $C$ has the interpretation as the vev of the operator perturbing the UV QFT. For a CFT no such perturbing operator exists and hence there cannot be a finite value for the vev. If we wish to maintain the interpretation of $C$ as a vev, we should make sure that $C=0$ for a CFT.

The choice $C=0$ for $\f_- = 0$ then fixes $W_2$ uniquely. Consistency with the CFT result \eqref{eq:WCFT4dapp} for $W$ then implies that
\begin{align}
W_2 = - \frac{1}{192 \, \ell} \, .
\end{align}
This is the choice we shall use throughout this work in $d=4$.

\noindent \textbf{Summary}: Here we collect our results from this section, specifying to $d=4$. Writing
\begin{align}
\f_\L = \f_-^\L \, \Lambda^{- \Delta_-} \, , \quad \textrm{and} \quad T(\f_\L) = \ell^{-2} \, R \, \Lambda^{-2}
\end{align}
our findings from this section imply:
\begin{align}
W(\f_\L) &=\frac{1}{\ell} \Big[ \, 6 + \frac{\Delta_-}{2} \, |\f_-^{\L}|^2 \, \Lambda^{-2 \Delta_-} + \frac{R \Lambda^{-2}}{4} - \frac{R^2 \Lambda^{-4}}{192} + C(\mathcal{R}_\L) \, |\f_-^\L|^{\frac{4}{\Delta_-}} \Lambda^{-4} + \ldots \Big] \, , \\
U(\f_\L) &= \ell \Big[ \, \frac{1}{4} + B(\mathcal{R}_\L) \, |\f_-^{\L}|^{\frac{2}{\Delta_-}} \, \Lambda^{-2} + \frac{R \Lambda^{-2}}{48 \Delta_-} \, \ln \big(\f_-^\L \, \Lambda^{- \Delta_-} \big)+ \ldots \Big] \, .
\end{align}

\section{Dictionary}
\label{app:dictionary}
In this section we relate expression appearing in sections \ref{sec:strategy}, \ref{sec:ResultsQFTren} and \ref{sec:ResultsQFTcutoff} to expressions obtained from the holographic dual and hence employing notation exclusively used in section \ref{sec:holo}.

\vspace{0.3cm}

\noindent \textbf{CFT with a UV cutoff.} Here we consider the UV CFT, i.e.~the CFT associated with the UV fixed point of a RG flow. Control parameters on the field theory side are the UV cutoff $\Lambda$ in units of energy and the scalar curvature $R$. These are related to quantities in the holographic dual as
\begin{align}
\label{dict1} \Lambda \equiv \frac{e^{A(u_\L)}}{\ell_{\textsc{uv}}} \, , \qquad R \Lambda^{-2}= \ell_{\textsc{uv}}^2 T(u_\L) \, ,
\end{align}
where $\ell_{\textsc{uv}}$ is defined through \eqref{elluv}, $u_\L$ is the location of the cutoff-surface in the holographic coordinate and $T$ was defined in \eqref{eq:defTc}. The effective action of the QFT is then written in terms of a function $\mathcal{F} (R \Lambda^{-2})$, which is given by
\begin{align}
\nonumber \tilde{a}_{\textsc{uv}} \Lambda^d \mathcal{F} (R \Lambda^{-2}) &= M^{d-1} \, e^{dA(u_\L)} \, \big[ W(T(u_\L)) + T(u_\L) U(T(u_\L)) \big] \\
\label{dict3} &= (M \ell_{\textsc{uv}})^{d-1} \, \Lambda^d \, \ell_{\textsc{uv}} \big[ W(T(u_\L)) + T(u_\L) U(T(u_\L)) \big] \, ,
\end{align}
where $W(T)$ and $U(T)$ are given in \eqref{eq:WCFT} and (\ref{eq:UCFTdS}, \ref{eq:UCFTAdS}). We then introduce the functions $\mathcal{W}(R \Lambda^{-2})$ and $\mathcal{U}(R \Lambda^{-2})$ as
\begin{align}
\label{dict2} \mathcal{W}(R \Lambda^{-2}) \equiv \ell_{\textsc{uv}} W(T(u_\L)) \, , \qquad \mathcal{U}(R \Lambda^{-2}) \equiv \ell_{\textsc{uv}}^{-1} U(T(u_\L)) \, ,
\end{align}
In $d=4$ we also have
\begin{align}
\label{dict4} \tilde{a}_{\textsc{uv}} \underset{d=4}{=} (M \ell_{\textsc{uv}})^3 \, ,
\end{align}
and hence
\begin{align}
\label{dict5} \mathcal{F} (R \Lambda^{-2}) \underset{d=4}{=} \mathcal{W}(R \Lambda^{-2}) +  R \Lambda^{-2} \, \mathcal{U} (R \Lambda^{-2}) \, .
\end{align}
The corresponding results for the IR CFT can be obtained from the above by replacing every subscript `UV' by `IR'.

\vspace{0.3cm}

\noindent \textbf{QFT with a UV cutoff.} We now have three control parameters, the UV energy cutoff $\Lambda$, the scalar curvature $R$ and the scale associated with the UV coupling $m$. These are now given by
\begin{align}
\label{dict6} \Lambda \equiv & \frac{e^{A}(u_\L)}{\tilde{\ell}} \, , \qquad R \Lambda^{-2} = \tilde{\ell}^2 \, T_\L \, , \qquad m \Lambda^{-1} \equiv |\f_\L|^{1/ \Delta_-} \, , \\
\label{dict7} &\textrm{with} \quad \tilde{\ell}^{-2} \equiv -\frac{V(\f_\L)}{d(d-1)} \, .
\end{align}
Here $T_\L$ (and later $W_\L$ and $U_\L$) are the functions $T$ (and $W$, $U$), evaluated at the UV cutoff $\f_\L$, for a family of RG flow solutions in a potential $V$, as described in section \ref{sec:onshell}. There is a maximal value for $m \Lambda^{-1}$, which is given by
\begin{align}
m \Lambda^{-1} \big|_{\textrm{max}} = |\f_{\textsc{ir}}|^{1/ \Delta_-} \, .
\end{align}
The effective action is written in terms of a function $\mathcal{F}(\tfrac{m}{\Lambda} , \tfrac{R}{\Lambda^2})$, which is given by
\begin{align}
\nonumber \tilde{a}_{\textsc{uv}} \Lambda^d \mathcal{F} \Big(\tfrac{m}{\Lambda} , \tfrac{R}{\Lambda^2} \Big) &= M^{d-1} \, e^{dA(u_\L)} \, \big[ W_\L(\f_\L, T_\L) + T_\L U_\L(\f_\L, T_\L) \big] \\
\label{dict8} &= (M \tilde{\ell})^{d-1} \, \Lambda^d \, \tilde{\ell}(\f_\L) \, \big[ W_\L(\f_\L, T_\L) + T_\L U_\L(\f_\L, T_\L) \big] \, .
\end{align}
The contribution to the equation for constant-curvature solutions due to the QFT is written with the help of a function $\mathcal{W} (\tfrac{m}{\Lambda}, \tfrac{R}{\Lambda^2} )$, which is given by
\begin{align}
\mathcal{W} \Big( \tfrac{m}{\Lambda}, \tfrac{R}{\Lambda^2} \Big) \equiv \tilde{\ell} \, W_\L (\f_\L, T_\L) \, .
\end{align}
Furthermore, for $d=4$ we also define the function
\begin{align}
\tilde{a} \Big( \tfrac{m}{\Lambda} \Big) \underset{d=4}{\equiv} \big( M \tilde{\ell}(\f_\L) \big)^3  \, .
\end{align}
As before, for $d=4$ we also have
\begin{align}
\label{dict9} \tilde{a}_{\textsc{uv}} \underset{d=4}{=} (M \ell_{\textsc{uv}})^3  \, .
\end{align}

\vspace{0.3cm}

\noindent \textbf{UV complete QFT.} In this case the free parameters on the field theory side are the scalar curvature $R$ and the scale $m$ associated with the UV coupling of the relevant operator $\mathcal{O}$. These are related to quantities in the holographic analysis as
\begin{align}
\label{dict10} m \equiv |\f_-|^{1 / \Delta_-} \, , \qquad R m^{-2} = \mathcal{R} \, .
\end{align}
In the following, we restrict to $d=4$. We again have
\begin{align}
\label{dict11} \tilde{a}_{\textsc{uv}} \underset{d=4}{=} (M \ell_{\textsc{uv}})^3 \, .
\end{align}
The (scheme-independent part of) effective action for the QFT is written in terms of a function $\mathcal{G}(R m^{-2})$, which is related to quantities on the gravity side as
\begin{align}
\mathcal{G}(R m^{-2}) \underset{d=4}{=} C(\mathcal{R}) + \mathcal{R} B(\mathcal{R}) - \tfrac{1}{96} \mathcal{R}^2 \Big( 1+ \log \big( \tfrac{\mathcal{R}}{48}\big) \Big) \, .
\end{align}
In the equation for constant-curvature solutions, the QFT contribution is proportional to the function $C(Rm^{-2})$. This is the same as $C(\mathcal{R})$ appearing above and has the interpretation as the vev of the operator $\mathcal{O}$ in units of its UV source:
\begin{align}
C(Rm^{-2}) \underset{d=4}{=} \tilde{a}_{\textsc{uv}}^{-1} \, \frac{4 - \Delta}{4} \, \langle \mathcal{O} \rangle \, m^{-\Delta} = (M \ell_{\textsc{uv}})^{-3} \, \frac{\Delta_-}{4} \, \langle \mathcal{O} \rangle \, |\f_-|^{- \frac{\Delta_+}{\Delta_-}} \, .
\end{align}

\section{Large and small curvature results}
\label{app:largesmallR}
In this appendix we present analytic expressions for the quantities $C(\mathcal{R})$ and $B(\mathcal{R})$ introduced in section \ref{sec:holoRGreview}, as expansions valid for large ($|\mathcal{R}| \rightarrow \infty$) and small ($|\mathcal{R}| \rightarrow 0$) values of the dimensionless curvature $\mathcal{R}$, respectively. The expressions are obtained using mainly analytical methods, together with the identity from thermodynamics \eqref{eq:CBidapp}. In one instance we need to employ an educated guess, which is confirmed by all numerical examples studied.

\subsubsection*{Analysis for $|\mathcal{R}| \rightarrow \infty$}
The derivation of $C(|\mathcal{R}| \rightarrow \infty)$ for the case $d=4$ has already been presented in \cite{R} and hence we will only collect the results and sketch the analysis. The computation of $B(|\mathcal{R}| \rightarrow \infty)$ has only been performed for $d=3$ and $\mathcal{R} >0$ in \cite{F}. Thus, here we will present the corresponding analysis for $d=4$ in some detail and give the results for both $\mathcal{R} >0$ and $\mathcal{R} <0$. The ansatz for $\f(u)$ and $A(u)$ for $\mathcal{R} \rightarrow \infty$ and the resulting expressions for $d=3$ can also be found in \cite{1604.06809}.

Consider holographic RG flows between a UV fixed point at a maximum of $V$ at $\f_\textsc{uv}=0$ and an IR end/turning point at $\f_0$. As was shown in \cite{R,F}, letting $\f_0$ approach the UV fixed point the corresponding value of $\mathcal{R}$ for the flow increases, so that in the limit $\f_0 \rightarrow \f_{\textsc{uv}}=0$ one finds $|\mathcal{R}| \rightarrow \infty$, with the precise relation $\f_0 \sim |\mathcal{R}|^{- \Delta_- / 2}$ in that limit.

Flows with $|\mathcal{R}| \rightarrow \infty$ can also occur for finite $|\f_0-\f_{\textsc{uv}}|$. However, for $R \geq0$ the existence of such a solution requires a carefully chosen potential and does not appear in `generic' potentials (see e.g.~\cite{R}). Hence we do not consider this further in this section. For $R <0$ such solutions exist generically for certain values of $\Delta_-$. This case will be discussed in the main text when the need arises, but here we neglect this possibility. Thus, in the following we focus on flows with $|\mathcal{R}| \rightarrow \infty$ that occur for $\f_0 \rightarrow \f_{\textsc{uv}}$.

As presented in \cite{R,F}, for $\f_0 \rightarrow \f_{\textsc{uv}}=0$ we can make the following ansatz for the scale factor $A(u)$ and the dilaton $\f(u)$. At leading order (order $\mathcal{O} \big( (\f_0)^0 \big)$) the scale factor $A(u)$ is just given by the scale factor associated with the UV fixed point, i.e.~the AdS scale factor recorded in \eqref{eq:AdSScaleFactor}. The next correction appears at order $\mathcal{O} \big( (\f_0)^2 \big)$. At leading order $\f(u)= \f_{\textsc{uv}}=0$ is constant, but its first non-trivial modification appears at order $\mathcal{O} \big(\f_0 \big)$. Thus, the ansatz for $A(u)$ and $\f(u)$ can be written as:
\begin{align}
\label{eq:largeRansatzA} A(u) &= A_0(u) +  \mathcal{O} \big( (\f_0)^2 \big) \, , \\
\f(u) &= \f_1(u) + \mathcal{O} \big( (\f_0)^2 \big) \, ,
\end{align}
where the subscript denotes the order in $\f_0$ of the respective term. Here $A_0(u)$ corresponds to the AdS scale factor recorded in \eqref{eq:AdSScaleFactor}. This ansatz is to be inserted into the equations of motion \eqref{eq:EOM4}--\eqref{eq:EOM6}, which can be written as
\begin{align}
\label{eq:EOM1} 2(d-1) \ddot{A} + \dot{\f}^2 + \frac{2}{d} e^{-2A} R &=0 \, , \\
\label{eq:EOM2} d(d-1) \dot{A}^2 - \frac{1}{2} \dot{\f}^2 + V - e^{-2A} R &=0 \, , \\
\label{eq:EOM3} \ddot{\f} +d \dot{A} \dot{\f} - V' &= 0 \, ,
\end{align}
where we have used the definitions \eqref{eq:defWc}--\eqref{eq:defTc}. We can then obtain an expression for $\f_1(u)$ by solving \eqref{eq:EOM3} for $\f(u)$ to order $\mathcal{O}(\f_0)$, as has been done in \cite{R} (for $d=4$) and \cite{F}  (for all $d$).

From the expression for $\f_1(u)$ we can then extract an expression for $C(\mathcal{R})$ at leading order in $|\mathcal{R}| \rightarrow \infty$. As  shown in \eqref{eq:vevdef}, $C$ corresponds to the vev $\langle \mathcal{O} \rangle$ in units of the source $\f_-$. Thus, to get an expression for $C$ we need to read off the values of $\langle \mathcal{O} \rangle$ and $\f_-$ corresponding to the solution $\f_1(u)$ and insert into \eqref{eq:vevdef}. As shown in \eqref{eq:phinearboundary}, the source $\f_-$ and the vev $\langle \mathcal{O} \rangle$ (rewritten in terms of $\f_+$ via \eqref{eq:jandvevdef}) appear as the leading and sub-leading coefficients in an expansion of $\f(u)$ in the vicinity of the boundary. Thus, we can read them off by expanding $\f_1(u)$ for $u \rightarrow - \infty$ and comparing with \eqref{eq:phinearboundary}. For $d=4$ this can be shown to give (see \cite{R} for details):
\begin{align}
\label{eq:ClargeRapp} C(\mathcal{R}) & \overset{|\mathcal{R}| \rightarrow \infty}{=} \frac{\Delta_-}{2} \, \frac{(3 - \Delta_-)(2 - \Delta_-)}{(1 - \Delta_-)} \, {\left( \frac{|\mathcal{R}|}{48} \right)}^{2 - \Delta_-} \Big(1 + \mathcal{O} \big( |\mathcal{R}|^{- \Delta_-/2} \big) \Big) \, .
\end{align}

We now turn to $B(\mathcal{R})$. The parameter $B$ appears in the near-boundary expansion of the function $U$ as (see \eqref{eq:U0sol} and \eqref{eq:U1sol}):
\begin{align}
\label{eq:Unearboundaryappendix} U(\f_\L) \overset{d=4}{=} \ell_{\textsc{uv}} \bigg[ \frac{1}{4} + \frac{\Delta_-}{48(\Delta_- -1)} \f_\L^2 + B \, |\f_\L|^{2 / \Delta_-} + \frac{\ell_{\textsc{uv}}^2 T(\f_\L)}{96} \log |\f_\L|^{2 / \Delta_-} + \ldots \bigg] \, .
\end{align}
To obtain an expression for $U$ at leading order in $|\mathcal{R}| \rightarrow \infty$ we can insert the ansatz \eqref{eq:largeRansatzA} into the expression \eqref{eq:UfromA}. This gives $U$ as a function of $u$. To later match with the near-boundary result in \eqref{eq:Unearboundaryappendix}, we compute $U(u_\L)$ where $\f(u_\L)=\f_\L$.
\begin{align}
U(u_\L) \overset{d=4}{=} - \frac{1}{2} e^{-2 A_0(u_\L)} \int_{u_0}^{u_\L} d \tilde{u} \, e^{2 A_0(\tilde{u})} \, \Big( 1 + \mathcal{O} \big((\f_0)^2 \big)\Big) \, .
\end{align}
In fact, we have already computed the leading term in the above, which is just the expression for $U$ for a CFT associated with the UV fixed point. The resulting expressions, albeit as a function of $T_\L = R e^{-2A(u_\L)}$, are given in \eqref{eq:UCFTdS} for $R >0$ and in \eqref{eq:UCFTAdS} for $R <0$. As $T_\L \rightarrow 0$ when approaching the boundary, we can expand these in powers of $T_\L$. The resulting expression valid for both $\mathcal{R}>0$ and $\mathcal{R}<0$ is given by
\begin{align}
\label{eq:Unearboundaryappendix2} U(\L) \overset{d=4}{=} \ell_{\textsc{uv}} \bigg[ \frac{1}{4} + \frac{ R \Lambda^{-2} }{96} \Big(1 + \log \frac{|R| \Lambda^{-2}}{48} \Big) + \mathcal{O} \big(R^2 \Lambda^{-4} \big) \bigg] \, \Big( 1 + \mathcal{O} \big((\f_0)^2 \big)\Big) ,
\end{align}
and where we wrote $\ell_{\textsc{uv}}^2 T_\L = \ell_{\textsc{uv}}^2 R e^{-2 A(u_\L)} = R \Lambda^{-2}$.

This is now to be compared with the near-boundary expression \eqref{eq:Unearboundaryappendix}. To this end we note that for $u_\L \rightarrow - \infty$ we have
\begin{align}
\f_\L = \f(u_\L) = \ell^{\Delta_-} \, \f_- \, e^{\Delta_- u_\L / \ell_{\textsc{uv}}} + \ldots = \ell^{\Delta_-} \, \f_- \, e^{- \Delta_- A(u_\L)} + \ldots = \f_- \Lambda^{- \Delta_-} + \ldots \, .
\end{align}
Using this, the expression in \eqref{eq:Unearboundaryappendix} can be written as
\begin{align}
\label{eq:Unearboundaryappendix3} U(\L) \overset{d=4}{=} \ell_{\textsc{uv}} \bigg[ \frac{1}{4} + \frac{\Delta_-}{48(\Delta_- -1)} \f_-^2 \Lambda^{-2 \Delta_-} + B \, |\f_-|^{2 / \Delta_-} \Lambda^{-2} + \frac{R \Lambda^{-2}}{96} \log \big( |\f_-|^{2 / \Delta_-} \Lambda^{-2} \Big) + \ldots \bigg] \, .
\end{align}
We can then read off $B$ by comparing the coefficients of the terms with an explicit power of $\Lambda^{-2}$ in \eqref{eq:Unearboundaryappendix2} and \eqref{eq:Unearboundaryappendix3}. This gives
\begin{align}
\label{app:BRlogRlargeR} B (\mathcal{R}) \overset{|\mathcal{R}| \rightarrow \infty}{=} \frac{\mathcal{R}}{96} \bigg( 1 + \log \frac{|\mathcal{R}|}{48} \bigg) + \ldots \, .
\end{align}
where the ellipsis denotes the subleading terms that will enter with an additional power of $(\f_0)^2 \sim |\mathcal{R}|^{- \Delta_-}$. Interestingly, we can determine these subleading terms easily by using the result for $C(\mathcal{R})$ in \eqref{eq:ClargeRapp} and the identity \eqref{eq:CBidapp} between $C(\mathcal{R})$ and $B(\mathcal{R})$. This fixes the subleading terms uniquely, so that we are left with
\begin{align}
B(\mathcal{R}) & \underset{\mathcal{R} >0}{\overset{|\mathcal{R}| \rightarrow \infty}{=}} \frac{\mathcal{R}}{96} \Big( 1+ \log \frac{\mathcal{R}}{48} \Big) + \frac{(3 - \Delta_-)(2-\Delta_-)^2}{96 (1 - \Delta_-)} \, {\left( \frac{\mathcal{R}}{48} \right)}^{1 - \Delta_-} \Big(1 + \mathcal{O} \big( \mathcal{R}^{- \Delta_-/2} \big) \Big) \, \\
B(\mathcal{R}) & \underset{\mathcal{R} <0}{\overset{|\mathcal{R}| \rightarrow \infty}{=}} \frac{\mathcal{R}}{96} \Big( 1+ \log \frac{-\mathcal{R}}{48} \Big) - \frac{\Delta_- (3 - \Delta_-)(2 - \Delta_-)}{96 (1 - \Delta_-)} \, {\left( \frac{-\mathcal{R}}{48} \right)}^{1 - \Delta_-} \Big(1 + \mathcal{O} \big( |\mathcal{R}|^{- \Delta_-/2} \big) \Big) \, .
\end{align}

\subsubsection*{Analysis for $|\mathcal{R}| \rightarrow 0$}
For our ansatz \eqref{eq:ansatz} RG flows with $\mathcal{R}=0$ correspond to solutions where $\f(u)$ interpolates between two extrema of the bulk potential $V$. Without loss of generality, here we can consider such a flow between a UV fixed point at a maximum of $V$ at $\f=\f_{\textsc{uv}}=0$ and a minimum at $\f=\f_{\textsc{ir}}$. In contrast, a RG flow solution with $\mathcal{R} \neq 0$ has its IR end or turning point at a generic point $\f_0$, i.e.~not an extremum. As observed in e.g.~\cite{R,F}, as one moves the IR end or turning point $\f_0$ towards the minimum, the corresponding RG flow solutions exhibit a decreasing value of $|\mathcal{R}|$ with the limit $|\mathcal{R}| \rightarrow 0$ corresponding to the instance when the IR end/ turning point approaches the minimum, i.e.~$\f_0 \rightarrow \f_{\textsc{ir}}$.

Following the ideas in \cite{F} we will use two different expansions to describe a flow for $|\mathcal{R}| \rightarrow 0$, one for the immediate vicinity of the IR end/ turning point and one for the remaining part of the flow all the way to the UV fixed point. In the latter region we can write $A(u), \f(u)$ as expansions in powers of $\ell_{\textsc{uv}}^2 R$ about the flat RG flow solution $A_{\textrm{flat}}(u), \f_{\textrm{flat}}(u)$. This expansion is identical to the near-boundary expansion reviewed in section \ref{sec:holoRGreview}. In the remaining region we exploit the fact that in the vicinity of the IR end/ turning point the bulk geometry
asymptotes to AdS$_{d+1}$ (see \cite{R} for a detailed discussion of this). The expansion there is then about the exact AdS solution associated with the minimum at $\f_{\textsc{ir}}$ and the expansion parameter is $\f_{\star} \equiv \f_{\textsc{ir}} - \f_0$.\footnote{Here we assume that $\f_0 < \f_{\textsc{ir}}$.} The two expansions are thus given by
\begin{align}
\label{app:Aexpflat} A(u) & \underset{u \ll u_0}{=} A_{\textrm{flat}}(u) + \mathcal{O} (\ell_{\textsc{uv}}^{2} R) \, , \\
\f(u) & \underset{u \ll u_0}{=} \f_{\textrm{flat}}(u) + \mathcal{O} (\ell_{\textsc{uv}}^{2} R) \, ,
\end{align}
and
\begin{align}
\label{app:AexpIRAdS} A(u) & \underset{u \rightarrow u_0}{=} A_{0, \textsc{ir}}(u) + \mathcal{O} (\f_{\star}^2) \, , \\
\f(u) & \underset{u \rightarrow u_0}{=} \f_{\textsc{ir}} +  \mathcal{O} (\f_{\star}) \, ,
\end{align}
with
\begin{align}
\label{app:AdSScaleFactorIR}
e^{A_{0, \textsc{ir}} (u)} = \left\{
  \begin{array}{l l l}
   \displaystyle{\frac{\ell_{\textsc{ir}}}{\alpha} \sinh \left(-\frac{u-u_0}{\ell_\textsc{ir}}\right)}, & \qquad
   \mathcal{R} >0 \, , \\
& \\
 \displaystyle{
   \frac{\ell_{\textsc{ir}}}{\alpha} \cosh
   \left(\frac{u-u_0}{\ell_{\textsc{ir}}}\right)}, & \qquad \mathcal{R} < 0 \, , \\
  \end{array} \right. \qquad \textrm{with} \qquad \ell_\textsc{ir}^2 = - \frac{d(d-1)}{V(\f_{\textsc{ir}})} \, .
\end{align}
It will be useful to record explicit expressions for $A_{\textrm{flat}}(u), \f_{\textrm{flat}}(u)$ in the vicinity of the IR fixed point at $\f_{\textsc{ir}}$. As $\f_{\textsc{ir}}$ corresponds to a minimum of $V$, we can write the potential in the vicinity of $\f_{\textsc{ir}}$ as
\begin{align}
V(\f) = - \frac{d(d-1)}{\ell_{\textsc{ir}}^2} - \frac{\Delta_-^{\textsc{ir}} (d- \Delta_-^{\textsc{ir}})}{2 \ell_{\textsc{ir}}^2} \, ( \f_{\textsc{ir}} - \f)^2 + \mathcal{O}\big(( \f_{\textsc{ir}} - \f)^3 \big) \, ,
\end{align}
with $\Delta_-^{\textsc{ir}} < 0$. Inserting this into \eqref{eq:EOM1}--\eqref{eq:EOM3} with $R=0$ one solve for $A_{\textrm{flat}}(u), \f_{\textrm{flat}}(u)$ as
\begin{align}
\label{app:AexpflatIR} A_{\textrm{flat}}(u) & \underset{u \rightarrow \infty}{=} \bar{\bar{A}} - \frac{u}{\ell_{\textsc{ir}}} + \mathcal{O} \big( e^{2 \Delta_-^{\textsc{ir}} u / \ell_{\textsc{ir}}} \big) \, , \\
\label{app:phiexpflatIR} \f_{\textrm{flat}}(u) & \underset{u \rightarrow \infty}{=} \f_{\textsc{ir}} - \bar{\f}_- \, e^{\Delta_-^{\textsc{ir}} u / \ell_{\textsc{ir}}}  + \mathcal{O} \big( e^{2 \Delta_-^{\textsc{ir}} u / \ell_{\textsc{ir}}} \big) \, ,
\end{align}
with $\bar{\bar{A}}, \bar{\f}_-$ integration constants.

We can now obtain useful expressions by matching our two expansions (at leading oder) at an intermediate value $u=u_1$. Here we choose $u_1 \gg \ell_{\textsc{ir}}$, (i.e.~in the IR part of the flow), but also $u_1 \ll u_0$. The latter condition can be understood a posteriori, as we will find that for $|R| \rightarrow 0$ we have $u_0 \rightarrow + \infty$, while we wish $u_1$ to remain finite. From \eqref{app:AexpIRAdS} with \eqref{app:AdSScaleFactorIR} we find
\begin{align}
\left. A(u_1) \right|_{\substack{u_1 \gg \ell_{\textsc{ir}} \\ u_1 \ll u_0}} = \ln \frac{\ell_{\textsc{ir}}}{2 \alpha} + \frac{u_0}{\ell_{\textsc{ir}}} - \frac{u_1}{\ell_{\textsc{ir}}} + \mathcal{O} \big( e^{-2(u_0-u_1) / \ell_{\textsc{ir}}} \big) + \mathcal{O} (\f_{\star}^2) \, .
\end{align}
This is to be matched with \eqref{app:Aexpflat}, where the relevant expression for $A_{\textrm{flat}}(u)$ is given by \eqref{app:AexpflatIR}. Thus we have
\begin{align}
\left. A(u_1) \right|_{\substack{u_1 \gg \ell_{\textsc{ir}} \\ u_1 \ll u_0}} = \bar{\bar{A}} - \frac{u_1}{\ell_{\textsc{ir}}} + \mathcal{O} \big( e^{2 \Delta_-^{\textsc{ir}} u_1) / \ell_{\textsc{ir}}} \big) + \mathcal{O} (\ell_{\textsc{uv}}^2 R) \, .
\end{align}
Matching the leading terms one thus finds
\begin{align}
\label{app:u0IRidentity} \frac{u_0}{\ell_{\textsc{ir}}} = \ln \frac{2 \alpha}{\ell_{\textsc{ir}}} + \bar{\bar{A}} = \frac{1}{2} \ln \frac{4d(d-1)}{\ell_{\textsc{ir}}^2 |R|} + \bar{\bar{A}} \, .
\end{align}
Thus, for $|R| \rightarrow 0$ we indeed find $u_0 \rightarrow \infty$. We can now confirm that an expansion in powers of $\f_{\star}$ is indeed an expansion for small $|R|$. Using $\f_0= \f(u_0)$ and approximating $\f(u)$ by \eqref{app:phiexpflatIR} one finds
\begin{align}
\f_{\star} = \f_{\textsc{ir}} - \f_0 \sim \bar{\f}_- e^{\Delta_-^{\textsc{ir}} u_0 / \ell_{\textsc{ir}}} \sim \big( \ell_{\textsc{ir}}^2 |R| \big)^{- \frac{\Delta_-^{\textsc{ir}}}{2}} \, ,
\end{align}
where in the last step we have used \eqref{app:u0IRidentity}.

We now employ the expansions introduced above to extract information about $B(\mathcal{R})$ and $C(\mathcal{R})$ for $|\mathcal{R}| \rightarrow 0$. As for the case $|\mathcal{R}| \rightarrow \infty$ discussed above, we will read off contributions to $B$ from the function $U$ defined in \eqref{eq:UfromA}. To be specific, we now restrict to $d=4$. Using \eqref{eq:UfromA} the function $U$ is thus given by
\begin{align}
\label{app:UforsmallR} U(u_\L) \overset{d=4}{=} \frac{1}{2} e^{-2 A(u_\L)} \int_{u_\L}^{u_0} d \tilde{u} \, e^{2 A(\tilde{u})} \, .
\end{align}
Here we are free to consider $u_\L \rightarrow - \infty$ in which case we can identify $e^{-2 A(u_\L)} = \ell_{\textsc{uv}}^{-2} \Lambda^{-2}$. The insight from our discussion above is that we can split the integration in \eqref{app:UforsmallR} over $[u_\L, u_0]$ into two integrals over the intervals $[u_\L, u_1]$ and $[u_1, u_0]$. In the first region $[u_\L, u_1]$ we can use the expansion \eqref{app:Aexpflat} for $A(u)$ while in the other region $[u_1, u_0]$ we will employ the expansion \eqref{app:AexpIRAdS}. Thus we write \eqref{app:UforsmallR} as
\begin{align}
\nonumber U(u_\L) &= \frac{1}{2} \ell_{\textsc{uv}}^{-2} \Lambda^{-2} \int_{u_\L}^{u_1} d \tilde{u} \, e^{2 A_{\textrm{flat}}(\tilde{u}) + \mathcal{O}(\ell_{\textsc{uv}}^2 R)} + \frac{1}{2} \ell_{\textsc{uv}}^{-2} \Lambda^{-2} \int_{u_1}^{u_0} d \tilde{u} \, e^{2 A_{0, \textsc{ir}}(\tilde{u}) + \mathcal{O}(\f_{\star}^2)} \\
\nonumber & = \frac{1}{2} \ell_{\textsc{uv}}^{-2} \Lambda^{-2} \int_{u_\L}^{u_1} d \tilde{u} \, e^{2 A_{\textrm{flat}}(\tilde{u})} + \frac{1}{2} \ell_{\textsc{uv}}^{-2} \Lambda^{-2} \int_{u_\L}^{u_1} d \tilde{u} \,  \mathcal{O}(\ell_{\textsc{uv}}^2 R) \\
\nonumber & \hphantom{\overset{d=4}{=}} + \frac{1}{2} \ell_{\textsc{uv}}^{-2} \Lambda^{-2} \int_{u_1}^{u_0} d \tilde{u} \, e^{2 A_{0, \textsc{ir}}(\tilde{u})} + \frac{1}{2} \ell_{\textsc{uv}}^{-2} \Lambda^{-2} \int_{u_1}^{u_0} d \tilde{u} \, e^{2 A_{0, \textsc{ir}}(\tilde{u})} \mathcal{O}(\f_{\star}^2) \\
\label{app:UI1I2I3I4} & = I_1 + I_2 + I_3 +I_4 \, .
\end{align}
We rewrite the first integral $I_1$ in the above further as
\begin{align}
I_1 = \frac{1}{2} \ell_{\textsc{uv}}^{-2} \Lambda^{-2} \int_{u_\L}^{\infty} d \tilde{u} \, e^{2 A_{\textrm{flat}}(\tilde{u})} - \frac{1}{2} \ell_{\textsc{uv}}^{-2} \Lambda^{-2} \int_{u_1}^{\infty} d \tilde{u} \, e^{2 A_{\textrm{flat}}(\tilde{u})} = I_{1a} + I_{1b} \, .
\end{align}
Using our expressions in \eqref{app:AdSScaleFactorIR} we can evaluate $I_3$ explicitly. Here we only show the calculation for $R > 0$, but the analysis for $R < 0$ can be performed analogously. For $R >0$ we thus have
\begin{align}
\nonumber I_3 &= \frac{1}{2} \ell_{\textsc{uv}}^{-2} \Lambda^{-2} \int_{u_1}^{u_0} d \tilde{u} \, e^{2 A_{0, \textsc{ir}}(\tilde{u})} = \frac{1}{2} \ell_{\textsc{uv}}^{-2} \Lambda^{-2} \int_{u_1}^{u_0} d \tilde{u} \, \frac{\ell_{\textsc{ir}}^2}{\alpha^2} \, \sinh^2 \Big( \frac{u-u_0}{\ell_{\textsc{ir}}} \Big) \\
&= \frac{\ell_{\textsc{ir}}^2}{\ell_{\textsc{uv}}^2} \frac{R \Lambda^{-2}}{48} \left[ u_1 -u_0 - \frac{\ell_{\textsc{ir}}}{2} \sinh \Big( 2 \frac{u_1-u_0}{\ell_{\textsc{ir}}} \Big) \right] \, .
\end{align}
Next we consider $I_{1b}$, where we use \eqref{app:AexpflatIR} for $A_{\textrm{flat}}$:
\begin{align}
\nonumber I_{1b} &= - \frac{1}{2} \ell_{\textsc{uv}}^{-2} \Lambda^{-2} \int_{u_1}^{\infty} d \tilde{u} \, e^{2 A_{\textrm{flat}}(\tilde{u})} = - \frac{1}{2} \ell_{\textsc{uv}}^{-2} \Lambda^{-2} e^{2 \bar{\bar{A}}} \int_{u_1}^{\infty} d \tilde{u} \, e^{- \frac{2 u}{\ell_{\textsc{ir}}}} \Big( 1 + \mathcal{O} \big( e^{2 \Delta_-^{\textsc{ir}} u / \ell_{\textsc{ir}}} \big) \Big) \\
& = - \frac{1}{4} \frac{\ell_{\textsc{ir}}}{\ell_{\textsc{uv}}^{2}} \Lambda^{-2} e^{2 \bar{\bar{A}}} \, e^{- \frac{2u_1}{\ell_{\textsc{ir}}}} \Big( 1 + \mathcal{O} \big( e^{2 \Delta_-^{\textsc{ir}} u_1 / \ell_{\textsc{ir}}} \big) \Big) \, .
\end{align}
Adding the results for $I_3$ and $I_{1b}$ and using the expression \eqref{app:u0IRidentity} for $u_0$ we arrive at the following expression:\footnote{Note that $u_1$ is unphysical and will disappear from the full expression for $U$. Thus here we can ignore any terms that come with an explicit dependence on $u_1$.}
\begin{align}
\nonumber I_3 + I_{1b} &=  \frac{\ell_{\textsc{ir}}^3}{\ell_{\textsc{uv}}^2} \frac{R \Lambda^{-2}}{96} \, \ln \frac{\ell_{\textsc{ir}}^2 R}{48} \\
& \hphantom{=} + \frac{\ell_{\textsc{ir}}^3}{\ell_{\textsc{uv}}^2} \frac{R \Lambda^{-2}}{96} \left[ \frac{2 u_1}{\ell_{\textsc{ir}}} +  e^{2 u_1 / \ell_{\textsc{ir}}} \mathcal{O} \big( \ell_{\textsc{ir}}^2 R \big) \right] +  \frac{\ell_{\textsc{ir}}}{\ell_{\textsc{uv}}^{2}} \Lambda^{-2} \mathcal{O} \big( e^{2 (\Delta_-^{\textsc{ir}}-1) u_1 / \ell_{\textsc{ir}}} \big) \, .
\end{align}
We are now in a position to read off contributions to $B$ from $U$. Following the analysis for $|\mathcal{R}| \rightarrow \infty$ we can identify contributions to $B(\mathcal{R})$ with terms at order $\Lambda^{-2}$ in $U$. As $B(\mathcal{R})$ only depends on the curvature $R$ via the dimensionless combination $\mathcal{R}$, we can promote any $R$ appearing to $\mathcal{R}$. Note that every term in \eqref{app:UI1I2I3I4} comes with a factor $\Lambda^{-2}$, so in principle every term will contribute to $B$. We can organize the various contributions according to their $\mathcal{R}$-dependence:
\begin{align}
\nonumber I_{1a}: & \qquad B(\mathcal{R}) \supset B_0 = \textrm{const.} \, , \\
\nonumber I_{2}: & \qquad B(\mathcal{R}) \supset \mathcal{O} (\ell_{\textsc{ir}}^2 R) \sim \mathcal{O} (\mathcal{R}) \, , \\
\nonumber I_{2}: & \qquad B(\mathcal{R}) \supset \ell_{\textsc{ir}}^2 \alpha^{-2} \mathcal{O} (\f_{\star}^2) \sim \mathcal{O} (|\mathcal{R}|^{1- \Delta_-^{\textrm{IR}}})  \, , \\
\nonumber I_3 + I_{1b}: & \qquad B(\mathcal{R}) \supset \frac{\ell_{\textsc{ir}}^3}{\ell_{\textsc{uv}}^3} \frac{\mathcal{R}}{96} \, \ln \frac{|\mathcal{R}|}{48} \, .
\end{align}
The above findings are also valid for $R < 0$ as can be verified explicitly. Thus, we find that for $\mathcal{R} \rightarrow 0$ the quantity $\mathcal{B}(\mathcal{R})$ takes the following form:
\begin{align}
\label{app:BresultsmallR} B(\mathcal{R}) \overset{d=4}{=} B_0 + B_1 \mathcal{R} + \frac{\ell_{\textsc{ir}}^3}{\ell_{\textsc{uv}}^3} \frac{\mathcal{R}}{96} \, \ln \frac{|\mathcal{R}|}{48} + \mathcal{O} \big( \mathcal{R}^2 \big) + \mathcal{O} \big( \mathcal{R}^{1 - \Delta_-^{\textrm{IR}}} \big) \, .
\end{align}
We will now make an educated guess. Recall that for $|\mathcal{R}| \rightarrow \infty$ the coefficient of the terms $\sim \mathcal{R}$ and $\sim \mathcal{R} \ln \tfrac{\mathcal{R}}{48}$ were identical, see \eqref{app:BRlogRlargeR}. Hence one is tempted to conjecture that this may also hold for $|\mathcal{R}| \rightarrow 0$. While we cannot extract $B_1$ analytically here, numerical analysis shows that this indeed holds for all cases employed in this work. Thus, for the purposes of this work we can set
\begin{align}
B_1 = \frac{\ell_{\textsc{ir}}^3}{\ell_{\textsc{uv}}^3} \frac{1}{96} \, .
\end{align}
Last, we turn to $C(\mathcal{R})$ for $|\mathcal{R}| \rightarrow 0$. Given our result for $B(\mathcal{R})$ in \eqref{app:BresultsmallR}, we can then extract $C(\mathcal{R})$ up to a constant using the relation \eqref{eq:CBidapp} between $C(\mathcal{R})$ and $B(\mathcal{R})$. Here we just record the result:
\begin{align}
C(\mathcal{R}) &\overset{d=4}{=} C_0 + C_1 \mathcal{R} + C_2 \mathcal{R}^2 + \mathcal{O} \big( \mathcal{R}^3 \big) + \mathcal{O} \big( \mathcal{R}^{2 - \Delta_-^{\textrm{IR}}} \big) \, , \\
\nonumber \textrm{with} & \qquad C_1 = B_0 \, , \qquad B_1= \frac{\ell^3_{\textsc{ir}}}{\ell_{\textsc{uv}}^3} \frac{1}{96} \, , \qquad C_2 = \frac{1}{192} \Big( 1- \frac{\ell^3_{\textsc{ir}}}{\ell_{\textsc{uv}}^3} \Big) \, .
\end{align}
The constant contribution $C_0$ persists for $\mathcal{R}=0$ and can be identified with the dimensionless vev for the RG flow for a theory on flat space-time.

\subsubsection*{Summary: Results for $d=4$:}

For $|\mathcal{R}| \rightarrow \infty$ one finds:
\begin{align}
C(\mathcal{R}) & \overset{|\mathcal{R}| \rightarrow \infty}{=} \frac{\Delta_-}{2} \, \frac{(3 - \Delta_-)(2 - \Delta_-)}{(1 - \Delta_-)} \, {\left( \frac{|\mathcal{R}|}{48} \right)}^{2 - \Delta_-} \Big(1 + \mathcal{O} \big( |\mathcal{R}|^{- \Delta_-/2} \big) \Big) \, , \\
B(\mathcal{R}) & \overset{|\mathcal{R}| \rightarrow \infty}{=} \frac{\mathcal{R}}{96} \Big( 1+ \log \frac{|\mathcal{R}|}{48} \Big)  + \mathcal{O} \big( |\mathcal{R}|^{1- \Delta_-} \big) \, .
\end{align}

For $|\mathcal{R}| \rightarrow 0$ one finds:
\begin{align}
C(\mathcal{R}) & \overset{|\mathcal{R}| \rightarrow 0}{=} C_0 + C_1 \mathcal{R} + C_2 \mathcal{R}^2 + \mathcal{O} \big( \mathcal{R}^3 \big) + \mathcal{O} \big( |\mathcal{R}|^{2 - \Delta_-^{\textrm{IR}}} \big) \, , \\
B(\mathcal{R}) & \overset{|\mathcal{R}| \rightarrow 0}{=} B_0 + B_1 \mathcal{R} \Big( 1+ \log \tfrac{\mathcal{|R|}}{48} \Big) + \mathcal{O} \big( \mathcal{R}^2 \big) + \mathcal{O} \big( |\mathcal{R}|^{1 - \Delta_-^{\textrm{IR}}} \big) \, , \\
\nonumber \textrm{with} & \qquad C_1 = B_0 \, , \qquad B_1= \frac{\ell^3_{\textsc{ir}}}{\ell_{\textsc{uv}}^3} \frac{1}{96} \, , \qquad C_2 = \frac{1}{192} \Big( 1- \frac{\ell^3_{\textsc{ir}}}{\ell_{\textsc{uv}}^3} \Big) \, .
\end{align}

\section{Higher-derivative terms and constant-curvature solutions}
\label{app:higherderivative}
In this work we study backreaction of a QFT on a 4-dimensional gravitational system described by the action \eqref{eq:S0def4d}, which we here reproduce for convenience:
\begin{align}
\label{eq:S0def4dapp} S_0[g] = \int d^4 x \sqrt{|g|} \, \mathcal{L}_0 = \int d^4 x \sqrt{|g|} \left(\frac{M_{0}^{2}}{2} R - M_0^2 \lambda_0 + \frac{a}{2} R^2  \right)  \, .
\end{align}
Here we have chosen the gravitational theory to contain higher-derivative terms of the form $\mathcal{L}_0 \supset \tfrac{a}{2} R^2$, i.e.~terms which can be written solely in terms of the curvature scalar. The reason for including such a 4-derivative term is that it will be generated anyway by the backreacting QFT. However, at the same level of derivatives $R^2$ is not the only curvature scalar we could have included. To be fully general, to include all curvature-invariants at a 4-derivative level we should have included in the Lagrangian density a term of the form
\begin{align}
\label{eq:higherderapp} \mathcal{L}_0 \supset a_1 R_{\mu \nu \rho \sigma} R^{\mu \nu \rho \sigma} + a_2 R_{\mu \nu} R^{\mu \nu} + a_3 R^2 \, ,
\end{align}
and hence started with a gravitational theory described by
\begin{align}
\label{eq:S0def4dapp2} \overline{S}_{0}[g] &= \int d^4 x \sqrt{|g|} \, \overline{\mathcal{L}}_0 \\
\nonumber &= \int d^4 x \sqrt{|g|} \left(\frac{M_{0}^{2}}{2} R - M_0^2 \lambda_0 + a_1 R_{\mu \nu \rho \sigma} R^{\mu \nu \rho \sigma} + a_2 R_{\mu \nu} R^{\mu \nu} + a_3 R^2  \right)  \, .
\end{align}
As we will argue presently, as long as we are interested in classical solutions to the above theory with constant scalar curvature $R$, the corresponding equations of motion arising from the action \eqref{eq:S0def4dapp} and \eqref{eq:S0def4dapp2} will be identical. The following argument will hold as long as the background space-time does not have any boundaries.

First, we use the fact that for a background without boundaries, adding the combination of curvature-invariants known as the Gauss-Bonnet term does not change the equations of motion. Thus, here add to $\overline{\mathcal{L}}_0$ in \eqref{eq:S0def4dapp2} the combination
\begin{align}
- a_1 \mathcal{L}_G = - a_1 \Big( R_{\mu \nu \rho \sigma} R^{\mu \nu \rho \sigma} -4 R_{\mu \nu} R^{\mu \nu} + R^2 \Big) \, .
\end{align}
This results in a new action, which we will denote by $\overline{\overline{S}}_{0}$, but which has the same classical equations of motion as $\overline{S}_{0}$:
\begin{align}
\label{eq:S0def4dapp3} \overline{\overline{S}}_{0}[g] = \int d^4 x \sqrt{|g|} \left(\frac{M_{0}^{2}}{2} R - M_0^2 \lambda_0 + \bar{\bar{a}}_2 R_{\mu \nu} R^{\mu \nu} + \bar{\bar{a}}_3 R^2  \right)  \, ,
\end{align}
where we also defined
\begin{align}
\bar{\bar{a}}_2 = a_2 + 4 a_1 \, , \qquad \bar{\bar{a}}_3 = a_3 - a_1 \, .
\end{align}
The observation now is that the action $\overline{\overline{S}}_{0}$ as given in \eqref{eq:S0def4dapp3} only differs from $S_0$ in \eqref{eq:S0def4dapp} by the appearance of an additional the term of the form $\sim \bar{\bar{a}}_2 R_{\mu \nu} R^{\mu \nu}$, also identifying $a= \bar{\bar{a}}_3$. Hence the only difference in the equations of motion arising from the two actions $\overline{\overline{S}}_{0}$ and $S_0$ will arise from the variation of the term $\sim \bar{\bar{a}}_2 R_{\mu \nu} R^{\mu \nu}$. This can be shown to be given by (see e.g.~\cite{Woolliams}):
\begin{align}
\label{eq:Riccisquaredvar} \delta \, \bar{\bar{a}}_2 \int d^4x \sqrt{|g|} \, R_{\mu \nu} R^{\mu \nu} = \bar{\bar{a}}_2 \int d^4x \sqrt{|g|} \, \bigg[ & - \frac{1}{2} R_{\rho \sigma} R^{\rho \sigma} g_{\mu \nu} + 2 R_{\rho \mu \sigma \nu} R^{\rho \sigma} \\
\nonumber & - \nabla_\mu \nabla_{\nu} R + \frac{1}{2} g_{\mu \nu} \Box R + \Box R_{\mu \nu} \bigg] \delta g^{\mu \nu} \, ,
\end{align}
where again it has been assumed that boundary terms vanish. Then, restricting to a 4d Einstein space, i.e.~$R_{\mu \nu} = \tfrac{1}{4} R g_{\mu \nu}$, with constant curvature $R$ we make the following observation:
\begin{itemize}
\item The terms on the second line of \eqref{eq:Riccisquaredvar} vanish.
\item The two terms on the first line of \eqref{eq:Riccisquaredvar} cancel one another.
\end{itemize}
Hence we are left with
\begin{align}
\label{eq:Riccisquaredvar2} \delta \, \bar{\bar{a}}_2 \int d^4x \sqrt{|g|} \, R_{\mu \nu} R^{\mu \nu} = 0 \, , \quad \textrm{for} \quad R_{\mu \nu} = \frac{R}{4} g_{\mu \nu} \, , \quad \textrm{with} \quad R = \textrm{const.}
\end{align}
As the variation of the term $\sim \bar{\bar{a}}_2 R_{\mu \nu} R^{\mu \nu}$ does not contribute at all to the equations of motion for constant-curvature solutions, it follows that, as far as constant-curvature solutions are concerned, there is no difference between considering the actions $\overline{\overline{S}}_{0}$ and $S_0$, and hence between $\overline{S}_{0}$ and $S_0$. However, note that this argument assumed that boundary terms do not contribute.



\addcontentsline{toc}{section}{References}

\begin{thebibliography}{99}

\bibitem{AIT}
  I.~Antoniadis, J.~Iliopoulos and T.~N.~Tomaras,
  {\em ``Quantum Instability of De Sitter Space,''}
  \hree{10.1103/PhysRevLett.56.1319}{Phys.\ Rev.\ Lett.\  {\bf 56} (1986) 1319};\\
E.~G.~Floratos, J.~Iliopoulos and T.~N.~Tomaras,
  {\em ``Tree Level Scattering Amplitudes in De Sitter Space Diverge,''}
  \hree{10.1016/0370-2693(87)90403-5}{Phys.\ Lett.\ B {\bf 197} (1987) 373}.


  \bibitem{AM}
  I.~Antoniadis and E.~Mottola,
  {\em ``Graviton Fluctuations in De Sitter Space,''}
  \hree{10.1063/1.529381}{J.\ Math.\ Phys.\  {\bf 32} (1991) 1037}.

\bibitem{Sasa}
  M.~Sasaki, H.~Suzuki, K.~Yamamoto and J.~Yokoyama,
  {\em ``Superexpansionary divergence: Breakdown of perturbative quantum field theory in space-time with accelerated expansion,''}
 \href{https://doi.org/10.1088/0264-9381/10/5/003}{Class.\ Quant.\ Grav.\  {\bf 10} (1993) L55};\\
{\em ``Probability distribution functional for equal time correlation functions in curved space,''}
  \href{https://doi.org/10.1142/S0217751X9400011X}{Int.\ J.\ Mod.\ Phys.\ A {\bf 9} (1994) 221}.


  \bibitem{TW2}
  N.~C.~Tsamis and R.~P.~Woodard,
  {\em ``The Physical basis for infrared divergences in inflationary quantum gravity,''}
  \hree{10.1088/0264-9381/11/12/012}{Class.\ Quant.\ Grav.\  {\bf 11} (1994) 2969}.

 \bibitem{Staro}
  A.~A.~Starobinsky,
  {\em ``Stochastic De Sitter (inflationary) Stage In The Early Universe,''}
  \hree{10.1007/3-540-16452-9\_6}{Lect.\ Notes Phys.\  {\bf 246} (1986) 107};\\
  A.~A.~Starobinsky and J.~Yokoyama,
  {\em ``Equilibrium state of a selfinteracting scalar field in the De Sitter background,''}
\hree{10.1103/PhysRevD.50.6357}{Phys.\ Rev.\ D {\bf 50} (1994) 6357}
\hre{astro-ph}{9407016}.

  \bibitem{STW}
  N.~C.~Tsamis and R.~P.~Woodard,
  {\em ``Stochastic quantum gravitational inflation,''}
  \hree{doi:10.1016/j.nuclphysb.2005.06.031}{Nucl.\ Phys.\ B {\bf 724} (2005) 295},
  \hre{gr-qc}{0505115}.

  \bibitem{GS}
  V.~Gorbenko and L.~Senatore,
  {\em ``$\lambda \phi^4$ in dS,''}
\hri{1911.00022}{[hep-th]}.

  \bibitem{Marolf}
  D.~Marolf and I.~A.~Morrison,
  {\em ``The IR stability of de Sitter QFT: results at all orders,''}
  \hree{doi:10.1103/PhysRevD.84.044040}{Phys.\ Rev.\ D {\bf 84} (2011) 044040},
  \hri{1010.5327}{[gr-qc]}.

  \bibitem{Hollands2}
  S.~Hollands,
  {\em ``Correlators, Feynman diagrams, and quantum no-hair in deSitter spacetime,''}
  \hree{10.1007/s00220-012-1653-2}{Commun.\ Math.\ Phys.\  {\bf 319} (2013) 1},
  \hri{1010.5367}{[gr-qc]}.



  \bibitem{Mukha}
  V.~F.~Mukhanov, L.~R.~W.~Abramo and R.~H.~Brandenberger,
  {\em ``On the Back reaction problem for gravitational perturbations,''}
  \hree{10.1103/PhysRevLett.78.1624}{Phys.\ Rev.\ Lett.\  {\bf 78} (1997) 1624},
  \hre{gr-qc}{9609026};\\
  {\em ``The Energy - momentum tensor for cosmological perturbations,''}
  \hree{10.1103/PhysRevD.56.3248}{Phys.\ Rev.\ D {\bf 56} (1997) 3248},
  \hre{gr-qc}{9704037}.

  \bibitem{AWoo}
  L.~R.~W.~Abramo and R.~P.~Woodard,
  {\em ``One loop back reaction on chaotic inflation,''}
  \hree{10.1103/PhysRevD.60.044010}{Phys.\ Rev.\ D {\bf 60} (1999) 044010},
  \hre{astro-ph}{9811430}.

  \bibitem{unruh}
  B.~Losic and W.~G.~Unruh,
  {\em ``Long-wavelength metric backreactions in slow-roll inflation,''}
  \hree{10.1103/PhysRevD.72.123510}{Phys.\ Rev.\ D {\bf 72} (2005) 123510},
  \hre{gr-qc}{0510078};\\
  {\em ``Cosmological Perturbation Theory in Slow-Roll Spacetimes,''}
  \hree{10.1103/PhysRevLett.101.111101}{Phys.\ Rev.\ Lett.\  {\bf 101} (2008) 111101},
  \hri{0804.4296}{[gr-qc]}.






\bibitem{Mottola}
E.~Mottola,
  {\em ``Particle Creation in de Sitter Space,''}
\hree{10.1103/PhysRevD.31.754}{Phys.\ Rev.\ D {\bf 31} (1985) 754};\\
  E.~Mottola,
  {\em ``Thermodynamic Instability Of De Sitter Space,''}
  \hree{10.1103/PhysRevD.33.1616}{Phys.\ Rev.\ D {\bf 33} (1986) 1616}.






  \bibitem{TW}
  N.~C.~Tsamis and R.~P.~Woodard,
  {\em ``Quantum gravity slows inflation,''}
\hree{10.1016/0550-3213(96)00246-5}{Nucl.\ Phys.\ B {\bf 474} (1996) 235},
  \hre{hep-ph}{9602315};\\
  {\em ``The Quantum gravitational back reaction on inflation,''}
  \hree{10.1006/aphy.1997.5613}{Annals Phys.\  {\bf 253} (1997) 1},
  \hre{hep-ph}{9602316}.

  \bibitem{Polyakov}
  A.~M.~Polyakov,
  {\em ``Infrared instability of the de Sitter space,''}
  \hri{1209.4135}{[hep-th]}.



  \bibitem{OW}
  V.~K.~Onemli and R.~P.~Woodard,
  {\em ``Quantum effects can render $w < -1$ on cosmological scales,''}
  \hree{doi:10.1103/PhysRevD.70.107301}{Phys.\ Rev.\ D {\bf 70} (2004) 107301},
  \hre{gr-qc}{0406098}.



   \bibitem{MM}
P.~Mazur and E.~Mottola,
  {\em ``Spontaneous Breaking of De Sitter Symmetry by Radiative Effects,''}
   \hree{10.1016/0550-3213(86)90058-1}{Nucl.\ Phys.\ B {\bf 278} (1986) 694.}








  \bibitem{BAllen}
  B.~Allen,
  {\em ``Vacuum States in de Sitter Space,''}
  \hree{10.1103/PhysRevD.32.3136}{Phys.\ Rev.\ D {\bf 32} (1985) 3136};\\
  B.~Allen and A.~Folacci,
  {\em ``The Massless Minimally Coupled Scalar Field in De Sitter Space,''}
  \hree{10.1103/PhysRevD.35.3771}{Phys.\ Rev.\ D {\bf 35} (1987) 3771.}


  \bibitem{Hollands1}
  S.~Hollands,
  {\em ``Massless interacting quantum fields in deSitter spacetime,''}
  \hree{10.1007/s00023-011-0140-1}{Annales Henri Poincare {\bf 13} (2012) 1039},
  \hri{1105.1996}{[gr-qc]}.

  \bibitem{Weinberg}
  S.~Weinberg,
  {\em ``Quantum contributions to cosmological correlations,''}
  \hree{10.1103/PhysRevD.72.043514}{Phys.\ Rev.\ D {\bf 72} (2005) 043514},
  \hre{hep-th}{0506236};\\
  {\em ``Quantum contributions to cosmological correlations. II. Can these corrections become large?,''}
  \hree{10.1103/PhysRevD.74.023508}{Phys.\ Rev.\ D {\bf 74} (2006) 023508}
  \hre{hep-th}{0605244}.

  \bibitem{Cha}
  K.~Chaicherdsakul,
  {\em ``Quantum Cosmological Correlations in an Inflating Universe: Can fermion and gauge fields loops give a scale free spectrum?,''}
 \hree{10.1103/PhysRevD.75.063522}{Phys.\ Rev.\ D {\bf 75} (2007) 063522}
  \hre{hep-th}{0611352}.

  \bibitem{Sloth}
  A.~Riotto and M.~S.~Sloth,
  {\em ``On Resumming Inflationary Perturbations beyond One-loop,''}
  \hree{10.1088/1475-7516/2008/04/030}{JCAP {\bf 0804} (2008) 030},
  \hri{0801.1845}{[hep-ph]}.

  \bibitem{Sena}
  L.~Senatore and M.~Zaldarriaga,
  {\em ``On Loops in Inflation,''}
  \hree{10.1007/JHEP12(2010)008}{JHEP {\bf 1012} (2010) 008},
  \hri{0912.2734}{[hep-th]};\\
  {\em ``On Loops in Inflation II: IR Effects in Single Clock Inflation,''}
 \hree{10.1007/JHEP01(2013)109}{JHEP {\bf 1301} (2013) 109},
  \hri{1203.6354}{[hep-th]};\\
  {\em ``The constancy of $\zeta$ in single-clock Inflation at all loops,''}
  \hree{10.1007/JHEP09(2013)148}{JHEP {\bf 1309} (2013) 148}
  \hri{1210.6048}{[hep-th]};\\
  G.~L.~Pimentel, L.~Senatore and M.~Zaldarriaga,
  {\em ``On Loops in Inflation III: Time Independence of zeta in Single Clock Inflation,''}
  \hree{10.1007/JHEP07(2012)166}{JHEP {\bf 1207} (2012) 166},
  \hri{1203.6651}{[hep-th]}.




  \bibitem{Seery}
  D.~Seery,
  {\em ``Infrared effects in inflationary correlation functions,''}
  \hree{10.1088/0264-9381/27/12/124005}{Class.\ Quant.\ Grav.\  {\bf 27} (2010) 124005},
  \hri{1005.1649}{[astro-ph.CO]}.

  \bibitem{julien}
  G.~Moreau and J.~Serreau,
  {\em ``Stability of de Sitter spacetime against infrared quantum scalar field fluctuations,''}
 \hree{10.1103/PhysRevLett.122.011302}{Phys.\ Rev.\ Lett.\  {\bf 122} (2019) no.1,  011302},
  \hri{1808.00338}{[hep-th]};\\
  {\em ``Backreaction of superhorizon scalar field fluctuations on a de Sitter geometry: A renormalization group perspective,''}
  \hree{10.1103/PhysRevD.99.025011}{Phys.\ Rev.\ D {\bf 99} (2019) no.2,  025011},
  \hri{1809.03969}{[h.ep-th]}.



  \bibitem{R}
  J.~K.~Ghosh, E.~Kiritsis, F.~Nitti and L.~T.~Witkowski,
  {\em ``Holographic RG flows on curved manifolds and quantum phase transitions,''}
  \hree{10.1007/JHEP05(2018)034}{JHEP {\bf 1805} (2018) 034},
  \hri{1711.08462}{[hep-th]}.

  \bibitem{F}
  J.~K.~Ghosh, E.~Kiritsis, F.~Nitti and L.~T.~Witkowski,
  {\em ``Holographic RG flows on curved manifolds and the $F$-theorem,''}
  \hree{10.1007/JHEP02(2019)055}{JHEP {\bf 1902} (2019) 055},
  \hri{1810.12318}{[hep-th]}.

  \bibitem{book}
  E.~Kiritsis,
  {\em ``String theory in a nutshell,''}
  second edition, \href{https://press.princeton.edu/titles/13376.html}{Princeton University Press, 2019}

  \bibitem{fR}
  T.~P.~Sotiriou and V.~Faraoni,
  {\em ``f(R) Theories Of Gravity,''}
  \hree{10.1103/RevModPhys.82.451}{Rev.\ Mod.\ Phys.\  {\bf 82} (2010) 451},
  \hri{0805.1726}{[gr-qc]};\\
  A.~De Felice and S.~Tsujikawa,
  {\em ``f(R) theories,''}
  \hree{10.12942/lrr-2010-3}{Living Rev.\ Rel.\  {\bf 13} (2010) 3},
  \hri{1002.4928}{[gr-qc]};\\
  S.~Nojiri and S.~D.~Odintsov,
  {\em ``Unified cosmic history in modified gravity: from F(R) theory to Lorentz non-invariant models,''}
\hree{10.1016/j.physrep.2011.04.001}{  Phys.\ Rept.\  {\bf 505} (2011) 59},
\hri{1011.0544}{[gr-qc]}.

  \bibitem{Obied}
  U.~H.~Danielsson and T.~Van Riet,
  {\em ``What if string theory has no de Sitter vacua?,''}
  \hree{10.1142/S0218271818300070}{Int.\ J.\ Mod.\ Phys.\ D{\bf 27} (2018) no.12, 1830007},
  \hri{1804.01120}{[hep-th]};\\
  G.~Obied, H.~Ooguri, L.~Spodyneiko and C.~Vafa,
  {\em ``De Sitter Space and the Swampland,''}
  \hri{1806.08362}{[hep-th]}.

  \bibitem{duff}
  M.~J.~Duff,
  {\em ``Twenty years of the Weyl anomaly,''}
  Class.\ Quant.\ Grav.\  {\bf 11} (1994) 1387
  doi:10.1088/0264-9381/11/6/004
  \hre{hep-th}{9308075}.

  \bibitem{St}
  A.~A.~Starobinsky,
  {\em ``A New Type of Isotropic Cosmological Models Without Singularity,''}
  \hree{10.1016/0370-2693(80)90670-X}{Phys.\ Lett.\ B {\bf 91} (1980) 99},
   {Adv.\ Ser.\ Astrophys.\ Cosmol.\  {\bf 3} (1987) 130}.

   \bibitem{BB}
  E.~Kiritsis,
  {\em ``Holography and brane-bulk energy exchange,''}
  \hree{10.1088/1475-7516/2005/10/014}{JCAP {\bf 0510} (2005) 014},
  \hre{hep-th}{0504219}.


\bibitem{2003.05501}
  P.~M.~Chesler and A.~Loeb,
  {\em ``Holographic duality and mode stability of de Sitter space in semiclassical gravity,''}
  \hri{2003.05501}{[hep-th]}.


  \bibitem{1106.4826}
  I.~Papadimitriou,
  {\em ``Holographic Renormalization of general dilaton-axion gravity,''}
  \hree{10.1007/JHEP08(2011)119}{JHEP {\bf 1108} (2011) 119},
  \hri{1106.4826}{[hep-th]}.

  \bibitem{SkenderisTownsend1}
  K.~Skenderis and P.~K.~Townsend,
  {\em ``Hidden supersymmetry of domain walls and cosmologies,''}
  \hree{10.1103/PhysRevLett.96.191301}{Phys.\ Rev.\ Lett.\  {\bf 96} (2006), 191301},
  \hre{hep-th}{0602260}.

  \bibitem{SkenderisTownsend2}
  K.~Skenderis and P.~K.~Townsend,
  {\em ``Hamilton-Jacobi method for curved domain walls and cosmologies,''}
  \hree{10.1103/PhysRevD.74.125008}{Phys.\ Rev.\ D {\bf 74} (2006), 125008},
  \hre{hep-th}{0609056}.

  \bibitem{KarchRandall}
  A.~Karch and L.~Randall,
  {\em ``Locally localized gravity,''}
  \hree{10.1088/1126-6708/2001/05/008}{JHEP {\bf 0105} (2001) 008},
  \hre{hep-th}{0011156}.

\bibitem{exotic}
E.~Kiritsis, F.~Nitti and L.~Silva Pimenta,
  {\em ``Exotic RG Flows from Holography,''}
  Fortsch.\ Phys.\  {\bf 65}, no. 2, 1600120 (2017)
  doi:10.1002/prop.201600120,
  \hri{1611.05493}{[hep-th]}.

  \bibitem{1401.0888}
  E.~Kiritsis, W.~Li and F.~Nitti,
  {\em ``Holographic RG flow and the Quantum Effective Action,''}
  \hree{10.1002/prop.201400007}{Fortsch.\ Phys.\ {\bf 62} (2014) 389-454},
  \hri{1401.0888}{[hep-th]}.

\bibitem{VilenkinStaro}
  A.~Vilenkin,
  {\em ``Classical and Quantum Cosmology of the Starobinsky Inflationary Model,''}
  \hree{10.1103/PhysRevD.32.2511}{Phys.\ Rev.\ D {\bf 32} (1985) 2511}.

\bibitem{anomalyinflation1}
  S.~W.~Hawking, T.~Hertog and H.~S.~Reall,
  {\em ``Trace Anomaly Driven Inflation,''}
  \hree{10.1103/PhysRevD.63.083504}{Phys.\ Rev.\ D {\bf 63} (2001) 083504},
  \hre{hep-th}{0010232}.

\bibitem{anomalyinflation2}
  J.~C.~Fabris, A.~M.~Pelinson and I.~L.~Shapiro,
  {\em ``Anomaly-induced Effective Action and Inflation,''}
  \hree{10.1016/S0920-5632(01)01060-X}{Nucl.\ Phys.\ Proc.\ Suppl.\  {\bf 95} (2001) 78},
  \hre{hep-th}{0011030}.

  \bibitem{anomalyinflation3}
  I.~L.~Shapiro,
  {\em ``An Overview of the Anomaly-induced Inflation,''}
  \hree{10.1016/S0920-5632(03)02431-9}{Nucl.\ Phys.\ Proc.\ Suppl.\  {\bf 127} (2004) 196},
  \hre{hep-ph}{0311307}.

  \bibitem{0612068}
  I.~Antoniadis, P.~O.~Mazur and E.~Mottola,
  {\em ``Cosmological dark energy: Prospects for a dynamical theory,''}
  \hree{10.1088/1367-2630/9/1/011}{New J.\ Phys.\ {\bf 9} (2007) 11},
  \hre{gr-qc}{0612068}.

 \bibitem{Koksma1}
  J.~F.~Koksma and T.~Prokopec,
  {\em ``The Effect of the Trace Anomaly on the Cosmological Constant,''}
  \hree{10.1103/PhysRevD.78.023508}{Phys.\ Rev.\ D {\bf 78} (2008) 023508},
  \hri{0803.4000}{[hep-th]}.

  \bibitem{anomalyinflation4}
  T.~Netto, A.~M.~Pelinson, I.~L.~Shapiro and A.~A.~Starobinsky,
  {\em ``From stable to unstable anomaly-induced inflation,''}
  \hree{10.1140/epjc/s10052-016-4390-4}{Eur.\ Phys.\ J.\ {\bf C76} (2016) 544},
  \hri{1509.08882}{[hep-th]}.

  \bibitem{Smolkin}
  O.~Ben-Ami, D.~Carmi and M.~Smolkin,
  {\em ``Renormalization group flow of entanglement entropy on spheres,''}
  \hree{10.1007/JHEP08(2015)048}{JHEP {\bf 1508} (2015) 048},
  \hri{1504.00913}{[hep-th]}.

  \bibitem{Birrell:1982ix}
  N.~D.~Birrell and P.~C.~W.~Davies,
  {\em ``Quantum Fields in Curved Space,''}
  \hree{10.1017/CBO9780511622632}{Cambridge Univ.\ Press, Cambridge, UK, 1982}.

  \bibitem{1404.7349}
  I.~Ben-Dayan, S.~Jing, M.~Torabian, A.~Westphal and L.~Zarate,
  {\em ``$R^2 \log R$ quantum corrections and the inflationary observables,''}
  \hree{10.1088/1475-7516/2014/09/005}{JCAP {\bf 1409} (2014) 005},
  \hri{1404.7349}{[hep-th]}.

  \bibitem{1604.06809}
  M.~Taylor and W.~Woodhead,
  {\em ``The holographic F theorem,''}
  \hri{1604.06809}{[hep-th]}.

  \bibitem{Woolliams}
  S.~A.~Woolliams,
  {\em ``Higher Derivative Theories of Gravity,''}
  \href{https://www.imperial.ac.uk/media/imperial-college/research-centres-and-groups/theoretical-physics/msc/dissertations/2013/LI_INCOMPACT3D2014.pdf}{Master's thesis, Imperial College London, 2013}





\end{thebibliography}
\end{document}